\documentclass[twocolumn]{aastex631}
\usepackage{amsmath}

\begin{document}

\title{Radial Velocity and Astrometric Evidence for a Close Companion to Betelgeuse}

\author[0000-0002-1417-8024]{Morgan MacLeod}
\affiliation{Center for Astrophysics $\vert$ Harvard $\&$ Smithsonian 60 Garden Street,  Cambridge, MA 02138, USA}
\affiliation{Institute for Theory and Computation}

\author[0000-0002-3199-2888]{Sarah Blunt}
\affiliation{Department of Astronomy \& Astrophysics, University of California, Santa Cruz, Santa Cruz, CA, USA, 95060}
\affiliation{Center for Interdisciplinary Exploration and Research in Astrophysics (CIERA), Northwestern University, Evanston, IL 60208, USA}
\affiliation{NSF Astronomy and Astrophysics Postdoctoral Fellow}

\author[0000-0002-4918-0247]{Robert J. De Rosa}
\affiliation{European Southern Observatory, Alonso de C\'{o}rdova 3107, Vitacura, Santiago, Chile}

\author[0000-0002-8985-8489]{Andrea K. Dupree}
\affiliation{Center for Astrophysics $\vert$ Harvard $\&$ Smithsonian 60 Garden Street Cambridge, MA 02138, USA}

\author[0000-0002-5310-1521]{Thomas Granzer}
\affiliation{Leibniz-Institut für Astrophysik Potsdam (AIP), An der Sternwarte 16, 14482, Potsdam, Germany}

\author[0000-0002-7042-4541]{Graham M. Harper}
\affiliation{Center for Astrophysics and Space Astronomy (CASA), University of Colorado, Boulder, CO 80309-0389, USA}

\author[0000-0001-6169-8586]{Caroline D. Huang}
\affiliation{Center for Astrophysics $\vert$ Harvard $\&$ Smithsonian 60 Garden Street Cambridge, MA 02138, USA}
\affiliation{NSF Astronomy and Astrophysics Postdoctoral Fellow}

\author[0000-0002-3944-8406]{Emily M. Leiner}
\affiliation{Department of Physics, Illinois Institute of Technology, Chicago, IL 60616, USA}
\affiliation{Center for Interdisciplinary Exploration and Research in Astrophysics (CIERA), Northwestern University, 1800 Sherman Avenue, Evanston, IL 60201, USA}

\author[0000-0003-4330-287X]{Abraham Loeb}
\affiliation{Center for Astrophysics $\vert$ Harvard $\&$ Smithsonian 60 Garden Street, MS-16, Cambridge, MA 02138, USA}
\affiliation{Institute for Theory and Computation}

\author[0000-0001-6975-9056]{Eric L. Nielsen}
\affiliation{Department of Astronomy, New Mexico State University, P.O. Box 30001, MSC 4500, Las Cruces, NM 88003, USA}

\author[0000-0002-6192-6494]{Klaus G. Strassmeier}
\affiliation{Leibniz-Institut für Astrophysik Potsdam (AIP), An der Sternwarte 16, 14482, Potsdam, Germany}

\author[0000-0003-0774-6502]{Jason J. Wang}
\affiliation{Center for Interdisciplinary Exploration and Research in Astrophysics (CIERA), Northwestern University, Evanston, IL 60208, USA}
\affiliation{Department of Physics and Astronomy, Northwestern University, Evanston, IL 60208, USA; Department of Astronomy, California Institute of Technology, Pasadena, CA 91125, USA}

\author[0000-0002-7032-2967]{Michael Weber}
\affiliation{Leibniz-Institut für Astrophysik Potsdam (AIP), An der Sternwarte 16, 14482, Potsdam, Germany}

\begin{abstract}
We examine a century of radial velocity, visual magnitude, and astrometric observations of the nearest red supergiant, Betelgeuse, in order to reexamine the century-old assertion that Betelgeuse might be a spectroscopic binary. These data reveal Betelgeuse varying stochastically over years and decades due to its boiling, convective envelope, periodically with a $ 5.78$~yr long secondary period, and quasi-periodically from pulsations with periods of several hundred days. We show that the long secondary period is consistent between astrometric and RV datasets, and argue that it indicates a low-mass companion to Betelgeuse, less than a solar mass, orbiting in a 2,110 day period at a separation of just over twice Betelgeuse's radius. The  companion star would be nearly twenty times less massive and a million times fainter than Betelgeuse, with similar effective temperature, effectively hiding it in plain sight near one of the best-studied stars in the night sky.  The astrometric data favor an edge-on binary with orbital plane aligned with Betelgeuse's measured spin axis.  Tidal spin-orbit interaction drains angular momentum from the orbit and spins up Betelgeuse,  explaining the spin--orbit alignment and Betelgeuse's anomalously rapid spin. In the future, the orbit will decay until the companion is swallowed by Betelgeuse in the next 10,000 years. 
\end{abstract}


\section{Introduction}

Betelgeuse, $\alpha$ Orionis, is the nearest red supergiant, and the brightest source in the infrared sky. Its distinctive location and luminosity in the night sky means that human records of the star stretch back tens of thousands of years \citep[e.g.][]{2014JAHH...17..180L,2018AuJAn..29...89H,2018JAHH...21....7S,2022MNRAS.516..693N,2023A&G....64.1.38N,2023mgm..conf.3519S}. Perhaps because of the sustained attention the star has received, it is also deeply enigmatic \citep{2023A&G....64.3.11W}. 

We know now that Betelgeuse's several astronomical-unit radius dwarfs our own inner solar system \citep{1921ApJ....53..249M,1996ApJ...463L..29G,2014A&A...572A..17M,2017ars..book.....L}. Indeed, Betelgeuse's extreme size has rendered its distance difficult to measure because the annual parallax due to Earth's orbit is smaller than its resolvable size on the sky. Betelgeuse is also vigorously convective \citep[e.g.][]{1998Natur.392..575L,2008AJ....135.1450G}, boiling with ever changing hot spots and eddies probed, for example, by spatially resolved imaging \citep{1996ApJ...463L..29G,2009A&A...508..923H,2018A&A...609A..67K} and spectroscopy \citep{2000ApJ...545..454L,2020ApJ...899...68D} and modeling of photocenter and RV variations \citep{2012JCoPh.231..919F,2017A&A...600A.137F,2022ApJ...929..156G,2023A&A...669A..49A}. Betelgeuse pulsates radially, varying in V-band brightness by several tenths of a magnitude with a roughly $\sim 400$~d primary periodicity \citep{2020ApJ...902...63J,2023NewA...9901962J} and in temperature by $\sim 30$~K \citep{2020ApJ...905...34H,2022JAVSO..50..205W}. In addition, visual brightness estimates show evidence of a $\sim 2000$~d variation, also observed in the system's radial velocity, driving the suggestion that it might be a spectroscopic binary \citep{1908PASP...20..227P,1911AN....187...33B,1916ApJ....44..250L,1928MNRAS..88..660S}. Indeed, $\alpha$ Orionis was included in the Lick Observatory's Second Catalog of Spectroscopic Binary Stars  -- or it was intended to be, and was included a year later in an erratum \citep{1910LicOB...6...17C,1911LicOB...6..154C}. 

This longer-period variability is among the many known ``Long Secondary Periods" (LSPs) of unknown origin in other red supergiants \citep[e.g.][]{1963AJ.....68..253H,1999IAUS..191..151W,2003ApJ...584.1035O,2004ApJ...604..800W,2006ApJ...650L..55D,2006MNRAS.372.1721K,2007ApJ...660.1486S,2009JRASC.103...11P,2009MNRAS.399.2063N,2009ApJ...707..573W,2010AJ....139.1909N,2010ApJ...725.1170S,2014ApJ...788...13S,2015MNRAS.448..464T,2015MNRAS.452.3863S,2016JAVSO..44...94P,2020MNRAS.492.1348T,2021ApJ...911L..22S,2021A&A...649A.110P,2023JAVSO..51..237P,2023StGal...6....6T,2024A&A...682A..88P}. The mechanism behind the LSPs has been a source of debate because their periods, being longer than that of the fundamental mode of oscillation, appear to rule out pulsation. Other suggestions have included the overturn of convective cells \citep[e.g.][]{2010ApJ...725.1170S}, a pulsation in the outermost radiative layers \citep{2015MNRAS.452.3863S,2020MNRAS.492.1348T}  or binary companions \citep{1999IAUS..191..151W,2004ApJ...604..800W,2007ApJ...660.1486S,2014ApJ...788...13S,2021ApJ...911L..22S}, with some discussion over whether the LSP could be a mistaken fundamental mode \citep{2023MNRAS.526.2765S,2023RNAAS...7..119M}. Particularly compelling evidence has been offered by a recent multi-wavelength survey of many red supergiants by \citet{2021ApJ...911L..22S} using a combination of optical data from OGLE with infrared data from WISE. \citet{2021ApJ...911L..22S} show that many red supergiants exhibiting LSPs in the optical show evidence of secondary eclipses in the infrared, demonstrating that the optical variability is driven by the presence of a close companion -- perhaps through its concentration of the star's dusty wind. 

In this paper, we analyze historic radial velocity and visual magnitude measurements to trace Betelgeuse's LSP. We demonstrate that the LSP is also observed as on-sky motion in astrometric data. 
Taken together, these lines of evidence reveal a close, and much lower mass, binary companion to Betelgeuse. The companion orbits with a period of $\sim 2100$~d, affecting Betelgeuse's radial velocity, position on the sky, visual magnitude, and -- through tidal interaction -- spin. 

This companion is revealed by the past century of monitoring of Betelgeuse; appreciating its presence moves us a step closer to unwrapping the enigmatic behavior of our nearest red supergiant. Indeed, the hypothesis that Betelgeuse is a spectroscopic binary is not new \citep{1908PASP...20..227P,1911LicOB...6..154C,1916ApJ....44..250L,1928MNRAS..88..660S,1933ApJ....77..110S}, but fell out of favor because of the complexity of the star's stochastic variability \citep[e.g.][]{1984PASP...96..366G}.  As we were preparing this manuscript \citet{2024arXiv240809089G} independently reached the conclusion that Betelgeuse hosts a low-mass companion. Our complimentary analyses take different approaches yet reach the same result: \citet{2024arXiv240809089G} review Betelgeuse's LSP in the more general context of LSPs in optical and radial velocity data, and rule out other possible LSP mechanisms. Our paper focuses on reanalyzing Betelgeuse's long observational history to test the hypothesis that a companion object exists.

In Section \ref{sec:RV} we describe the radial velocity and visual magnitude datasets, and describe modeling these data to determine possible properties of an unseen companion object. In Section \ref{sec:astrometry} we describe the astrometric data available for Betelgeuse and conduct a parallel analysis. In Section \ref{sec:LF} we describe the possible properties of a companion star, and how this star might be currently affecting Betelgeuse through tidal interaction. In Section \ref{sec:discussion} we discuss possible next steps and in Section \ref{sec:conclusions} we conclude.

\section{LSP in Photometry and Radial Velocity}\label{sec:RV}
\subsection{Datasets}

\subsubsection{Visual Magnitudes}
Visual magnitudes of Betelgeuse are recorded in the American Association of Variable Star Observers (AAVSO) database reaching back one hundred years to approximately 1920. Photoelectric photometry in the V-band from a variety of sources is also available spanning a similar period \citep[e.g.][]{2024arXiv240809089G}. Though earlier observers certainly noted Betelgeuse's changing brightness \citep[e.g.][]{1869ABSBo...7..315A,1913AN....194...81O,1913PA.....21....5S}, we restrict ourselves to the AAVSO database for the current analysis. We bin individual reports into two-week averages. In general, visual magnitudes come with a large $\sim 1$~mag observer-to-observer dispersion, but can be quite precise when averaged in large numbers.

In more recent decades, multicolor photometry of Betelgeuse has been collected \citep[e.g.][]{2022OEJV..233....1O,2022NatAs...6..930T}. Analyzing possible color changes and or IR-Vis correlations will be an intriguing path for follow up study, but is not considered at present, favoring the longer-span of the visual magnitudes. 

\subsubsection{Radial Velocities}

We have collected published radial velocity (RV) measurements derived from metallic absorption lines in Betelgeuse’s spectrum. These records span 1896 – 2024, representing a remarkable 128 year baseline, and the work of many observers. With this baseline we are able separate stochastic variability caused, for example, by Betelgeuse's convection, from a periodic signal associated with the LSP.   The raw spacings between nearest observations vary hugely -- from a few minutes in the STELLA dataset to nearly 7 years around 1950. In order to mitigate some differences in reporting (for example, whether several spectrograms were averaged into a single reported RV or not), we employ 10-day bins of RV points, which leaves us with a total of 795 RV samples.  We performed parallel analyses with 30 and 365 day bins of data and found that these choices didn't substantively affect our conclusions. An annotated list of our data sources is given in Appendix \ref{sec:RVsources}. RV points include either reported or estimated errors as detailed in  Appendix \ref{sec:RVsources}.

\subsection{The Long Secondary Period}

\begin{figure*}
    \centering
    \includegraphics[width=\textwidth]{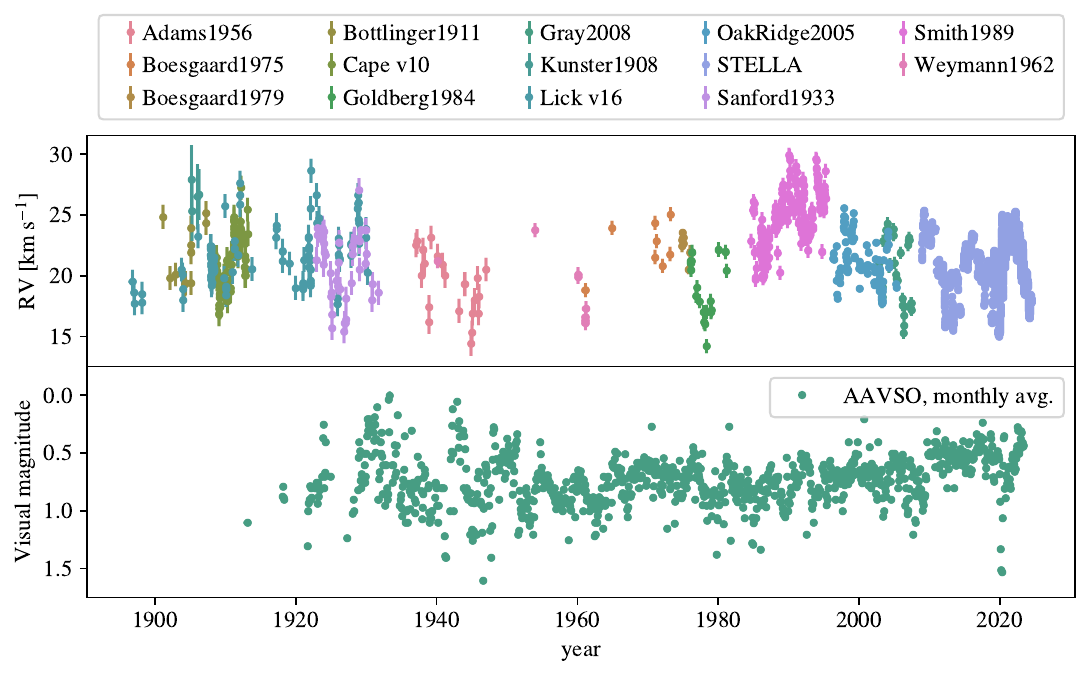}
    \caption{More than a century of observations of Betelgeuse reveal multiple trends. This figure summarizes the history and sources of RV measurements and visual magnitude estimates reported to the AAVSO. RV points are colored by their data source, with a full compilation given in Appendix \ref{sec:RVsources}. Observed across the past century, Betelgeuse varies at a range of timescales from days to decades. Some of this variability appears periodic, while other portions appear uncorrelated. }
    \label{fig:rvsummary}
\end{figure*}

\begin{figure}
    \centering
    \includegraphics[width=\linewidth]{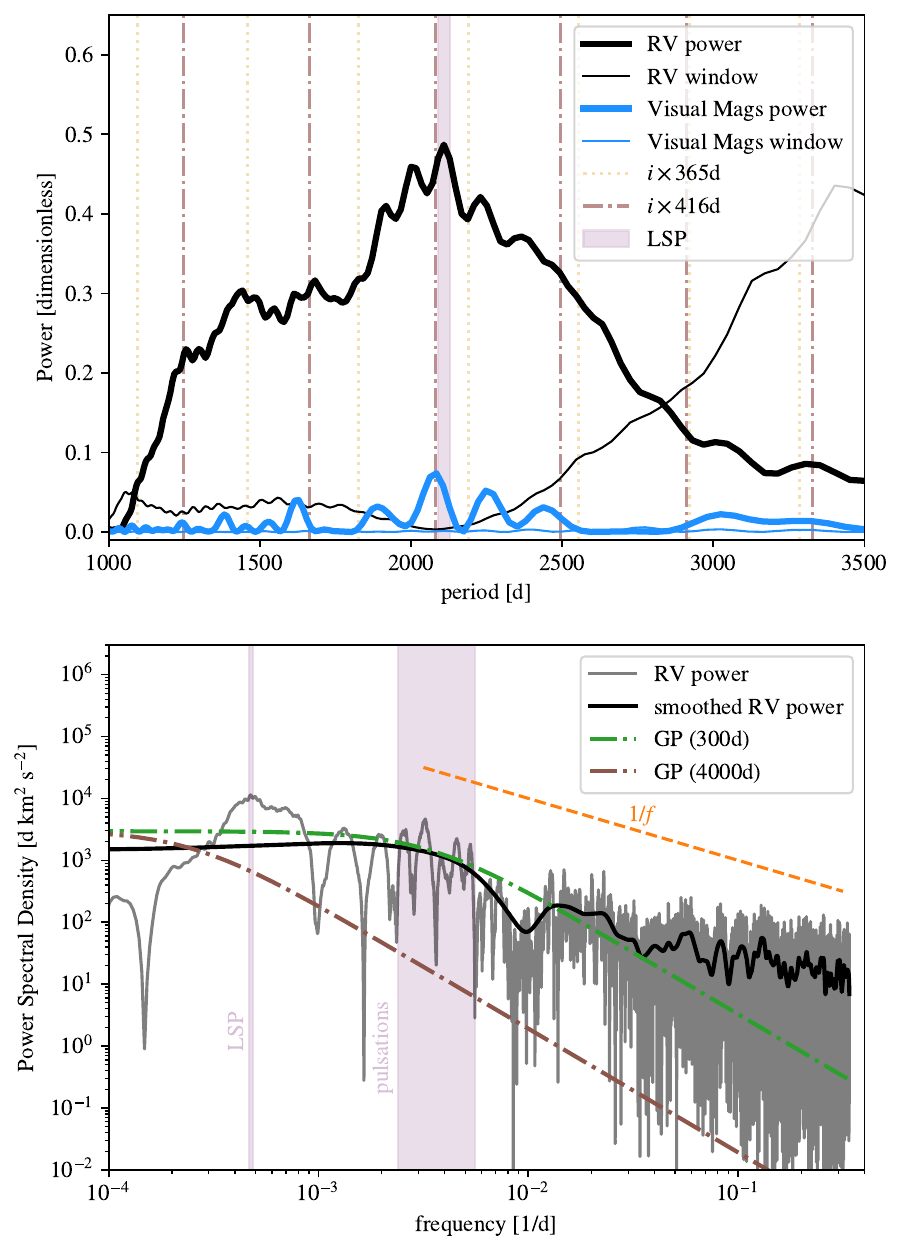}
    \caption{Dimensionless periodogram (top) and dimensional power spectral density (bottom) of Betelgeuse's RV variability. The periodogram and power spectrum both show features of Betelgeuse's multiple forms of variability, with peaks near the LSP and at aliases of the fundamental mode of pulsation period. The power spectrum favors a $f^{-1}$ ``pink noise" at frequencies higher than the pulsation frequency, while at lower frequencies the spectrum flattens to a white noise profile.   }
    \label{fig:periodogram}
\end{figure}

Figure \ref{fig:rvsummary} shows the history of RV and visual magnitude estimates accumulated for our analysis. The RV measurements are colored according to their source, from the earliest measurement in 1986 reported in volume 16 of the Annals of the Lick observatory, to the most recent, and extensive, monitoring from STELLA \citep{2022csss.confE.185G}.

These RVs show several overlapping effects, all with similar amplitude on the order of 1--few~km~s$^{-1}$. From low to high frequency these are:  
\begin{enumerate}
    \item Long term ``drifts" in RV and visual magnitude that we associate with the migration of convective cells around the stellar surface relative to our line of sight. The star rotates with a similar, $\sim 36 $~yr period \citep{2018A&A...609A..67K}, contributing to this drift. We note that in any given decade, the apparent baseline RV of Betelgeuse wanders by several km~s$^{-1}$, but without a consistent periodicity or timescale. Similarly, the visual magnitude changes by a few tenths over similar, but not completely correlated, timescales. This lack of periodicity is consistent with what is expected from the random motions of turbulent convection \citep[e.g.][]{2006MNRAS.372.1721K}.  
    
    \item The $\sim 2100$~d LSP variation. The LSP is visible as a 5-6 year sine-wave cycle of  radial velocity, most obvious in periods of denser sampling like 1900-1930 and 2000-present. The LSP is also present in the AAVSO visual magnitudes, though it is important to point out that its appearance changes significantly from cycle to cycle. 
    
    \item The $\sim 400$~d (and shorter) pulsation variations.  These are best seen in the most recent data, from roughly 1985 onward, where the sampling frequency is high \citep[see, for example, the presentation of the STELLA data by][]{2022csss.confE.185G}.  Analysis by several authors indicates the presence of multiple modes, including the presumed radial fundamental mode at $\sim 416$~d, and overtones at shorter periods \citep{2020ApJ...902...63J,2022csss.confE.185G,2023NewA...9901962J,2023ApJ...956...27M}.  
\end{enumerate}

The sum of these signals in Figure \ref{fig:rvsummary} implies that Betelgeuse is varying -- both in apparent RV and magnitude -- at a wide range of timescales and in both periodic and aperiodic ways. Though we have a particular richness of data for this star, this sort of behavior is broadly typical of red supergiants \citep[e.g.][]{2006MNRAS.372.1721K,2017ars..book.....L,2024A&A...682A..88P,2024arXiv240809089G}.  

We analyze the variability further by examining periodograms and power spectra of the data in Figure \ref{fig:periodogram}. The upper panel of Figure \ref{fig:periodogram} compares the Lomb-Scargle periodogram of all of the RV measurements to all of the visual magnitude estimates, with a focus around the $\sim 2100$~d approximate period of the LSP. The lower panel shows the power spectrum of RV measurements. 

The periodogram exhibits a broad peak of power in RV around $\sim 2100$~d. There is a similar, broad peak with several aliasing features in the visual magnitudes.  The window functions of these data -- computed by taking the periodograms of a constant at the same sampling as the datasets -- do not show equivalent peaks, implying that the signal seen is due to intrinsic variability rather than sampling. The periodograms do show a range of aliases at timescales both related to integer years and to multiples of the $\sim 416$~d fundamental mode pulsation. For example, the highest peak in the visual magnitude periodogram is at 2080~d$\approx 5\times 416$~d. Perhaps because of these confusion in the signal from these aliases, some previous works have identified slightly different LSP periods, for example 2190~d~$\approx6.00$~yr for photometric variation and 2510~d~$\approx 6.87$~yr for RV, as estimated by \citet{2023NewA...9901962J}. 

The power spectrum shown in the lower panel of Figure \ref{fig:periodogram} shows the amplitude of Betelgeuse's RV variability at a wide range of frequencies. The overall shape is a flat spectrum (characteristic of white noise) at low frequencies, coupled with a turnover to a $1/f$, pink noise scaling at higher frequencies. The $1/f$ slope is characteristic of variability associated with turbulently convective motions at a range of scales all contributing to the star's RV variability, and is similar to what is observed in visual magnitude power spectra \citep{2006MNRAS.372.1721K,2020ApJ...902...63J}, though we note that in high-cadence TESS data, \citet{2019ApJ...878..155D} found a variety of power spectra. 

We see some features in the power spectrum associated with the LSP (a broad, shallow peak) and with the range of fundamental mode and overtone pulsations of the star (near $f\sim 3\times 10^{-3}$~d$^{-1}$). The dip in the power spectrum at $\sim 10^{-2}$~d is likely associated with the roughly half-year window of visibility of Betelgeuse. The frequency of pulsations marks the inflection between a the flat and $1/f$ portions of the power spectrum. Physically, this could be because small, turbulent motions involving small regions of the star contribute at frequencies higher than the entire star's characteristic oscillation frequency. The smaller these regions are, the higher their local sound crossing frequency, but the lower their contribution to the star's overall RV, creating the declining envelope of the power spectrum.  Locally, a Kolomogorov spectrum of turbulence would impart a power density of $v^2/f \propto f^{-2}$, steeper than the observed $1/f$ slope.  At frequencies lower than the pulsations, we don't expect a dramatic increase in power due to convection, because variation would represent milder, subsonic motions across the star.

\subsection{Modeling and Analysis}

In this section, we describe modeling of Betelgeuse's LSP RV variability in conjunction with a possible companion driving the LSP. 

\subsubsection{Model}
Betelgeuse's LSP is intermediate in timescale between the rapid pulsations and long-timescale convective drift, thus our model must include flexibility to capture these other variability sources, along with the periodic signal of Betelgeuse's LSP. 

We make use of the {\tt RadVel} \citep{2018PASP..130d4504F} radial velocity fitting toolkit. Our model includes a single unseen companion object, an Exponential Gaussian Process (GP) kernel, and an RV ``jitter" to encompass otherwise unmodeled white noise. The GP Kernel is used to model both the long-term convective drift and the shorter-period pulsations and stochastic convective variability. The Exponential kernel is sometimes known as a damped random walk or Ornstein–Uhlenbeck kernel and has functional form 
\begin{equation}
    C_{ij} = \eta_1^2 \exp  \left(\frac{ -|t_i - t_j| }{ \eta_2 }\right),
\end{equation}
where $C$ is the covariance between elements $i$ and $j$ \citep{2023ARA&A..61..329A}. We implement this kernel in {\tt RadVel}\footnote{github.com/morganemacleod/radvel}.
We base our kernel on comparison to the star's variability power spectrum as shown in the lower panel of Figure \ref{fig:periodogram}. We found that the Quasi Periodic kernel often adopted to model stellar RV variation in exoplanet searches does not adapt well to this application because it has a  characteristic timescale, where Betelgeuse varies at a range of timescales. We adopt a fixed inflection timescale $\eta_2 = 300$~d, and allow the amplitude $\eta_1$ to be a free parameter, with units of velocity.  The choice of this inflection timescale is driven by the distinction between the white and pink portions of the power spectral density in Figure \ref{fig:periodogram}.  Figure \ref{fig:periodogram} also shows the power spectrum of this GP kernel with the fitted amplitude (Table \ref{tab:fit_params_rv}) of $\eta_1=2.26$~km~s$^{-1}$. This model compares favorably to the data but decreases faster at high frequencies $\propto 1/f^2$ as opposed to $1/f$. We found this was useful to limit overfitting in the GP regression given that there is intrinsic variability in Betelgeuse at shorter timescales than the sampling of (much of) the RV data, a fact compounded by the long gaps in parts of the historical time series. We also show the same Kernel but with $\eta_1=0.62$~km~s$^{-1}$ and $\eta_2=4000$~d, which we apply in the Appendix as an alternative model, where the GP only captures the long-term RV drift, not short-term jitter. 

We experimented with fitting the RV time series with both models including free eccentricity and ones restricted to circular. We find that when adequately fitting the pulsation variability with the inclusion of the GP, it is difficult to derive meaningful constraints on the system eccentricity. When we adopted models that underfit the pulsation variation ($\eta_2 = 4000$~d), eccentricities $e\lesssim 0.3$ were favored. Because of the model dependence of this inference, we cannot make any firm claims at present.  Despite the uncertain eccentricity, we found that the period and semi-amplitude inference were quite robust with regard to any constraint or lack thereof on $e$. Therefore, in what follows, we will  present the results of  circular-restricted RV fits. 

In the $e=0$ model, there are six free parameters, listed in Table \ref{tab:fit_params_rv}. 
After specifying priors on each of the  parameters, we find a maximum likelihood fit and then explore the parameter space of model fits using {\tt RadVel}'s Markov Chain Monte Carlo (MCMC) sampler, which is based on {\tt emcee} \citep{2013PASP..125..306F}. The following section presents results from this MCMC fitting.

\subsubsection{Results}

\begin{deluxetable*}{cccc}
    \tablewidth{\textwidth}
    \tablecaption{ Radial Velocity Model Parameters
    \label{tab:fit_params_rv}}
    \tablehead{Parameter & Unit & Prior & Posterior Median \& 68\% CI}
\startdata
\textbf{Fitted Parameters} & & & \\
$P$ & days & $\mathcal{U}[2000,2200]$ & $2109.2^{+9.2}_{-9.1}$\\ 
$T_{c}$ & year & $\mathcal{U}[2018.9,2027.1]$ & $2023.12^{+0.34}_{-0.35}$  \\
$\ln \left( K / {\rm km~s^{-1}} \right)$  & -- & $\mathcal{U}[-2.30,2.30]$ &  $0.39^{+0.20}_{-0.26}$\\
$\gamma$ & km~s$^{-1}$ & Gaussian, $\mu$=21.1, $\sigma=3$ &  $21.27^{+0.28}_{-0.29}$ \\
jitter & km~s$^{-1}$ & $\log \mathcal{U}[0.01,3]$ & $0.02^{+0.02}_{-0.01}$ \\
$\eta_1$ & km~s$^{-1}$ & $\log \mathcal{U}[0.01,20]$ & $2.22^{+0.07}_{-0.06}$ \\
\hline
\textbf{Derived Parameters} & & & \\
$K$ &  km~s$^{-1}$ &  & $1.48^{+0.33}_{-0.34}$ \\
$M_a$ & M$_{\odot}$ & $\mathcal{U}[10,25]$ & $17.5\pm5.1$ \\
$M_b \sin(i) / M_a$ & & & $0.035^{+ 0.009}_{-0.008}$  \\
$f$ & $M_\odot$ & & $7.1^{+5.8}_{-3.9}\times 10^{-4}$
\enddata
\tablecomments{ Free parameters, their priors,  inferences for the circular binary model applied to the RV time series. $P$ is the binary period, $T_c$ is the time of conjunction, $K$ is the RV semi-amplitude, $\gamma$ is the mean RV, jitter is an additional white noise term, and $\eta_1$ is the GP amplitude. Priors are described by $\mathcal{U}$ for uniform, Gaussian, $\log \mathcal{U}$ for uniform in the logarithm.  }
\end{deluxetable*}

\begin{figure}
    \centering
    \includegraphics[width=\linewidth]{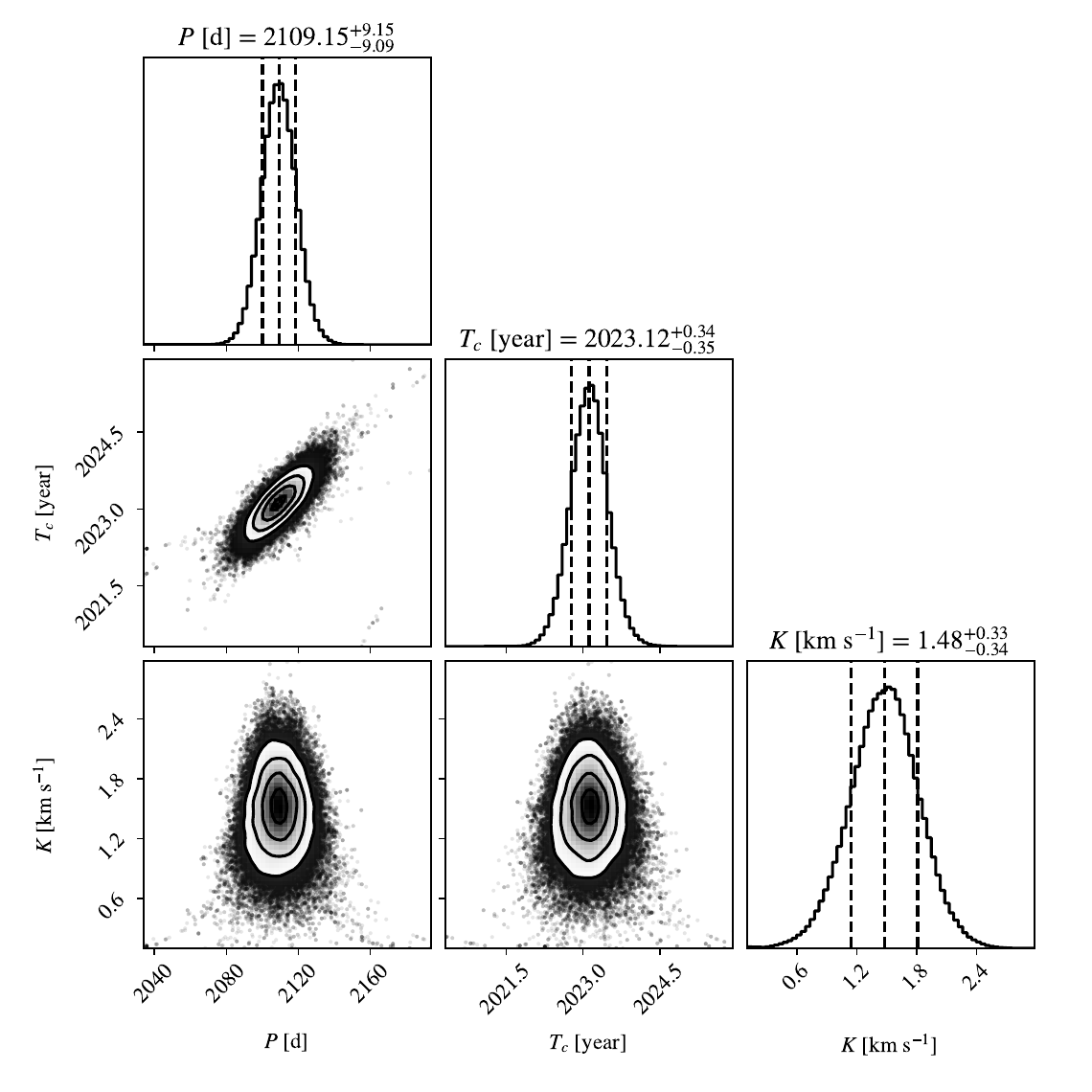}
    \caption{Posterior parameter distributions of LSP period $P$, time of conjunction, $T_c$, and semi-amplitude, $K$, of RV models sampled from broad, uninformative priors using {\tt RadVel}. The period is tightly constrained to $\sim 2110$~d or 5.78~yr. The semi-amplitude is more uncertain, but is larger than 0.66~km~s$^{-1}$ at 95\% confidence.  A full table of six fitted parameters is given in Table \ref{tab:fit_params_rv}.  }
    \label{fig:RVcorner}
\end{figure}

\begin{figure*}
    \centering
    \includegraphics[width=\textwidth]{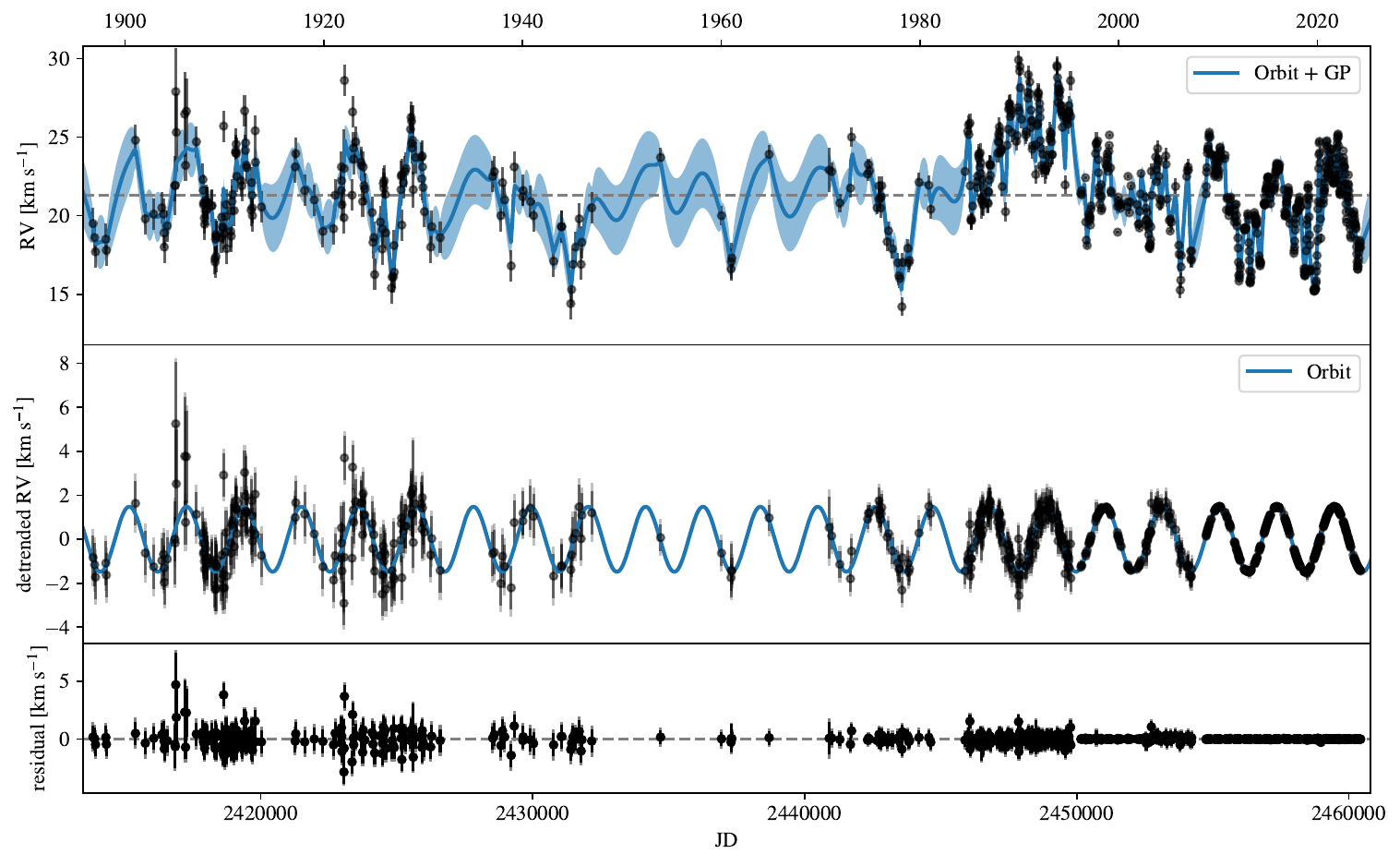}
    \caption{ RV time series from 1896 to present showcasing 10-day averaged RV measurements of Betelgeuse. The panels compare to a model including an exponential GP -- see the power spectrum of Figure \ref{fig:periodogram} -- and an orbital component. The upper panel shows the full RV time series, the middle removes the median GP prediction to only show the orbital component of the model and data, while the final panel removes all model predictions and shows the residual.   Errors on points are the one-sigma errors (top panel) and the one sigma errors including the one-sigma uncertainty of the GP prediction (lighter grey errorbars in middle  and lower panels). } Notably, the phase and period of the LSP remain constant through the $\sim 22$ cycles of our 127-year baseline -- strongly suggestive that a strictly-periodic solution like an orbit is appropriate. 
    \label{fig:RV}
\end{figure*}

\begin{figure*}
    \centering
    \includegraphics[width=\textwidth]{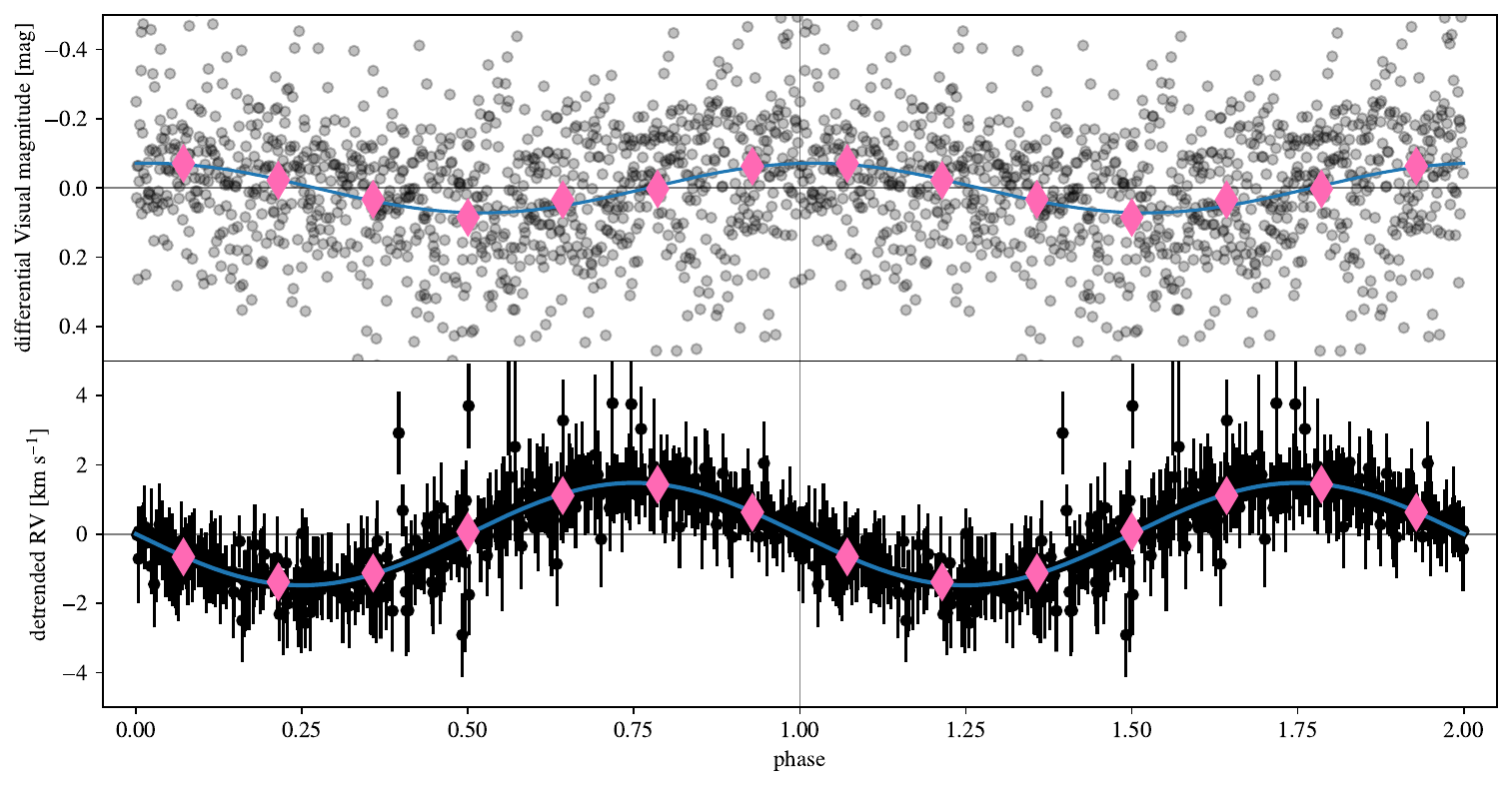}
    \caption{Phased, detrended visual magnitude and RV curves. The visual magnitude baseline is modeled and subtracted off with a third-order polynomial that captures long-term drift but not pulsation. The RV detrending is done as in Figure \ref{fig:RV}, where the median GP prediction is removed.  Pink points show the mean of the phased data, while lines show the fitted models. While both the RV and visual magnitudes vary with the LSP, their phases are different.  }
    \label{fig:phase}
\end{figure*}

From uninformative priors on period, $P$, semi-amplitude $K$, and time of conjunction, $T_c$, a narrow range of posterior solutions arise in our MCMC experiments. We show these key fitting parameters and their covariances in Figure \ref{fig:RVcorner}. The data are best described by a narrow range of periodic solutions, with a tightly constrained period of $P\approx 2110 \pm 9$~d. The Julian year of conjunction is $T_c \approx 2023.12 \pm 0.35$, which represents an approximately $\pm 7$\% uncertainty relative to the 5.78~yr period. The semi-amplitude is $K\approx 1.48 \pm 0.34$~km~s$^{-1}$ (asymmetric credible intervals are reported in Table \ref{tab:fit_params_rv}).   As we show in Appendix \ref{sec:RVmodelingAppendix}, fits including a periodic component are more predictive of unseen data than models without. 

The period is so tightly constrained (to $\sim 0.5$\%) because of the $\sim 22$ full cycles sampled over the 128-year span of the RV data. However, our derived period is very similar to what has been reported in previous studies, particularly that of \citet{1928MNRAS..88..660S} who derived $P=5.79$~yr and $K=2.13$~km~s$^{-1}$ from data spanning 1896 to 1926. Studies examining smaller time windows derive a wider range of LSP periods. For example, differences in RV ephemeris between early studies called into question the original claim that Betelgeuse was a spectroscopic binary \citep[e.g.][]{1908PASP...20..227P,1908ApJ....27..301K,1911AN....187...33B,1916ApJ....44..250L,1928MNRAS..88..660S,1933ApJ....77..110S}. At the time, there was concern over stability of spectrographs and differences between instruments. In hindsight, we suggest that this is not a shortcoming of these works, but is instead at the root of the star's intrinsic convective activity and variability. 

We note that $K\approx 2$~km~s$^{-1}$, similar to the result of \citet{1928MNRAS..88..660S},  is also favored in the most-recent $\sim 15$~yr of data, which includes the high-cadence STELLA monitoring. For example, \citet{2022csss.confE.185G} mention a preferred semi-amplitude of 2.68~km~s$^{-1}$. Sinusoids fitted to noisy, undersampled data tend to systematically underestimate amplitudes because they do not necessarily capture the peaks that are observed in higher-cadence sampling.  Thus, while the period is robustly constrained given the long time window, the semi-amplitude is much more uncertain than the period ($\sim 25$\% uncertainty in the 68\% credible interval) in large part because of the challenge of predicatively modeling the multiple sources of RV variation from the star. 

Figure \ref{fig:RV} shows the RV time series and maximum likelihood model. The upper panel shows the full time series and the model, with the line and shaded region representing the median and 68\% credible interval of the GP. The middle panel is detrended in that the GP component is removed, showing only the periodic, or orbital, component. The lower panel shows the residual. We see from this -- particularly in the raw dataset of the upper panel -- that a single, periodic signal matches the phase of variations across the more than century of variability. And the phase predicted by our modeling agrees well with the phase modeled by \citet{1928MNRAS..88..660S} nearly a century ago. This history, though of course short compared to a stellar lifetime, does suggest that the the LSP period was stable over nearly 22 full cycles, pointing to a periodic (e.g. binary), rather than stochastic (e.g. convection) origin.

We note that some degree of overfitting is expected to be intrinsic to the GP component of our model because of undersampling of some of the range of fitted frequencies \citep[see][for a related discussion]{2023AJ....166...62B}. This is most clearly seen in the narrow dispersion of points in the residuals (relative to the errors). Even in the most-rapidly sampled portions of STELLA data, there are intrinsic variations between subsequent datapoints, implying that a GP that adequately matches the variability amplitude also tends to have the flexibility to thread through each datapoint. The most robust test of whether overfitting affects our model's utility is through cross validation of the model against unseen data. Therefore, to understand how overfitting affects our model inference, we test the predictive power of the periodic model in Appendix \ref{sec:RVmodelingAppendix} by leaving out the last decade of data and comparing models with and without an orbital component, mimicking the experiments described in \citet{2023AJ....166...62B}. As discussed in the Appendix, these tests show that a model with a periodic, orbital, component and a GP are significantly more predictive of unseen data than a model with a GP -- modeling the general, aperiodic variability -- alone (Figure \ref{fig:dropsampleGP}). 

All of these factors point to the LSP being a truly periodic, stable signal in the RV data.  If we adopt the RV fit as an orbital solution, the epoch of conjunctions of the secondary toward the observer is
\begin{equation}\label{RVephemeris}
    {\rm year} = 2023.12 + 5.775 E
\end{equation}
where $E$ is an integer number of orbits, or 
\begin{equation}
    {\rm JD} = 2459984 + 2109.2 E
\end{equation}
The model orbital component of the radial velocity of Betelgeuse is therefore
\begin{equation}\label{RVformula}
    RV =  - K \sin \left( 2\pi \frac{t-T_c }{P}  \right). 
\end{equation}
The sign reflects the fact that in the half-orbit following conjunction, the motion of Betelgeuse is toward the observer, which we designate a negative RV.

We phase the detrended RV curve (with the median GP prediction removed) and compare to the phased AAVSO visual magnitude curve in Figure \ref{fig:phase}.  
To examine the periodic component, we remove long-term trends by fitting a 3rd-order polynomial that removes the drifting mean visual magnitude.  The mean phased light and RV curves are shown with large pink diamonds.  Though the dispersion is large (both because of the uncertainty of visual measurements and the pulsational variation of Betelgeuse) the magnitude varies as with the same period as the radial velocities, with a curve approximately described by 
\begin{equation}\label{Magformula}
    \Delta {\rm mag} \sim A_{\rm vis} \sin \left( 2 \pi \frac{t-T_c }{P} + \phi_{\rm vis} \right)  , 
\end{equation}
where we find the amplitude $A_{\rm vis}\approx  0.07$~mag, and the phase offset $\phi_{\rm vis}\approx -1.7$~radian, implying that minimum light (maximum magnitude) occurs at a phase of $(3\pi/2 - \phi_{\rm vis}) \approx 3.01$~radian, or a phase of 0.48, which implies minimum light occurs approximately a half-orbit after conjunction,  $ 0.48 P \sim 1010$~d. This phase offset is similar (phase difference $\lesssim10$\%) to that identified by \citet{2024arXiv240809089G} focusing on the STELLA RV data and contemporaneous V-mag photometry. Either result suggests that minimum light is around a half-orbit after conjunction, where the putative companion is behind Betelgeuse itself.

\subsection{Implied Companion}

\begin{figure}
    \centering
    \includegraphics[width=\linewidth]{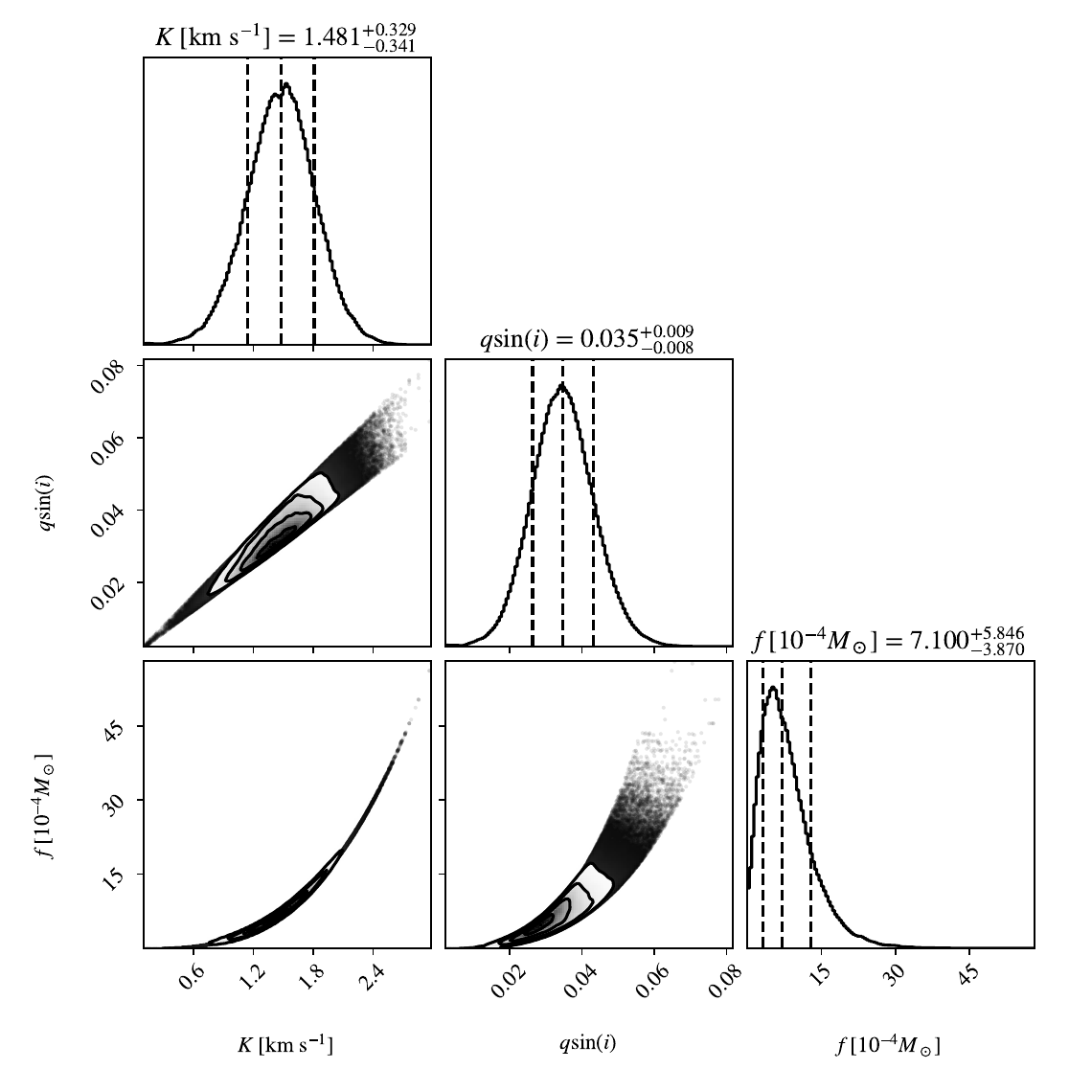}
    \caption{Posterior parameter distributions of companion properties inferred from {\tt RadVel} fit to the RV time series. Here we show the interdependence between semi-amplitude, $K$, mass ratio $q\sin(i) = M_b \sin(i)/M_a$, and mass function $f$. The solution favors a low binary mass ratio, of $M_b \sin(i)/M_a < 1/20$ (Table  \ref{tab:fit_params_rv}).  }
    \label{fig:Kqf}
\end{figure}

A priori, we do not know the mass of Betelgeuse \citep[e.g.][]{2020ApJ...902...63J}. We therefore leave this a loosely constrained parameter, selecting a uniform distribution between 10 and 25 $M_\odot$ in the derived parameters of Table \ref{tab:fit_params_rv}. In this section, we report the scaling of derived parameters with $M_a$.  We adopt $K=1.48^{+0.33}_{-0.34}$~km~s$^{-1}$ to solve for the properties of an unseen companion object causing Betelgeuse's RV motion. The system mass function is 
\begin{equation}
    f = \frac{ P K^3} {2 \pi G} \approx 7.1^{+5.8}_{-3.9}\times 10^{-4}  M_\odot ,
\end{equation}
and the mass ratio -- modified by the inclination factor of $\sin(i)$ -- is 
\begin{equation}
    q \sin (i) = \frac{M_b \sin(i)}{ M_a}  
    \approx 0.034 \pm 0.008  \left(M_a \over 17.5 M_\odot\right)^{-1/3}
\end{equation}
which scales linearly with $K$ at fixed $M_a$, because $K=(2\pi G / P M_a)^{1/3} q\sin(i)$. Thus, were $K= 2.68$ \citep[e.g. as determined based on the STELLA data by][]{2022csss.confE.185G} we would have $q \sin (i) = 0.062$ \citep{2024arXiv240809089G}. The relationship between $K$, $q\sin(i)$, and $f$ is shown in Figure \ref{fig:Kqf}. 

Given the this low mass ratio, the implied companion mass is
\begin{equation}
M_b \sin(i) \approx 0.60 \pm 0.14  \left(M_a \over 17.5 M_\odot\right)^{2/3} M_\odot.
\end{equation} 
The companion orbits at a semi-major axis of 
\begin{equation}
a \approx 1818 \pm 6 \left(M_a \over 17.5 M_\odot \right)^{1/3} R_\odot,
\end{equation}
Finally, the velocity semiamplitude of the secondary, $K_b$, is not a sensitive function of $K$, but depends on $M_a$ as
\begin{equation}
    K_b \approx 43.1\pm 0.1  \left(M_a \over 17.5 M_\odot\right)^{1/3} {\rm km~s}^{-1},
\end{equation}
and the instantaneous value of the secondary RV can be found from 
\begin{equation}
    RV_b =   K_b \sin \left( 2\pi \frac{t-T_c }{P}  \right),
\end{equation}
where the model parameters are identical to those of equation \eqref{RVformula}.

\section{LSP in Astrometry}
\label{sec:astrometry}

The RV fit described in the previous section, if observed at an inclination of 90 degrees, implies an astrometric motion with semi-amplitude of  $a_a \approx 1.6 \left( K / 1.5{\rm ~km~s}^{-1} \right)~{\rm mas} $
at a distance of 175~pc (Section \ref{sec:BGprop}), which though smaller than Betelgeuse's $\sim 21$~mas radius is comparable to its $\sim 5.8$~mas parallax and motivates a search for astrometric signatures of a LSP.  We do not pursue a joint fit between the RV and astrometric data at this point because the present RV dataset is much longer in baseline, higher in cadence, and higher relative precision than the astrometric data, meaning that any joint fit would be mostly driven by the RVs themselves. In addition, as we describe below, one of our main conclusions is that the astrometric variation requires a more complicated model than is warranted with our existing data. Rather than producing a ``final'' solution for the astrometry, we focus on pointing out similarities between the astrometric solution and the RV solution that argue for the presence of a companion. Following this philosophy, we independently model the astrometry and examine our results in light of the RV inferences. We describe that effort in this section, and all data and code needed to reproduce these results is available online\citep{blunt_zenodo_bg,blunt_zenodo_orbitize_bg}.\footnote{ Also at \url{https://github.com/sblunt/betelgeuse} and  \url{https://github.com/sblunt/orbitize_for_betelgeuse}} 

\subsection{Astrometric Data}

We compiled an astrometric dataset from the literature, consisting of absolute astrometry derived from three different radio instruments (e-MERLIN, ALMA, and the VLA), published in \cite{Harper:2008a} and \cite{2017AJ....154...11H}, as well as the 2021 re-reduction of the Hipparcos-2 intermediate astrometric data (IAD) catalog. The Hipparcos-2 catalog was originally published in a DVD accompanying the book \cite{van-Leeuwen:2007a}, and discussed in \cite{van-Leeuwen:2007b}. In 2014, an updated version of the Hipparcos-2 catalog IAD, incorporating slightly improved solutions for all stars, was made available as a Java Tool. More recently, in 2021, a machine readable version of this updated reduction was made available through the ESA.\footnote{We downloaded this version of the Hipparcos data from \url{https://www.cosmos.esa.int/web/hipparcos/hipparcos-2}.} In both the 2007 and 2014 reductions, a single Hipparcos scan measurement was marked as bad by the Hipparcos team. We excluded this datapoint from our analysis. Finally, Betelgeuse has a ``stochastic'' Hipparcos solution, which differs from a standard 5-parameter astrometric solution only in the addition of a ``cosmic noise'' astrometric jitter term.

\begin{deluxetable*}{cl}
    \tablewidth{\textwidth}
    \tablecaption{ Variables used in Section \ref{sec:astrometry}
    \label{variable-defs}}
    \tablehead{Parameter & Definition}
\startdata
$\chi^2_{\rm Hip}$ & $\chi^2$ computed for only Hipparcos data \\
$\mathcal{L}_{\rm Hip}$ & likelihood computed for only Hipparcos data\\
$d$ & distance between astrometric model prediction and Hipparcos scan measurement (\citealt{Nielsen:2020a} Eq 6)\\
$\epsilon$ & Hipparcos scan observational uncertainty\\
$\sigma_{\rm ast}$ & ``cosmic error''-- fixed white noise jitter term added in quadrature to Hipparcos scans \\
$\alpha_C^*$ & $R.A.\cos{\delta_0}$ model prediction\\
$\delta_C$ & dec model prediction\\
$\alpha_{H 0}^*$ & fitted parameter: $R.A.\cos{\delta_0}$ offset from published Hipparcos position at Hipparcos epoch \\
$\delta_{H 0}$ & fitted parameter: dec. offset from published Hipparcos position at Hipparcos epoch \\
$\pi$ & parallax \\
$X$ & ICRS Barycenteric position of Earth (Cartesian X component)\\
$Y$ & ICRS Barycenteric position of Earth (Cartesian Y component)\\
$Z$ & ICRS Barycenteric position of Earth (Cartesian Z component) \\
$\alpha_0$ & published Hipparcos R.A. position for Betelgeuse \\
$\delta_0$ & published Hipparcos dec position for Betelgeuse\\
$\tau$ & fractional time of periastron passage (\citealt{Blunt:2020a} Eq 1)\\
$P$ & orbital period \\
$\Omega$ & orbital position angle of nodes \\
$e$ & eccentricity \\
$M_{\rm b}$ & secondary mass \\
$M_{\rm a}$ & primary (Betelgeuse) mass \\
$i$ & orbital inclination; defined such that 90$^{\circ}$ is edge-on\\
$\mu_{\delta}$ & proper motion in dec. direction \\
$\mu_{\alpha^*}$ & proper motion in $R.A.\cos{\delta_0}$ direction \\
$\Delta {\rm dec}$ & measured dec. offset from $\delta_0$ \\
$\sigma_{\rm dec}$ & observational uncertainty in above \\
$\Delta {\rm R.A.}$ & measured R.A. offset from $\alpha_0$ \\
$\sigma_{\rm R.A.}$ & observational uncertainty in above\\
$K_{\delta}$ & Keplerian model prediction (dec. component) \\
$K_{\alpha^*}$ & Keplerian model prediction ($R.A.\cos{\delta_0}$ component) \\
\enddata
\tablecaption{Generally, the presence of a $*$ indicates multiplication by $\cos{delta_0}$}
\end{deluxetable*}

\subsection{Astrometric Model}

We use the astrometric orbit modeling software {\tt orbitize!} to model Betelgeuse's astrometric measurements \citep{Blunt:2020a,2024JOSS....9.6756B}. Our modeling does not introduce significant new methodology relative to \cite{2017AJ....154...11H} and \cite{Nielsen:2020a}. However, we describe our method below for completeness, defining all variables identically to \cite{Nielsen:2020a}.

Briefly, \cite{Nielsen:2020a} first reconstruct the raw Hipparcos 1D scans from the available data format, residuals to the best-fit Hipparcos solution. We follow \citet{Nielsen:2020a}'s equations 1-4 to reconstruct the raw Hipparcos scans, and their equation 6 to calculate a distance metric between a scan and model prediction, with the addition of removing the cosmic noise added to each point. Therefore, our $\chi^2$ is given by
\begin{equation}
    \chi^2_{\rm Hip} = \sum_t \frac{d^2(t)}{\epsilon(t)^2 - \sigma_{\rm ast}^2} .
\end{equation} where $d(t)$ is the distance between the model prediction and the tangent point on the Hipparcos scan (Eq 6 of \citealt{Nielsen:2020a}), $\epsilon$ is the observational scan error, and $\sigma_{\rm ast}$ is the constant cosmic variance term defined by the Hipparcos team (for Betelgeuse, 2.25 mas)\footnote{We noticed that in the Java Tool version of the Hipparcos IAD, which we used for this fit, the cosmic variance value (``var'' in the data file header) is 0.15, while in the DVD version, the cosmic variance value is 2.4. By plotting the cosmic variances for all stochastic solutions given by the DVD version of the IAD and the Java Tool version, we inferred that the Java Tool IAD cosmic variance is given in units of $\sqrt{\rm mas}$/10., and that there was no significant change in this value between the two reductions.}. The other extension we made to the \cite{Nielsen:2020a} framework was to jointly fit Hipparcos data along with 2D absolute astrometric points obtained from radio observations following the approach defined by \citet{Harper:2008a,2017AJ....154...11H}. 

A Hipparcos-only astrometric model has 5 parameters: {$\alpha_{H0}^*$, $\delta_{H0}$, $\mu_{\alpha^*}$, $\mu_{\delta}$, and $\pi$}, where $\alpha_{H0}^*$ and $\delta_{H0}$ are defined as offsets from the published Hipparcos position at J1991.25, $\mu_{\alpha^*}$, $\mu_{\delta}$ are the proper motions, and $\pi$ is the parallax (see Table \ref{variable-defs}). To  incorporate the new radio measurements into this framework, we define the radio data as spherical offsets from the J1991.25 Hipparcos epoch.\footnote{N.B. \texttt{orbitize!} expects offsets as $\Delta$R.A., \textit{not} $\Delta$R.A.$\cos{\delta_0}$.} 

We can thus fit the a model including a position, proper motion, and parallax in addition to a Keplerian orbital model prediction, $K(t)$. The joint model is
\begin{equation}
\begin{split}
\alpha_C^*(t) =& \alpha_{H 0}^*+\pi\left(X(t) \sin \left(\alpha_0\right)-Y(t) \cos \left(\alpha_0\right)\right) \\ +& (t-1991.25) \mu_\alpha^* + K_{\alpha^*}(t) \\
\delta_C(t) =& \delta_{H 0}+\pi [ X(t) \cos (\alpha_0) \sin (\delta_0) \\ 
+& Y(t) \sin (\alpha_0) \sin (\delta_0) \\
-& Z(t) \cos (\delta_0)]+(t-1991.25) \mu_\delta + K_{\delta}(t).
\end{split}
\end{equation} 
This model is fit jointly to the two different data sets via the likelihood, which we define as 
\begin{equation}
\begin{split}
    \log{\mathcal{L}} =& \log{\mathcal{L_{\rm Hip}}} \\ &-\frac{1}{2} \left( \frac{\frac{\alpha_C^*(t)}{\cos{\delta_0}} - \Delta R.A.}{\sigma_{R.A.}}\right)^2 \\ &- \frac{1}{2}\left(\frac{ \delta_C(t) - \Delta {\rm dec}}{\sigma_{\rm dec}}\right)^2 \\ &- \log{\sqrt{2\pi\sigma_{R.A.}^2}} - \log{\sqrt{2\pi\sigma_{\rm dec}^2}}.
\end{split}
\end{equation}
We modified \texttt{orbitize!} to perform this joint fit. Jointly fitting Hipparcos, Gaia, and arbitrary 2D absolute astrometry is available in \texttt{orbitize!} 3.0.1 (and this functionality is demonstrated in the \href{https://orbitize.readthedocs.io/en/latest/tutorials/abs_astrometry.html}{Absolute Astrometry Tutorial}).

\subsection{Results}

In total, we performed a suite of 10 model fit variations, some of which will be discussed in the sections below. Our main findings come from the ``standard'' fit, but we study variations on this fiducial model as a test of the robustness of our inferences. 

\begin{deluxetable*}{cccc}
    \tablewidth{\textwidth}
    \tablecaption{ Astrometric Binary Model Parameters
    \label{tab:fit_params}}
    \tablehead{Parameter & Unit & Prior & Posterior Median \& 68\% CI}
\startdata
\textbf{Fitted Parameters} & & & \\
$\pi$ & mas & $\mathcal{U}[0,15]$ & $5.8\pm{0.4}$ \\
$\mu_{\alpha^{*}}$ & mas yr$^{-1}$ & $\mathcal{U}[17.2,37.8]$ & $24.89\pm{0.08}$ \\
$\mu_{\delta}$ & mas yr$^{-1}$ & $\mathcal{U}[4.8,17.8]$ & $9.50^{+0.08}_{-0.07}$ \\
$\alpha^{*}_{H0}$ & mas & $\mathcal{U}[-7.10,7.10]$ & $0.05\pm{0.4}$ \\
$\delta_{H0}$ & mas & $\mathcal{U}[-5.80,5.80]$ & $0.1\pm{0.4}$ \\
$P$ & days & $\mathcal{U}[1000,2700]$ & $2190^{+84}_{-96}$\\ & & & $1652^{+46}_{-48}$ \\
$M_a$ & M$_{\odot}$ & $\mathcal{U}[10,25]$ & $18.7^{+4.5}_{-6.0}$ \\
$M_b$ & M$_{\odot}$ & log$\mathcal{U}[0.1,10]$ & $2.2\pm{0.5}$ \\
$e$ &  & =0 & \\
$\Omega$ & deg & $\mathcal{U}[0, 2\pi]$ & $60\pm{6}$ \\
$\tau$ &  & $\mathcal{U}[0, 1]$ & $0.76^{+0.11}_{-0.13}$ \\
$i$ & deg & $\sin{i}$ & $98\pm{5}$ \\
\textbf{Derived Parameters} & & & \\
\hline
\hline
$M_{b} \sin(i)$ & M$_{\odot}$ & & $2.1\pm{0.5}$ \\
q &  & & $0.12\pm{0.02}$ \\
$a_a$ & mas & &$5.7\pm0.7$ \\
$T_c$ & year & &$2022.9\pm0.8$ \\
\enddata
\tablecomments{Parameter inferences from the standard model fit applied to the Hipparcos and radio datasets. }
\end{deluxetable*}

\begin{figure*}
    \centering
    \includegraphics[width=\linewidth, trim={8cm 0cm 8cm 1cm}]{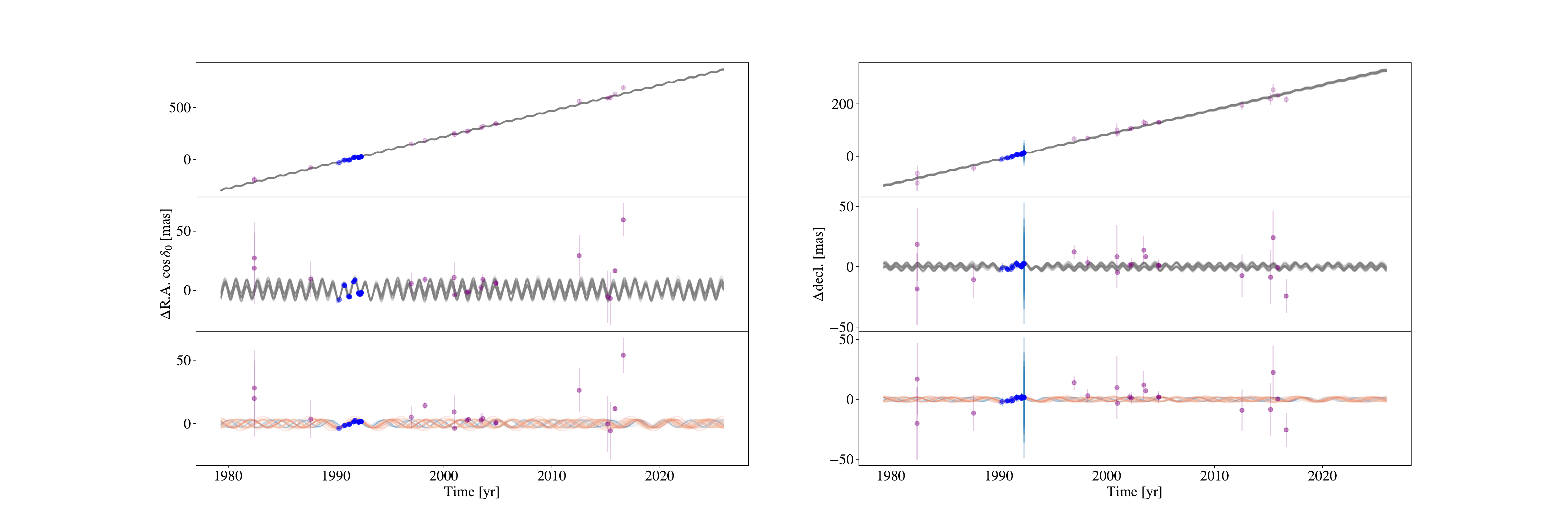}
    \includegraphics[width=\linewidth, trim={8cm 0cm 8cm 1cm}]{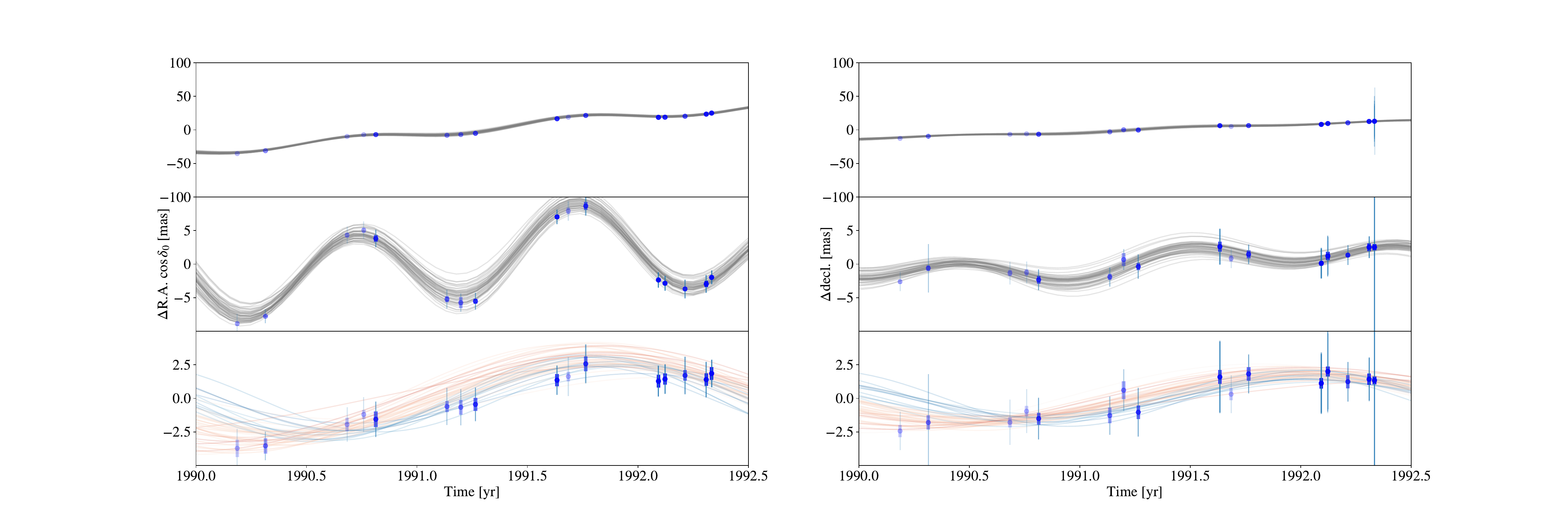}
    \caption{ Astrometric time series for the standard fit, covering the full dataset (upper panels) and  zoomed in on the Hipparcos epochs (lower panels). Within each section, bottom panel lines are colored by period, with periods greater than 2000 days shown in red, and periods less shown in blue. Overall, the top row of each dream plot shows the data along with complete model predictions for 50 random draws from the posterior. The middle row shows the proper motion component subtracted off, leaving the parallax and Keplerian astrometric signals. The bottom row shows the parallax component subtracted off, leaving only the Keplerian astrometric signal. The details of constructing this figure are described fully in Section \ref{sec:dream-math}.}
    \label{fig:dreamplot}
\end{figure*}

\subsubsection{A New Astrometric Solution for Betelgeuse}

In our standard fit, we fit the full model to the combined radio and Hipparcos datasets. We adopt broad, uninformative priors on each of the parameters, which are listed in Table \ref{tab:fit_params} along with posterior inferences of their values. 

Figure \ref{fig:dreamplot} shows the time series of right ascension and declination data and compares to 50 randomly-drawn samples from the standard fit model. The upper group of panels show the full data span, while the lower panels zoom in on the epoch of the Hipparcos mission.  Each panel of Figure \ref{fig:dreamplot} shows the data with a subset of the full model removed. The uppermost panels show the raw relative positions. The primary component visible in the data is the proper motion as a function of time that induces a linear drift. 

The center panels of Figure \ref{fig:dreamplot} remove the proper motion, leaving the parallax and Keplerian components of the full model. The annual cycle of parallax is the signature most-visible in these regions. Zooming in to the Hipparcos epoch, it is apparent why the higher cadence of these data are particularly constraining in modeling the parallax signal. Finally, the lowermost panels of each group show the data and model with both proper motion and parallax removed, leaving only any possible Keplerian motion as a residual.

\begin{figure}
    \centering
    \includegraphics[width=\linewidth, trim = {0.75cm 0 0.75cm 0}]{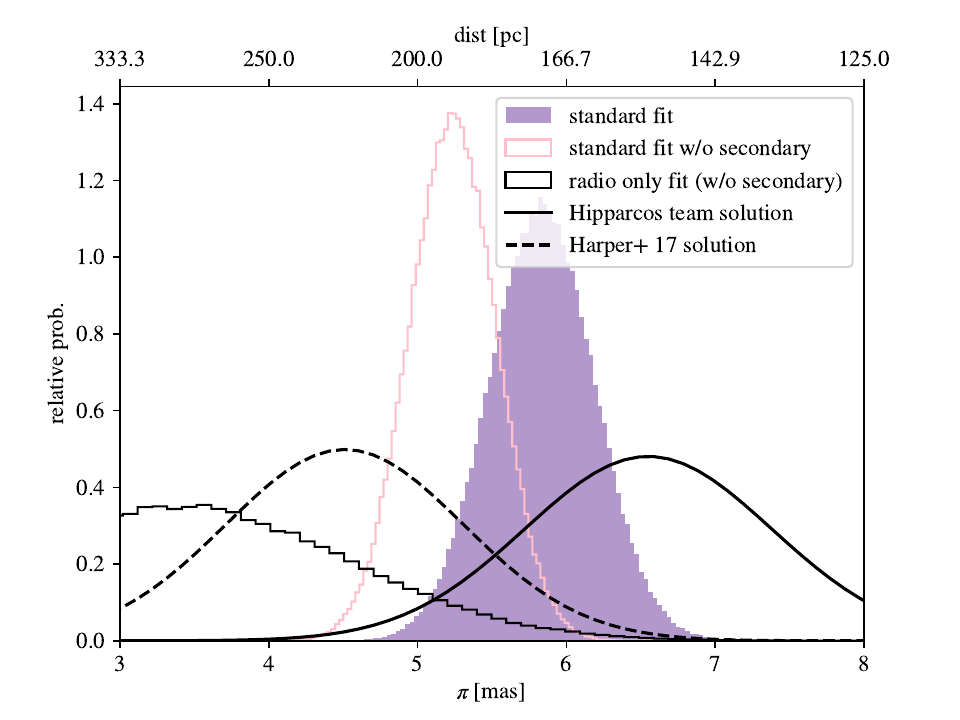}
    \includegraphics[width=\linewidth]{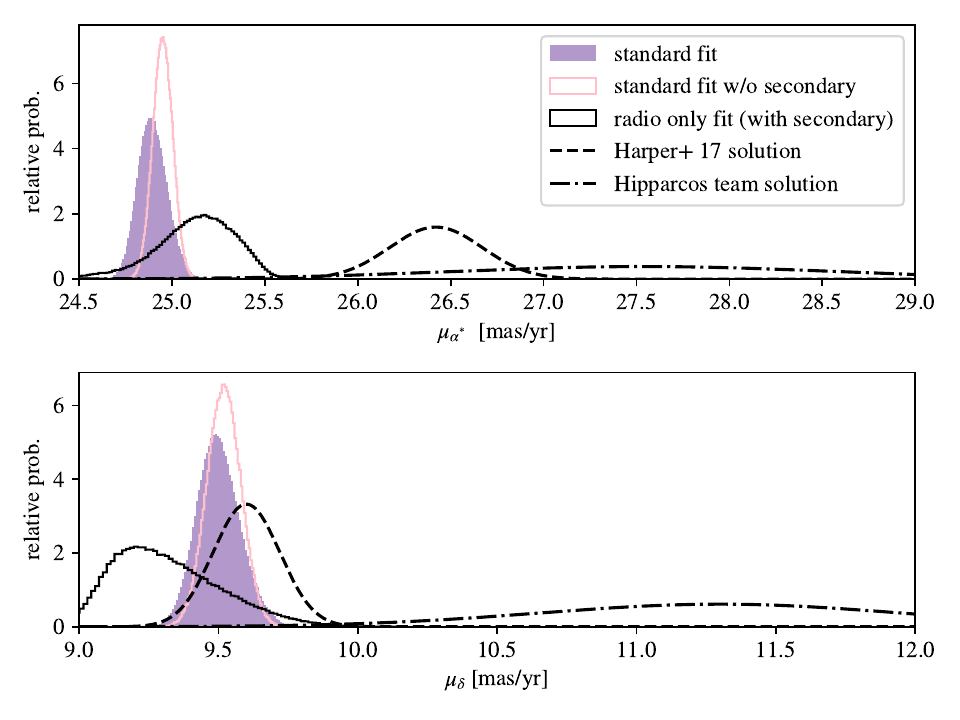}
    \caption{Parallax (top) and proper motion (middle, lower) posteriors for various fits of interest. This figure shows that overall, including a companion in the orbit fit does not significantly affect the parallax estimate. Our derived parallax is intermediate between the Hipparcos team solution and that of \citet{2017AJ....154...11H}. The presence of a companion somewhat assuages the tension between Hipparcos-derived and radio-derived proper motion estimates; since the Hipparcos baseline is about half of the LSP, the companion signal is approx linear over the Hipparcos baseline (see Figure \ref{fig:dreamplot}). This was interpreted as proper motion by the Hipparcos fit, resulting in a different PM estimate. }
    \label{fig:plx-comp}
\end{figure}

Figure \ref{fig:dreamplot} highlights that the position and parallax are quite precisely probed by both Hipparcos and and radio datasets \citep{2017AJ....154...11H}. 
The parallax and proper motion constraints are shown in Figure \ref{fig:plx-comp}. The major change from the last published astrometric solution \citep{2017AJ....154...11H} is a smaller proper motion, in both right ascension and declination, and a larger parallax. Over the Hipparcos baseline, the companion signal is approximately linear, and it can be modeled as the additional proper motion shown in Figure \ref{fig:plx-comp}. The major driver of differences relative to the \citet{2017AJ....154...11H} solution is the treatment of cosmic noise as described in Appendix \ref{sec:harperreproduction}. As discussed by \citet{2017AJ....154...11H}, a major difference between these datasets -- beside their wavelength -- is their cadence. The radio data are lower-cadence and more sporadically sampled, while the Hipparcos data are more concentrated with a carefully-planned 6 month cadence to optimize parallax measurements.

\subsubsection{Binary Model Inferences}

\begin{figure}
    \centering
    \includegraphics[width=\linewidth]{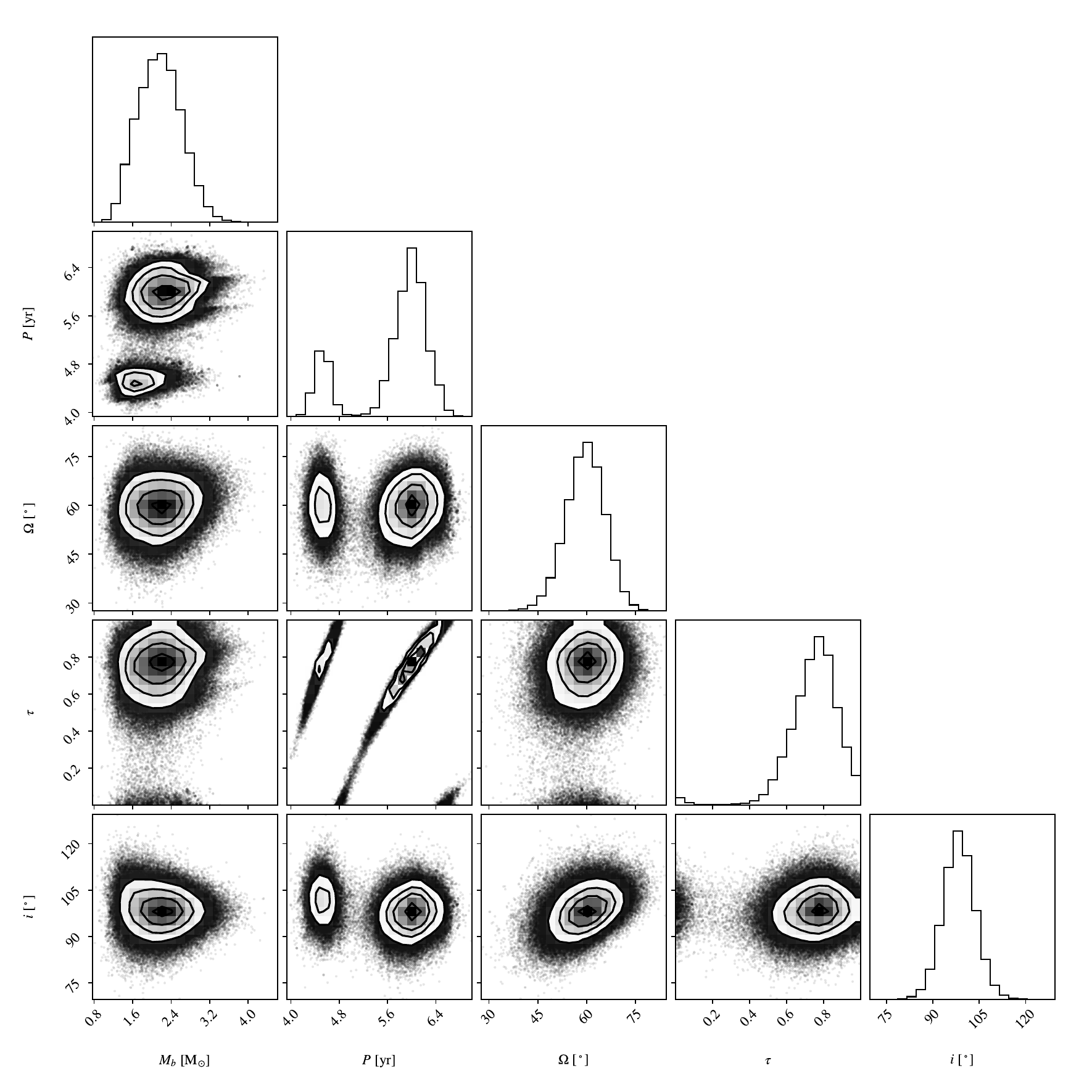}
    \caption{Corner plot of the standard fit focused on properties of the companion object. We see that constrained clusters of parameters emerge, favoring one of two clusters of orbital periods, constrained orientation ($\Omega$,$i$), and a constrained orbital phase, through the orbit-fraction $\tau$. }
    \label{fig:minimal-corner}
\end{figure}

\begin{figure}
    \centering
    \includegraphics[width=\linewidth]{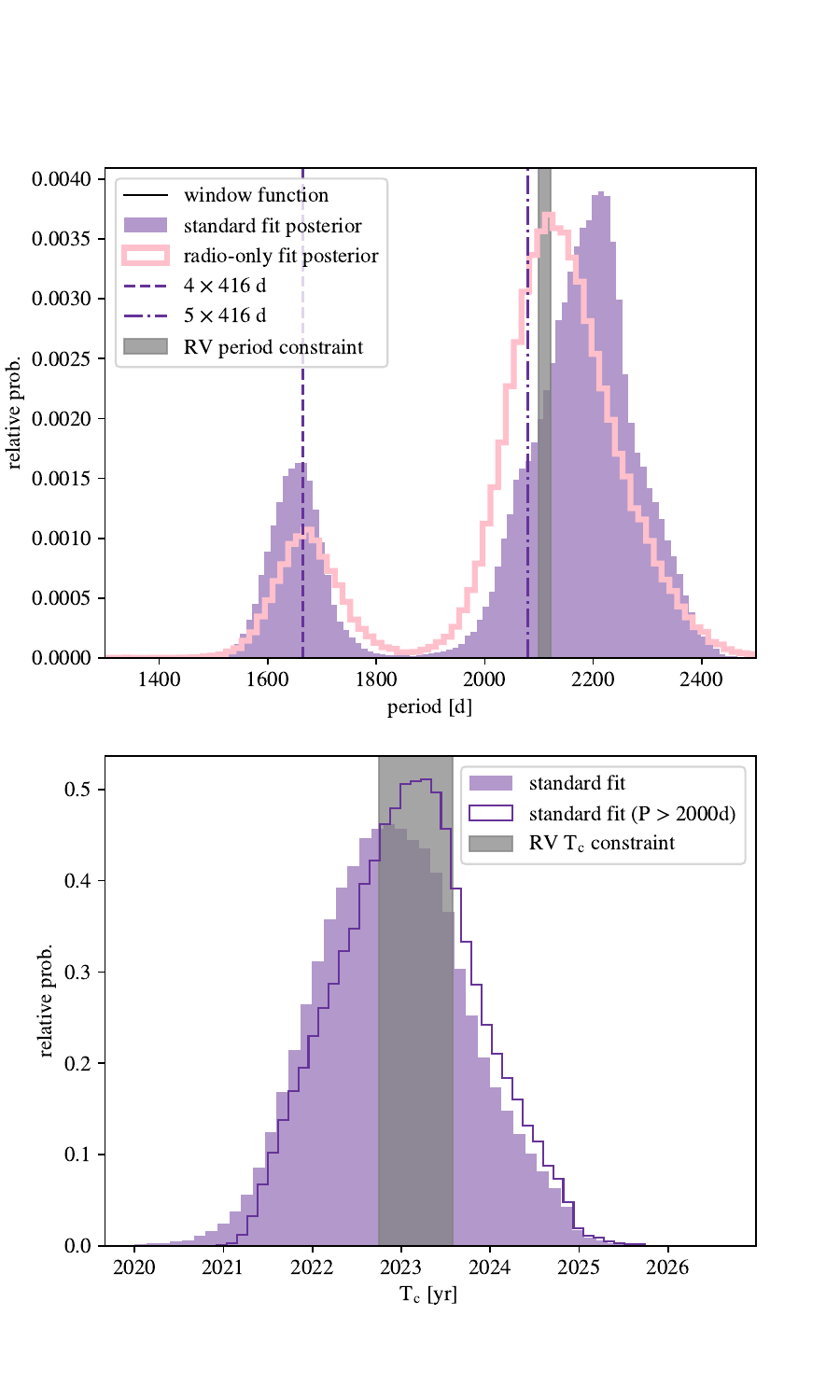}
    \caption{Period and time of conjunction posteriors shown compared to inferences from the RV fits. The multiple peaks of the period posterior are not related to the data's sampling window function. Instead the shorter period peak appears to be a multiple of the star's fundamental mode of pulsation. The longer period peak overlaps closely with the RV-derived LSP, as does the time of conjunction, indicating that the RV and astrometry are probing the same physical variation.  }
    \label{fig:window-func}
\end{figure}

\begin{figure}
    \centering
    \includegraphics[width=\linewidth]{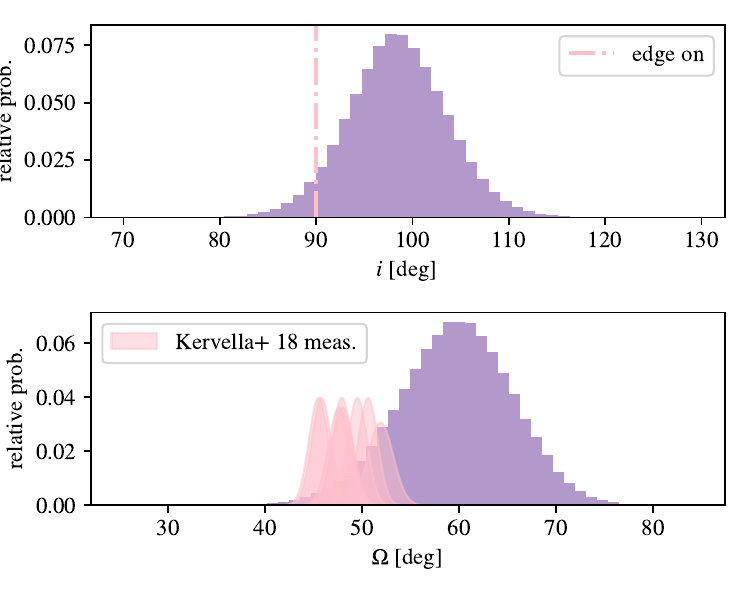}
    \caption{Inclination and on-sky position angle derived from the astrometric standard fit. Intriguingly, the inclination is constrained to be edge-on, as would be expected in the binary model of photometric variability, and $\Omega$ is oriented with the measured spin  \citep{2018A&A...609A..67K} of Betelgeuse -- indicative of a possible spin--orbit connection. } 
    \label{fig:angles}
\end{figure}

In addition to the baseline astrometric solution for Betelgeuse, our standard model implies a non-zero Keplerian component (Table \ref{tab:fit_params}), as we have highlighted in Figure \ref{fig:dreamplot}. The preferred solution has an astrometric amplitude of $a_a\sim 5.7$~mas, comparable to the annual parallax motion, but extending over a longer period. 

We find two clusters of possible periodicity for the the Keplerian model term, $P\sim 4.5$~yr or $\sim6$~yr (Table \ref{tab:fit_params}). In Figure \ref{fig:dreamplot} we color the lower-panel samples according to their orbital period -- blue lines representing shorter periods, while red represent longer periods. The Hipparcos zoom makes it clear that the multiplicity of periods can be traced, in part, back to the $\sim 2.5$~yr Hipparcos observing window. In sampling just a portion of an orbital cycle, these data allow multiple combinations of phase offset and period, which are partially constrained by the relatively precise radio measurements in the early 2000s. 

A corner plot of binary parameters from the astrometric solution is shown in Figure \ref{fig:minimal-corner}. In addition to the two period clusters, we see a robust constraint on on-sky inclination, $i$, and position angle $\Omega$. The period and fractional phase, $\tau$ are inter-related as hinted by our discussion of the brief  Hipparcos observing window. 

Figure \ref{fig:window-func} examines the periodicity and the derived time of conjunction in more detail. We compare the standard fit posterior to the radio-only model and to the RV-derived solutions. The longer of the two astrometric periods closely corresponds with the LSP that has been identified in the past and in our RV modeling. The shorter period is also present in every astrometric fit variation we ran. 
Since our data is  sparse, we  examined whether the observed periods could be artifacts of the data's sampling. Neither the $\sim 6$~yr LSP peak, nor the $\sim 4.5$~yr shorter-period peak appear to be driven by the window function of the dataset (see Figure \ref{fig:window-func}).  Nor does aliasing between the $\sim 1800$~d window function peak and the observed periods appear to explain any of the features of the periodicity. 
Instead, the shorter period appears to be best explained by aliasing with four times the fundamental mode period of $\sim 416$~d. A priori, we might not expect the radial pulsation of the fundamental mode to have an astrometric imprint, but departures from ideal spherical symmetry in Betelgeuse's chaotic and distorted outer layers could drive photocenter movement as the star pulsates \citep[e.g.][]{2009A&A...506.1351C,2010A&A...515A..12C,2011A&A...528A.120C,2012JCoPh.231..919F,2017A&A...600A.137F,2023A&A...669A..49A}.  We therefore posit that the fundamental mode is showing up as a quasi-periodic signal in the astrometry, and the Keplerian signal is also showing up as a peak at the LSP, $\sim 2100$~d. A more complete model could include contributions from each of these mechanisms, but more data will be needed to support this more sophisticated analysis.

As shown in Figure \ref{fig:angles}, the implied inclination of the companion is edge-on, which is physically necessary for model in which RV and photometric variation is driven by the companion moving ahead and behind the primary star. The orbital position angle of nodes, $\Omega$, is also consistent with measurements of the position angle of the star's rotation axis \citep{2018A&A...609A..67K}, implying that the putative companion's orbit is aligned with the star's measured rotation axis. 

Finally, and importantly, we discuss the signal amplitude. The recovered companion  from  astrometry alone is considerably more massive than inferred from RVs. The astrometric  measurement is in tension at a several-sigma level with the RV-derived companion mass (see Table \ref{tab:fit_params} as compared to Table \ref{tab:fit_params_rv}), even when accounting for the fact that the RVs only constrain $M_b \sin(i)$, because $i\sim 90 \deg$ such that $\sin(i)\approx 1$. 
 With an astrometric amplitude of $a_a \approx 5.7 \pm 0.7$~mas, we would expect an RV amplitude of $K\sim 5.3$~km~s$^{-1}$, which is ruled out by the RV data.  We discuss possible origins of this difference  in Section \ref{sec:ast-jitter}, but more data are likely needed to draw concrete conclusions.

\subsection{Interpretation and Caveats}

In presenting a possible identification of the LSP in Betelgeuse's astrometric data we note that there are concerns that may influence future analyses. Some of these points have already been raised and examined by \citet{2017AJ....154...11H}, but it is important to mention them in the context of our re-analysis. 

\subsubsection{A difference in the radio- and Hipparcos frames?}

One potential source of correlated error in our fits is astrometric calibration errors between the Hipparcos realization of the ICRS frame and the radio instruments' realization of the ICRS frame. This is a theme in the field of analyzing absolute astrometry from different instruments. For example, \citet{Brandt:2021a}  empirically correct for sky-position-dependent calibration errors between Gaia and Hipparcos. We note that Betelgeuse is so bright that it saturates Gaia, which means we cannot take advantage of these efforts (or the latest Gaia measurements themselves). However, \citet{Brandt:2023a} find, through a similar analysis of the residual distribution of Hipparcos IAD errors, that adding a 0.15 mas offset to all Hipparcos residuals improves their statistical properties. To investigate whether the addition of this offset affected our standard fit, we reran it with the addition of this offset, finding no change in the derived parameters greater than 1$\sigma$. 

It is absolutely possible that there remain uncalibrated frame differences between the Hipparcos and radio datasets, and that these could be affecting the results of our standard fit. However, performing a full analysis of the frame differences between the radio and Hipparcos frames is out of the scope of this paper, and would require more information than is available in the current dataset. Regardless, we argue that any possible frame differences are not the primary driver of our LSP solutions. Radio-only fits (i.e. fits excluding the Hipparcos data entirely) identify similar periodic signals, as shown in Figure \ref{fig:window-func}.

\subsubsection{Differences across radio instruments?}

This analysis uses  published radio astrometry from three different instruments (VLA, e-MERLIN, ALMA), \citep[as summarized by][]{2017AJ....154...11H}. Because the data we use is absolute astrometry, the position of Betelgeuse is only known as well as the positions of the phase calibrators. The VLA observations  used one of two phase calibrators, ALMA used the same one, and e-MERLIN used a different one \citep{Harper:2008a,2017AJ....154...11H}.  Relatedly, because observations are performed at a range of frequencies from several GHz to hundreds of GHz, there might be dependence on observing frequency of the material being probed and thus the photocenter as a function of frequency and observation epoch. Further, this effect could be exacerbated because radio and sub-mm data probe radii outside the optical photosphere where the complexity of the circumstellar region might come into play \citep[e.g. the continuum disk of Betelgeuse as observed by ALMA is 50\% larger than in the optical][]{2017A&A...602L..10O}.   It is therefore possible that systematic offsets between the respective radio measurements drive the observed periodicity. We suggest that this seems unlikely given the coincidence of the derived periods with the independently-expected periodicities and on-sky angles.

\subsubsection{Underestimation of astrometric jitter?}
\label{sec:ast-jitter}

Because Betelgeuse's radius of $\sim 21$~mas is much larger than any of the signals discussed, a major concern is whether we are truly observing a shifting stellar position or just a wandering photocenter \citep[e.g.][]{2011A&A...528A.120C,2019MNRAS.487.4832C,2020A&A...640A..23C,2022A&A...661L...1C,2023MNRAS.523.2369S}. This question has driven concerns, for example, about the inversion of Gaia parallaxes of AGB stars and red supergiants \citep{2022A&A...661L...1C}.  For example, \citet{2022A&A...661L...1C} suggest that optical photocenter displacements of 1-5\% of the radius are typical based on simulated red supergiants, which would be $\lesssim 1$~mas given Betelgeuse's angular size.  However, there is certainly evidence for more extreme excursions, including during Betelgeuse's great dimming, where apparent hot and cool regions dramatically shifted the star's on-sky appearance \citep[e.g.][]{2021Natur.594..365M}. Further, it is worth considering that these effects might  be observing-wavelength dependent, particularly in radio and sub-millimeter observations probing beyond the optical photosphere.    

 With these concerns in mind,  previous astrometric fits for Betelgeuse have included an astrometric ``jitter'' term-- adding an additional white noise to all the observational errors bars in quadrature. The assumption is that Betelgeuse's convective activity results in stochastic variations in astrometry that can potentially be modeled with an added white noise. \citet{2017AJ....154...11H}, for example, systematically examined how different amounts of jitter affect parameter inferences in their Table 3. 

\begin{figure}
    \centering
    \includegraphics[width=\linewidth]{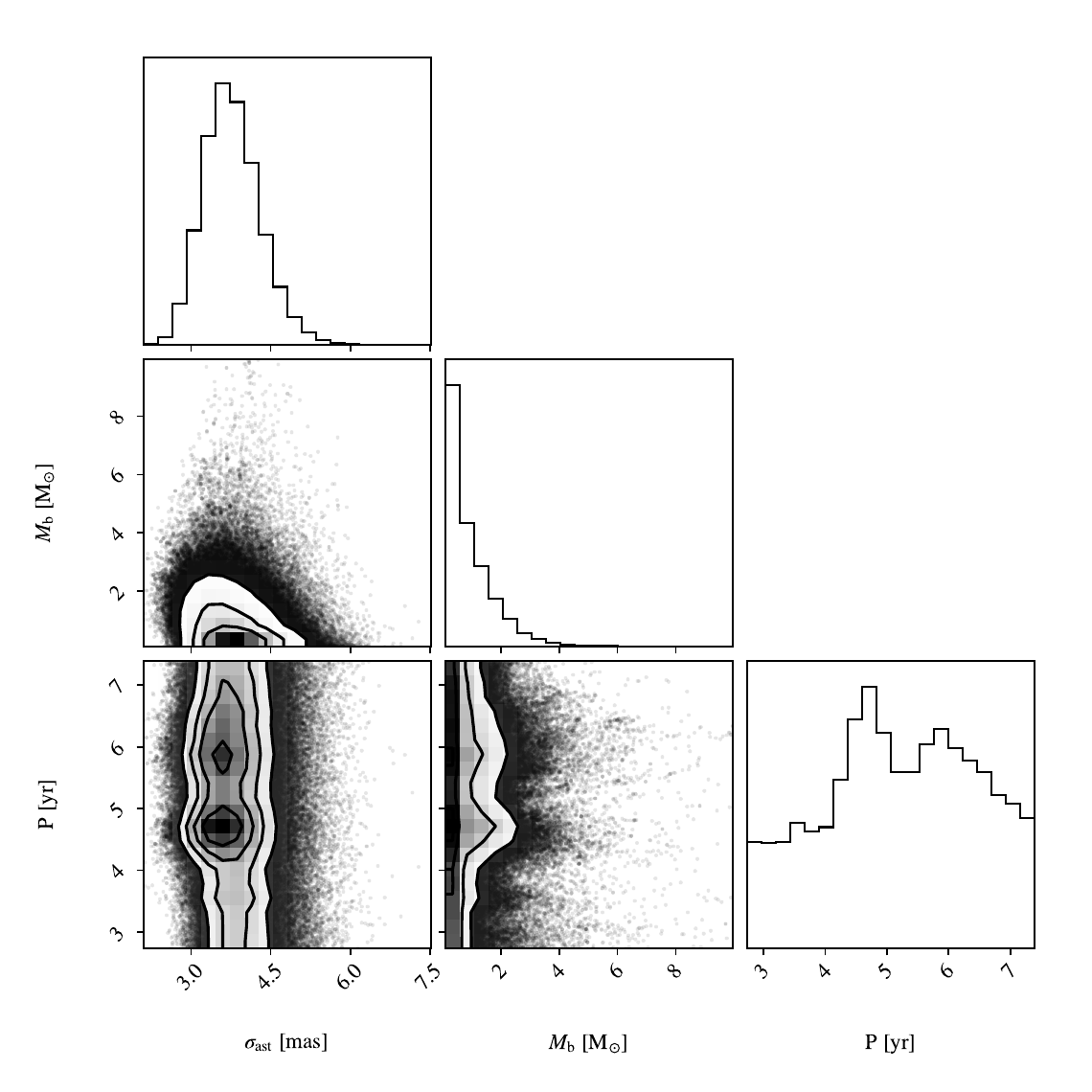}
    \caption{ Corner plot for a model including astrometric jitter. Here we focus on astrometric jitter versus companion mass. Although our fit prefers to treat the signal as white noise (which we expect given number of data points), the anticorrelation between jitter and mass is evidence for the reality of the signal. The inclusion of some degree of astrometric jitter alleviates the tension between companion mass in astrometry and RV. The most likely ``true'' model is a combination of astrometric jitter and a Keplerian signal. More data, collected at a range of sampling intervals, can delineate these possibilities. }
    \label{fig:jitter}
\end{figure}

In order to evaluate the restricted case of including an additional white noise parameter, we performed a fit allowing both the standard parameters and an additional jitter term, added in quadrature to both the Hipparcos and radio measurements, to float. The key quantities of the result are shown in Figure \ref{fig:jitter}. We make the following observations: 1) even when an arbitrary jitter is included, the periodic signals still appear in the fit posterior. The correlation between $P$ and $\sigma_{\rm ast}$ demonstrates that the periodic model is favored when $\sigma_{\rm ast}\lesssim 3.5$~mas (because there are preferred periods), and at higher $\sigma_{\rm ast}$, there are solutions in which the full variation can be explained by white noise. Relatedly, jitter is anticorrelated with companion mass, $M_{\rm b}$. Thus a small value of $\sigma_{\rm ast}$ requires a companion $M_{\rm b} \gtrsim 2 M_\odot$, to explain the full variation in the data, but a moderate $\sigma_{\rm ast}$ is compatible with a lower companion mass. In a model-selection sense, however, there is not a preference for the periodic solution relative to the jitter-solution, as can be seen by the secondary mass being consistent with zero in this fit. As a side note, this derived value of astrometric jitter exceeds the 2.4mas value recommended to add to all Hipparcos errors by \cite{Brandt:2023a}, and can be viewed as a more conservative version of a fit incorporating that correction.

We have evidence from photometry and RV that the stochastic, convective noise signature is more likely pink, $\propto 1/f$, than white at most timescales (e.g. as seen in Figure \ref{fig:periodogram} and described by \citet{2000A&AS..145..283K}). This means that a large jitter term probably overestimates short-term stochasticity if it captures long-term variations. For example, within the 2.5~yr Hipparcos window, there is some hint of ``slope" relative to the radio points, as discussed in the context of the proper motion inferrences. If this is not the signal of a LSP, it could be a slow, stochastic drift with a few-mas amplitude. On the other hand, in a given 6-month interval, there does not appear to be a need for a few-mas jitter (as, for example, we can see by examining Figure \ref{fig:dreamplot}). These factors hint at, but don't give us enough data to fully constrain, the astrometric power spectrum.  

We conclude by noting that the tension between the RV-derived mass and the astrometry-derived mass can be assuaged by picking a value of astrometric jitter between 0 and 4.5 mas. At present, though, this means that it isn't possible to conclude between a model with an astrometric LSP and one with $\sigma_{\rm ast}\sim$few~mas. More likely, a realistic signal might comprise contributions from all of these terms,  including a frequency-dependent photocenter displacement term, as adopted in modeling the RV data. More monitoring, particularly monitoring that probes a range of timescales from short to long, is needed to firmly trace the origin of the possible astrometric LSP. 

\subsubsection{Interpretation}
As we have shown, the current radio and Hipparcos dataset favors either a significant stochastic term or contains signatures of the LSP. If the LSP has been detected, it would be present in astrometry, RV, and photometry. The consistency of the inferred time of conjunction and period with RV determinations, as well as the consistency of the position angle with rotational measurements are suggestive that our detection favors the LSP. Though we have found that our analysis is generally robust with respect to our modeling choices, its sensitivity to the small number of radio positions (each painstakingly collected) is clear. Future monitoring is the most likely path to either confirming or ruling out the astrometric LSP.

\section{Betelguese And Its Companion}\label{sec:LF}

In this section we derive the physical properties of a companion implied by Betelgeuse's LSP.

\subsection{Betelgeuse's Properties}\label{sec:BGprop}

We begin with constraints on the physical properties of Betelgeuse itself. Because neither the distance, nor the mass, of Betelgeuse are known with any degree of precision, we must consider these properties free parameters in our analysis. However, as \citet{2016ApJ...819....7D} and \citet{2020ApJ...902...63J} have pointed out, the mass, radius, distance, and luminosity of the star are not independent. Several invariant measures are available: the presence of a fundamental-mode oscillation with a characteristic period of $\sim 400$~d, and the spectroscopic measurement of a characteristic effective temperature of $T_{\rm eff}\sim 3600$~K \citep{2020ApJ...902...63J,2020ApJ...891L..37L,2020ApJ...905...34H}. \citet{2016ApJ...819....7D} and \citet{2022ApJ...927..115L} additionally consider surface abundances \citep[reviewed by][]{2023A&G....64.3.11W}. The fundamental mode period scales with the properties of the star with $(R^3 / GM)$, with a dimensionless coefficient determined by the interior structure. Given Betelgeuse's nearly-completely convective envelope, this dimensionless coefficient is not particularly sensitive to the precise mass or radius of the star \citep[e.g.][]{2020ApJ...902...63J,2023ApJ...956...27M}. 

Meanwhile, the astrometric fitting of Section \ref{sec:astrometry} has updated the parallax and proper motion estimates for Betelgeuse, based on the latest Hipparcos IAD reductions. These new  inferences, shown in Figure  \ref{fig:plx-comp}, alleviate the previously-reported tension between the original Hipparcos team solutions for Betelgeuse's parallax and proper motion and the fits based on radio data by \citet{2017AJ....154...11H}.  Our updated  estimate of the distance to Betelgeuse for the standard astrometric model fit that includes a companion is $172^{+13}_{-11}$~pc, based on parallax inversion (Table \ref{tab:fit_params}). This is consistent within one sigma of \citet{2020ApJ...902...63J}'s inference of $168^{+27}_{-15}$~pc based on asteroseismology. Based on the observed $42.28\pm0.43$~mas apparent photospheric radius of \citet{2014A&A...572A..17M}, the estimated stellar radius is 
\begin{equation}\label{BGradius}
    R_a \approx 796 R_\odot \left( d \over 175~{\rm pc} \right),
\end{equation}
with similar $\sim 1$\% error to \citet{2014A&A...572A..17M}'s angular size measurement. We also note that equation \eqref{BGradius} is a linear scaling of \citet{2020ApJ...902...63J}'s estimate of $764 R_\odot$ at $168$~pc. Given a $3600$~K effective temperature \citep{2020ApJ...891L..37L}, the star's luminosity is 
\begin{equation}\label{BGluminosity}
    L_a = 9.6 \times 10^4 L_\odot \left( d \over 175~{\rm pc} \right)^2
\end{equation}
The implied mass can be computed based on the fundamental mode period \citep{2020ApJ...902...63J}, and its dimensionless frequency $\omega_{\rm fm} = (2\pi / P_{\rm fm}) / (GM / R^3)^{1/2}$, as 
\begin{equation}\label{BGmass}
    M_a \approx \left(\frac{2 \pi R^{3/2}}{\omega_{\rm fm} P_{\rm fm} G^{1/2}   } \right)^2  \approx 18.4 M_\odot  \left( d \over 175~{\rm pc} \right)^3 ,
\end{equation}
where in the second equality, we have adopted $\omega_{\rm fm} = 1.45$ for this model-dependent quantity \citep{2023ApJ...956...27M}. Based on {\tt gyre} modeling of fundamental mode frequencies, we estimate a possible range of $\omega_{\rm fm} \approx  1.4 - 1.5$ for broadly Betelgeuse-like stellar models. 

\begin{figure}
    \centering
    \includegraphics[width=\linewidth]{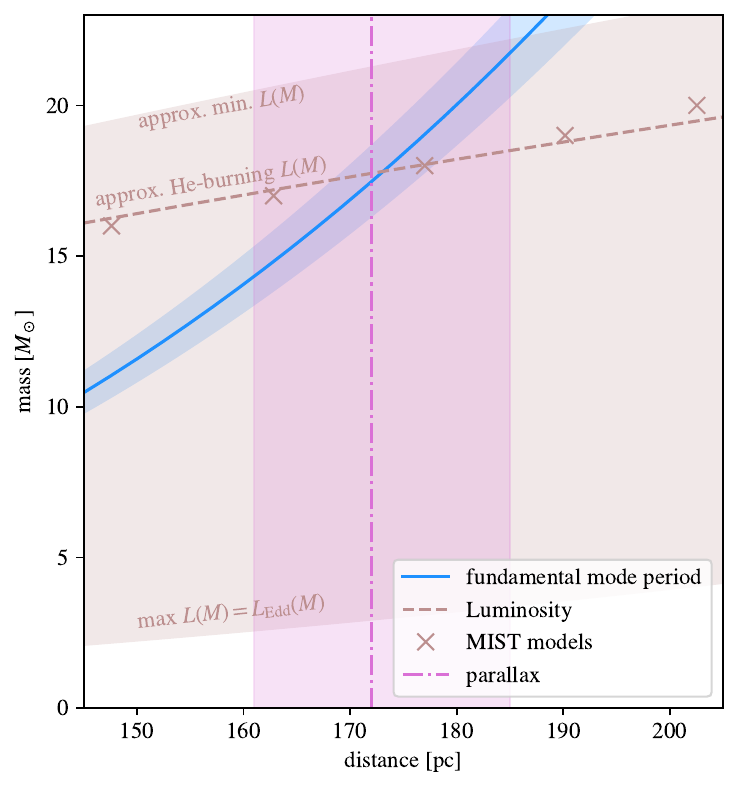}
    \caption{Joint constraints on Betelgeuse's mass and distance from stellar modeling and parallax measurement. The fundamental mode period constraint, equation \eqref{BGmass}, comes from matching the star's fundamental mode period of approximately $416$~d. The Luminosity constraint, equation \eqref{BGluminosity}, comes from matching the star's modeled luminosity as a function of mass and evolutionary stage to its distance. The dashed line shows the He-burning luminosity implied mass, $M_a \sim 18.5 (d/185~{\rm pc})^{2/3.5}$, and compares to the mean luminosity of solar-metalicity MIST models \citep{2016ApJ...823..102C}. We estimate that because supergiants can be a minimum of a factor $\sim 2$ dimmer as they evolve from core He burning toward C burning, the mass could be $\sim 20$\% higher. Because stripped-envelope supergiants retain their luminosity, there isn't a practical lower mass limit from luminosity. Taken together these constraints agree quite well at a distance $\sim 175$~pc and a mass $\sim 17.5M_\odot$.    }
    \label{fig:mass_distance}
\end{figure}

The stellar luminosity is a steep function of mass, roughly proportional to $M^{3.5}$, offering a parallel constraint based in equation \eqref{BGluminosity} \citep[e.g.][]{2023arXiv230614232C}. For example, a star that satisfies equation \eqref{BGmass} at a distance of 220~pc has a mass of $\sim 34.5M_\odot$, and would have a luminosity $\sim 6.5\times 10^5 L_\odot$, which is significantly more than the $1.5\times 10^5 L_\odot$ required by equation \eqref{BGluminosity} at 220~pc. Figure \ref{fig:mass_distance} compares the fundamental mode mass-radius constraint to the luminosity-mass constraint as a function of distance to Betelgeuse. We see that the parallax, fundamental mode, and luminosity constraints intersect around $d \sim 175$~pc and a mass of $\sim 17.5 M_\odot$, implying a radius $\sim 780R_\odot$ and luminosity $\sim 9.3 \times 10^4 L_\odot$ \citep[much as derived by][]{2020ApJ...902...63J}. We emphasize that because these constraints involve stellar modeling, they  are not completely empirical.

From the position, radial velocity, and proper motion, we derive Betelgeuse's velocity relative to the Local Standard of Rest (LSR) to be $(U_{\rm LSR},V_{\rm LSR},W_{\rm LSR}) = (-10.96^{+0.41}_{-0.47}, 0.69^{+0.68}_{-0.87}, 27.01^{+4.19}_{-3.22})$~km~s$^{-1}$, for the standard astrometric fit including a companion. This gives a net velocity relative to the LSR of $\sim 29_{-3}^{+4}$~km~s$^{-1}$. As discussed by \citet{2020ApJ...902...63J} and \citet{2023A&G....64.3.11W}, this high velocity is notable, since it quite discrepant from the usually-low $W_{\rm LSR}$ velocities of many young stars and is likely indicative of some dynamical kick in the evolutionary history of Betelgeuse. That this velocity is a large fraction of the orbital velocity of the potential companion indicates that the putative binary could have easily  bound, but that this kick this would have been an important episode in the system's orbital and evolutionary history.

\subsection{Binary System Properties}

We have approached fitting a binary model to Betelgeuse's LSP data in RV and astrometry in the previous sections. The LSP has been long-identified in RV \citep{1908PASP...20..227P,1910LicOB...6...17C,1911LicOB...6..154C,1928MNRAS..88..660S,1984PASP...96..366G} and photometry \citep{1869ABSBo...7..315A,1911MNRAS..71..701B,1913PA.....21....5S,1913AN....194...81O,1928PWasO..15..177S}. The signal identified in astrometry in Section \ref{sec:astrometry} is novel. As we synthesize potential system parameters from these binary models, we are able to combine constraints from each method. Where they overlap, the RV fit comprises many more measurements over a longer time baseline and is generally more tightly constraining. 

There are some key agreements between the datasets. In particular, both the period peaks identified and their phase -- as measured by the time of conjunction (Figure \ref{fig:window-func}) are consistent. This agreement is indicative that the same motion is being probed by the line-of-sight and on-sky motions. The inclination constraint from the astrometric data suggests $i\sim 90\deg$, consistent with being approximately edge-on, such that $\sin(i) \approx 1$. 

However, the amplitude of the motion is discrepant. The RV fitting favors a $K\sim 1.5$~km~s$^{-1}$, and strongly disfavors a velocity semi-amplitude of larger than a few km~s$^{-1}$. The best-fit astrometric amplitude would be consistent with an RV motion of $\sim 5.3$~km~s$^{-1}$. The implied mass ratios thus differ, from $q\sin(i) \approx 0.035 \pm 0.01$ in the RV case to $q \approx 0.12 \pm 0.02$ in the astrometric solution. It is important to point out that the RV mass ratio for $\sin(i)=1$ is consistent within roughly $3\sigma$ of the astrometric. 
We suggest that the realistic mass ratio must be $q\sin(i) \lesssim 0.07$ in order to be consistent with the RV time series -- given that the recent data with the highest precision and sampling favors $K\approx 2.68$ \citep{2022csss.confE.185G,2024arXiv240809089G}. 

For an adopted Betelgeuse mass  of $M_a \sim 17.5 M_\odot$, the companion mass is thus $M_b \lesssim 1.25 M_\odot$. The RV-fit preferred value is $0.6 M_\odot$. The companion semi-major axis would then be $a\approx 1818 R_\odot \approx 8.45 $~au. The orbital velocity of the companion is $\sim 43.1$~km~s$^{-1}$. The semi-major axis is just more than twice the stellar radius, $2.3 R_a$, a ratio which is not sensitive to the distance because it is set by the ratio of the fundamental mode period to the LSP period. 

Intriguingly, in addition to its edge-on orientation, the measurement of the on-sky position angle of the orbit, $\Omega$, overlaps closely with the measured orientation of Betelgeuse's spin by \citet{2018A&A...609A..67K}. This suggests the possibility of a spin-orbit interaction, which we explore more fully in Section \ref{sec:tides}.

\begin{figure}
    \centering
    \includegraphics[width=\linewidth]{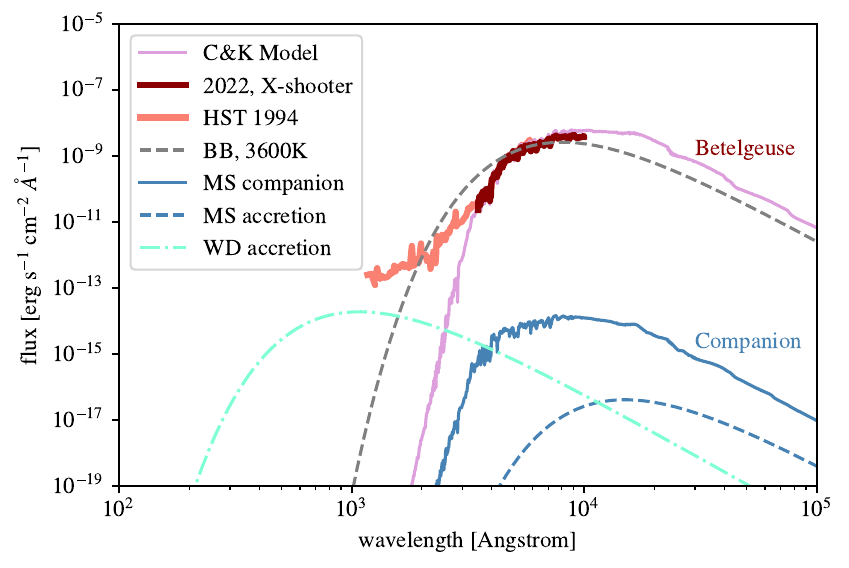}
    \caption{UV and optical spectra of Betelgeuse and its possible companion. For Betelgeuse we show data from the MAST archive, while for the companion we show a model spectrum for $T_{\rm eff}=3750$~K, $\log g=4$ from the library of \citet{1997A&A...318..841C}. The companion is nearly $10^6$ times less luminous than Betelgeuse, and, because of their similar effective temperatures, the companion's flux is substantially below that of Betelgeuse across the ultraviolet to near-infrared range.  }
    \label{fig:spectra}
\end{figure}

Given its mass and the approximate age of Betelgeuse system, the companion object would most likely be a G, K, or early M dwarf in spectral type. At $0.6 M_\odot$ and $\sim 8$~Myr age, a K5 dwarf might not have reached the H-burning main sequence yet, indicating that its radius could be somewhat inflated and its effective temperature could be lower than the $\sim 4400$~K such an object would have on the main sequence.  Binaries of similarly extreme mass ratio ($q\approx 0.07$) involving M dwarfs are known \citep{2020MNRAS.499.3775S} but lie at the limits of current detection strategies. Given the $\sim 8.5$~au separation, we do not believe that the coexistence of a young or pre-main sequence companion is incompatible with the bulk of Betelgeuse's 8~Myr main sequence. For the majority of its evolution, the primary star would be relatively compact, on the order of $\sim 10R_\odot$, with the similarity of Betelgeuse's radius and the orbital separation only coming with the star's post-main sequence growth. A solar mass pre-main sequence star is only inflated by a factor of a few in radius  \citep[$\lesssim 5 R_\odot$, as estimated from the MIST stellar tracks,][]{2016ApJ...823..102C} for ages greater than $10^5$ years, again much less than the orbital separation. The bolometric flux intercepted by the companion is low compared to the companions own luminosity, ($< L_\odot$ at $5R_\odot$ and $<0.03L_\odot$ at $1R_\odot$), so we do not expect dramatic energetic effects from the irradiation.

A low effective temperature and luminosity $L_b \sim 0.2 L_\odot$ implies that a companion would be effectively hidden in plain sight -- outshone by Betelgeuse at all wavelengths.  Figure \ref{fig:spectra} highlights this issue. We show UV and optical spectra of  Betelgeuse along with an 3750~K, $\log g=4$ model spectrum (normalized to $0.2L_\odot$) that mimics a slightly inflated pre main sequence low-mass companion \citep{1997A&A...318..841C}. Betelgeuse's high-energy flux originates in its luminous and dynamic chromosphere. We might wonder whether high-energy flux from a more-compact, young companion star might be visible. The X-ray limits on Betelgeuse are sensitively dependent on temperature but are stringent, for {\it Chandra}'s 0.2-10~keV band, $L_X/L_{\rm bol}\lesssim10^{-8}$, implying $L_X \lesssim 10^{-3}L_\odot$, if a CSM hydrogen column number density of $10^{23}$~cm$^{-2}$  is assumed \citep{2006astro.ph..6387P,2020ATel13501....1K}. Thus $L_X / L_{\rm bol, companion} \gg 10^{-3}$ for $L_{\rm bol, companion}\sim L_\odot$ is ruled out, levels that do not generally appear in even rapidly-rotating young stars \citep{2011ApJ...743...48W}. Thus a young, rotating companion with  $L_X / L_{\rm bol, companion} \sim 10^{-3}$) might be on the cusp of detection, making X-ray wavelengths the space where the putative companion would be similar in brightness to Betelgeuse. At lower energies in the UV-NIR, the companion is hidden at orders of magnitude lower flux than Betelgeuse. 

Table \ref{tab:joint} summarizes our findings with respect to the possible Betelgeuse binary system.

\begin{deluxetable}{cccc}
    \tablewidth{\textwidth}
    \tablecaption{ Betelgeuse System Parameters
    \label{tab:joint}}
    \tablehead{Parameter & Value & Unit & Notes}
\startdata
$d$ & $172^{+13}_{-11}$ & pc & 1,3 \\
$R_a$ & $782 \pm 55$ & $R_\odot$ &  3 \\
$L_a$ & $ 9.3\pm1.3 \times 10^4$ &  $L_\odot$ & 3 \\
$M_a$ & $17.5\pm 2$ &  $M_\odot$ & 3 \\
$M_b$ & $0.60\pm 0.14$ & $M_\odot$ & 2,4 \\
$P$ & $2109\pm 9$ & d & 2 \\ 
$T_{c}$ &  $2023.12^{+0.34}_{-0.35}$ & year & 2  \\
$\Omega$ & $60\pm 6$ & deg & 1 \\
$i$  & $98\pm 5$ & deg & 1 \\
\enddata
\tablecomments{1) Astrometric constraint 2) RV constraint 3)  from parallax inversion and stellar evolution modeling of fundamental mode frequencies and stellar luminosities. Uncertainties in these quantities are driven by the uncertain distance, but are directly correlated with distance as in eqs. \eqref{BGradius}, \eqref{BGluminosity}, and \eqref{BGmass}; 4) based on the RV time series, $q\lesssim0.07$ is an upper limit companion to primary mass ratio.  }
\end{deluxetable}

\subsection{Companion's Photometric Signatures}

A companion orbiting so close to Betelgeuse would interact with the star's extended chromosphere \citep{1984ApJ...284..238H,1987ApJ...314..690H,1987ApJ...317L..85D,2000ApJ...545..454L,2001ApJ...551.1073H,2009A&A...504..115K,2011A&A...531A.117K,2020A&A...638A..65O,2024ApJ...966L..13D} and also its wind, which has a mass loss rate in the range of $10^{-7}$--$10^{-6} M_\odot$~yr$^{-1}$ \citep{1986ARA&A..24..377D,1994ApJ...424L.127H,2022AJ....163..103H,2023ApJ...942...69M}. The radial  velocity is relatively low and inhomogenous \citep[e.g.][]{2017A&A...600A.137F} in the chromosphere region where the companion orbits, typically $\lesssim10$~km~s$^{-1}$. Betelgeuse's extended atmospshere is highly inhomogeneous and is of comprised of gas from hundreds to thousands of Kelvin  \citep[e.g.][]{2000ApJ...545..454L,2001ApJ...558..815L,2001ApJ...551.1073H,2017A&A...602L..10O,2020A&A...638A..65O,2022AJ....163..103H,2021A&A...650L..17K,2024ApJ...966L..13D}. The relative velocity of the companion star and the surrounding wind is, therefore, mostly due to the orbital velocity. The characteristic cross section for capture of gas from the wind is the ``accretion radius" \citep{1939PCPS...35..405H}, 
\begin{equation}
    R_{\rm acc} \approx \frac{2GM_b}{v_{\rm rel}^2},
\end{equation}
which for our nominal parameters gives $R_{\rm acc} \sim 10^2 R_\odot$. The cross section of captured gas is $\pi R_{\rm acc}^2$, thus the mass capture rate compared to Betelgeuse's mass loss rate, $-\dot M_a$, is 
\begin{equation}
\dot M_b \sim - \frac{\pi R_{\rm acc}^2 }{4\pi a^2} \dot M_a  \sim 10^{-9} \left(\frac{\dot M_a}{-10^{-6} M_\odot{\rm yr}^{-1}} \right)   M_\odot{\rm yr}^{-1},
\end{equation}
 or a fraction of $\sim 10^{-3}$ of the the mass lost from the primary. 
The accretion of this material releases a luminosity of 
\begin{equation}
    L_{\rm acc} \sim \epsilon \frac{G M_b \dot M_b}{R_b},
\end{equation}
where $\epsilon$ is the radiative efficiency. For $\epsilon\sim0.1$ and a main sequence companion, we find $L_{\rm acc}\sim 3\times 10^{-3} L_\odot$, and that if the emitting surface is $4\pi R_b^2$, then the effective temperature of the accretion flow is $\sim 1900$~K. Since both this luminosity and temperature are small compared to that of the companion star, this low mass capture rate would likely be unobservable. If, instead, we imagined a white dwarf companion with $R_b\sim 0.015R_\odot$, the accretion temperature rises to $2.6\times 10^4$~K, with a total luminosity of $\sim 0.1L_\odot$. 

Were the companion a black hole or neutron star, the accretion luminosity would be $L_{\rm acc}\sim \eta \dot M_b c^2$,  with $\eta \sim 0.1$, or $\sim 10^3 L_\odot \sim 10^{-2} L_{\rm bol}$, approximately 1\% Betelgeuse's bolometric luminosity.  Empirically, however, known neutron star -- red superigant binaries that accrete from a wind are under-luminous by up to two orders of magnitude compared to similar estimates \citep[e.g. see section 3.4 of ][]{2024MNRAS.528L..38D}. The known systems have X-ray luminosities around $L_\odot$ \citep[e.g.][]{2020ApJ...904..143H,2020ApJ...896...32G,2024MNRAS.528L..38D}.   Either the theoretical or empirical flux levels are ruled out by orders of mangitude by the soft X-ray limit of $L_X/L_{\rm bol}\lesssim10^{-8}$  \citep{2020ATel13501....1K}, and are additionally ruled out by a high-energy {\it NuSTAR} non-detection \citep{2021PhRvL.126c1101X}. 

The ratio of orbital velocity to radial velocity of the wind shapes the geometry of the shock cone and gravitationally-focused tail lagging behind the companion star. The trailing angle of the spiral will be 
\begin{equation}
    \Theta_{\rm tail} \sim \frac{v_{\rm wind}}{v_{\rm orb}} \sim 0.25
\end{equation}
or $\sim 22.5 \deg$, adopting a nominal radial velocity of $v_{\rm wind}\sim 10$~km~s$^{-1}$, but noting that it changes by a similar amount both spatially and in time  \citep{1987ApJ...317L..85D,2000ApJ...545..454L,2001ApJ...558..815L}. This means that the tail would be wandering, but extends largely in the orbital direction, trailing behind the orbiting companion \citep[e.g. on larger scales][]{2020Sci...369.1497D}. The opening angle of the shock cone is similar. For a representative sound speed of a few~km~s$^{-1}$, the Mach number of the orbital motion is $\sim 10$. The opening angle of the mach cone is $\Theta_{\rm cone}\sim \mathcal{M}^{-1} \approx c_s / v_{\rm orb} \sim 6 \deg$. For a simulation perspective on possibly similar gas dynamics, see \citet{2020ApJ...892..110C}.

The mass in the tail can be very roughly estimated to be 
\begin{equation}
    \Delta m_{\rm tail} \sim \dot M_b \frac{P}{2\pi} \sim 10^{-9} \left(\frac{\dot M_a}{-10^{-6} M_\odot{\rm yr}^{-1}} \right)   M_\odot,
\end{equation}
the captured mass accumulated over the inverse orbital frequency, $P/2\pi$. Given the cross section of $\pi R_{\rm acc}^2$, we can estimate a surface density of $\Sigma_{\rm tail} \sim \Delta m_{\rm tail} / \pi R_{\rm acc}^2$. 

The minimum in optical brightness arrives approximately half an orbit after the companion conjunction. A possible secondary eclipse of the companion would have far to small of an amplitude to explain the light curve. 
If, instead, we attribute the LSP's photometric amplitude to varying obscuration from this tail as hypothesized by \citet{2021ApJ...911L..22S}, we can estimate its extinction based on the captured mass in the tail.
 In the half-orbit following conjunction, the spiral wake trails radially outward by $\sim 0.5 \Theta_{\rm tail} \times 2\pi a$, or about $0.8a$, therefore reaching $\sim 16$~au, or $\sim 4R_a$. If the tail material cools as it expands, it may reach the $\sim 1500$~K temperatures where dust condensation is efficient at this location. 
 
If a fraction $f_{\rm dust}$ of the tail forms dust with an optical depth of $\sigma_{\rm dust}$, then the dust optical depth is $A_V = 1.086 \tau$, or 
\begin{equation}
\begin{split}
    A_V &\sim f_{\rm dust} \Sigma_{\rm tail} \sigma_{\rm dust} \\
    &\sim 0.1 \left(\frac{f_{\rm dust}}{10^{-2}} \right) \left(\frac{\Sigma_{\rm tail}}{10^{-2} {\rm cm}^{2}  {\rm g}^{-1}} \right) \left(\frac{\sigma_{\rm dust}}{10^3 {\rm g \ cm}^{-2}} \right).  
\end{split}
\end{equation}
We note that this scaling also implies that dust forms with relatively high efficiency at this phase, such that the dust-to-gas fraction is on the order of 1\%. 
If a similar surface density extends to cover a large fraction (eg. half) of the disk of Betelgeuse -- as might be the case for an extended tail along the orbital axis, then this amplitude of obscuration could explain the order of magnitude of the visual amplitude of the LSP ($\sim 0.07$~mag). 

 While varying dust extinction is the core of the \citet{2021ApJ...911L..22S} hypothesis for varying optical and infrared brightness of LSP variables \citep[and for the dimming of Betelgeuse as argued by][]{2020ApJ...891L..37L}, a subset of papers have used the Wing TiO photometric index as a probe of effective temperature to argue that effective temperature variations alone drive Betelgeuse's changing brightness in the LSP and during the great dimming  \citep[e.g.][]{2020ApJ...905...34H,2022JAVSO..50..205W}. If this argument is correct, the model of a dusty tail would need to be revised.

\subsection{Tidal Evolution of the Betelgeuse Binary System}\label{sec:tides}

With a companion orbiting at just over twice Betelgeuse's radius, the gravitational interaction of the objects is bound to affect their joint evolution. In this section, we discuss the role of tides and stellar evolution in shaping the evolution of a Betelgeuse--companion pair. 

The presence of the companion raises tides in Betelgeuse's low-density envelope. The geometric and photometric effects of these tides have been estimated by \citet{2022MNRAS.516.5021A}.  The approximate amplitude of the tidal distortion is 
\begin{equation}
    A_{\rm tide} = \frac{\delta R}{R} \sim \frac{M_b}{M_a}\left(\frac{R_a}{a}\right)^3,
\end{equation}
given our estimated system parameters, an expected relative amplitude is $A_{\rm tide} \sim 0.03\times (1/2.27)^3 \approx 0.003$. The resulting flux variation would be proportionate to this amplitude, $\delta F/F \sim \alpha_{\rm flux} A_{\rm tide}$, with $\alpha_{\rm flux}\sim 13/5$ \citep{2022MNRAS.516.5021A}, implying $\delta F/F \sim0.007$. Detecting fluctuations of this amplitude in Betelgeuse's tumultuous light curve is, at the very least, challenging.  Indeed, the photometric variation associated with the LSP is an order of magnitude larger, and the phase is such that minimum light does not align with conjunction, suggesting that another mechanism must be at play. 

Even if the tide is not immediately evident in Betelgeuse's light curve, dissipation of the tide should couple the system's angular momentum evolution. In particular the spin angular momentum of Betelgeuse's envelope will exchange with the orbital angular momentum of the tide. Betelgeuse's envelope has a moment of inertia that can be estimated as $I_a \sim \eta M_a R_a^2$, where the coefficient depends on the stellar density distribution, because 
\begin{equation}
    I_a = {8\pi \over 3} \int_0^{R_a} \rho(r) r^4 dr, 
\end{equation}
where $\rho(r)$ is the density profile as a function of radius. For models of Betelgeuse, the dimensionless coefficient is $\eta = I_a/(M_a R_a^2) \sim 0.11$. Comparatively, the moment of inertia of the orbit of the secondary about the center of mass is $I_{\rm orb} = \mu a^2$, where $\mu = M_a M_b/(M_a+M_b) \sim M_b$. Given the properties derived earlier, we can estimate that $I_{\rm orb} \sim 0.15 M_a R_a^2$. Thus, the total moments of inertia of the orbiting companion and envelope are similar. 

In 1879, George Darwin derived what has become known as the Darwin tidal instability, showing that when $I_{\rm orb} < 3 I_1$, tidal dissipation cannot bring the system into spin-orbit synchronization because there is not enough angular momentum in the orbit to spin up the envelope to synchronization \citep{1879RSPS...29..168D,1880RSPT..171..713D,2014ARA&A..52..171O}. This is because the effective moment of inertia of the spin-orbit synchronized system is $I_a -\mu a^2 / 3$, thus when $I_a > \mu a^2 /3$, the total angular momentum would need to increase to for the orbit to tighten and remain synchronous.  Instead, in these cases of insufficient total angular momentum for synchronization, tidal dissipation leads to runaway orbital decay \citep{1879RSPS...29..168D,1980A&A....92..167H,2001ApJ...562.1012E,2006MNRAS.373..733S,2014ARA&A..52..171O}. Given the parameters derived, Betelgeuse clearly lies in this regime, indicating that a stable,  synchronous spin-orbit configuration is out of reach for such a close, unequal pairing.

Thus the orbit of the Betelgeuse system decays as tidal spin-orbit exchange of angular momentum drains angular momentum from the orbit while adding it to the spin. To understand the rate that tidal dissipation drives changes in the Betelgeuse system's orbit, we consider the dissipation of the equilibrium tide by the star's vigorous convective motions. In dissipation driven by turbulent convection, the characteristic overturn timescale of the largest eddies is also the time over which coherence is erased -- implying that this is the characteristic timescale over which energy in, for example, the tidal bulge is dissipated. Based on this idea, the convective dissipation timescale is estimated \citep{1977A&A....57..383Z,1995A&A...296..709V,2020MNRAS.496.3767V,2023ApJ...956...27M} to be 
\begin{equation}
    \gamma^{-1} \sim \left( \frac{L_a}{M_a R_a^2} \right)^{-1/3}
\end{equation}
where we derive $\gamma^{-1} \sim 809$~d. 
The tidal torque is \citep{2008EAS....29...67Z},
\begin{equation}
    \Gamma \sim - \gamma (\Omega_{\rm orb} - \omega_a)q^2 M_a R_a^2 \left( \frac{R_a}{a} \right)^{6},
\end{equation}
where $q=M_b/M_a$, $\Omega_{\rm orb}$ is the orbital frequency, and $\omega_a$ is the spin frequency of the star. 

The timescale for change of the orbital angular momentum, $L_{\rm orb}=I_{\rm orb} \Omega_{\rm orb}$, is $\tau_{\rm decay} =   L_{\rm orb}/|\Gamma|$ \citep[e.g.][]{1996ApJ...470.1187R}. Or, in the limit that $\omega_a \rightarrow 0$, 
\begin{equation}
    \tau_{\rm decay} \sim \gamma^{-1} \frac{1}{ q (1+q) } \left( \frac{a}{R_a} \right)^8 . 
\end{equation}
Because of the similarity of the tidal forcing period to the overturn timescale, $\gamma^{-1} \sim P/2$, the decay rate may be longer by a factor related to the inability of the tide to couple to these largest-scale eddies \citep{1996ApJ...470.1187R,2008EAS....29...67Z,2014ARA&A..52..171O}.
Calculating the instantaneous decay rate of the orbit given Betelgeuse's nominal current configuration (Table \ref{tab:joint}), we find $\tau_{\rm decay} \sim 4.3\times 10^4$~yr. 

The immediate conclusion from this calculation is that Betelgeuse's orbit is changing, and relatively rapidly because the tidal decay timescale is much less than the system age of $\sim 8-10$~Myr \citep{2020ApJ...902...63J}. However, on comparable timescales, Betelgeuse itself is evolving -- changing in radius and internal structure in ways that influence the tidal evolution. In particular, for most of its evolution, Betelgeuse was much smaller than its current radius, allowing the system to persist to the current epoch. 

\subsection{Evolutionary Model}

To capture the joint evolution of Betelgeuse and its companion's orbit we compute a binary evolution model in the stellar evolution code {\tt MESA} \citep{2011ApJS..192....3P,2013ApJS..208....4P,2015ApJS..220...15P,2018ApJS..234...34P,2019ApJS..243...10P,2023ApJS..265...15J}. Our calculation treats the coupled spin-orbit evolution of the pair, following from the tidal torques. To illustrate a possible history of a Betelgeuse-like binary system, we initialize a calculation with a Betelgeuse analog of $18.8M_\odot$ a companion of $0.7M_\odot$, both with solar metalicity. The orbit is circular, and initially has a period of 2650~d  (semi-major axis of $2169.1R_\odot$). The initial spin of the primary star is synchronized with the orbit. The companion is initialized as a pre-main sequence star with core-temperature below the H-burning limit. We adopt a ``Dutch" wind for the primary star and a mixing length coefficient of 2.1 \citep[see, e.g.][ for a more detailed discussion of Betelgeuse-like models in MESA]{2020ApJ...902...63J}. 

We evolve the system until the orbital period decreases to the observed 2110~d. At this point, the primary star has a radius of $799R_\odot$ and it has decreased in mass to $18.4M_\odot$ due to mass loss by stellar winds. At this point, the star has a $4.6M_\odot$ He-core and is core Helium burning. We compute the star's fundamental mode period using {\tt gyre} \citep{2013MNRAS.435.3406T} and find it is 423~d, quite similar to the observed $\sim 416$~d. 

\begin{figure*}
    \centering
    \includegraphics[width=\textwidth]{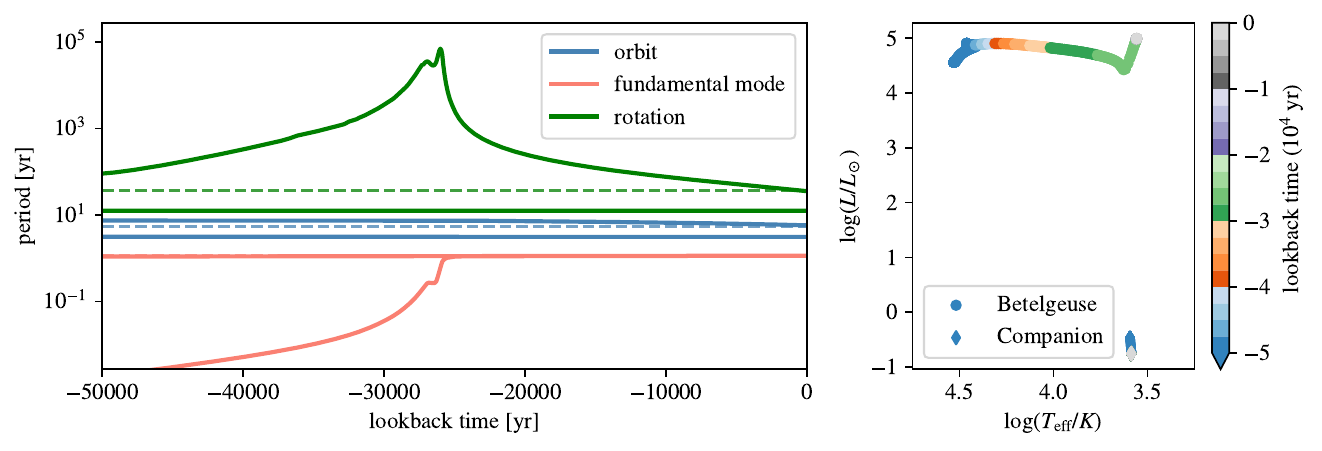}
    \caption{Evolution of an example Betelgeuse and companion system to the current date. We show quantities in terms of a lookback time relative to an epoch where the system properties match those observed. This model adopts a primary star initial mass of $18.8 M_\odot$, a companion of $0.7M_\odot$, and an initial period of $2650$~d. In particular, we show that it is possible for the tidal spin-orbit coupling to lead to  fundamental mode, orbital, and spin periods that all match those of Betelgeuse. Without tidal dissipation and a companion, Betelgeuse's spin period would be nearly $10^{5}$~yr as seen by the peak of rotation period. The right-hand panel shows the evolution of the pair in the Hertzsprung-Russell Diagram.   }
    \label{fig:mesa}
\end{figure*}

Figure \ref{fig:mesa} shows the system evolution. 
At first, the orbit widens as the primary star loses mass. But, as its radius grows, the tidal interaction strengthens and starts to spin up the primary, draining angular momentum from the orbit. When the orbital period reaches 2110~d, the spin period is 35.4~yr, similar to the 36-year observed spin period \citep{2018A&A...609A..67K}.  The action of tides on the Betelgeuse binary system provides a natural explanation for not only the system's LSP and radial velocities, but also for the rapid, $36$~yr spin. 

The solution presented here is not unique; especially given the large uncertainties in the strength of tidal torques and on the internal dissipation and transport of angular momentum in stellar envelopes. Indeed, with a slightly elevated tidal coupling coefficient (or increased secondary mass), models that reach the present spin, orbital, and fundamental mode periods are possible closer to the era of rapid growth of the convective envelope, when the star traverses the Hertzsprung-Russel Diagram (approximately $-2.6\times 10^4 $~yr in the model in Figure \ref{fig:mesa}), without needing to modify $M_a$. We note that tidal decay coefficients in this regime are likely orders-of-magnitude uncertain \citep{2014ARA&A..52..171O}, so making this adjust would be well within the realm of possibility. Such a model is consistent with the argument based on historical texts that Betelgeuse's color may have been significantly more yellow 2000 years ago \citep{2022MNRAS.516..693N,2023A&G....64.1.38N}. At present, we do not distinguish between these cases. Instead, we highlight that the presence of a companion, and the action of tides, can simultaneously account for  Betelgeuse's pulsation, rotation, and the presence of a LSP in visual magnitudes and radial velocities.

As our model system evolves forward in time, we find that orbit continues to decay until mass transfer from Betelgeuse toward the companion begins (in about 9000~yr), while the donor is still core helium burning. This mass transfer will be unstable, leading to runaway orbital decay and mass exchange \citep[e.g.][]{2018ApJ...863....5M,2020ApJ...893..106M}.  Eventually Betelgeuse swallows its companion. The likely outcome here is the ejection of a few-tenths to a solar mass of H-rich circumstellar medium (CSM), and the further spin-up of Betelgeuse's envelope as the pair merge into a single star \citep{2006MNRAS.373..733S,2024ApJ...962..168S}. Such a scenario has been considered and modeled in the specific case of Betelgeuse as a possible explanation for the star's spin \citep{2017MNRAS.465.2654W,2020ApJ...905..128S,2024ApJ...962..168S}.  When Betelgeuse explodes as a H-rich supernova in this case, the supernova blastwave would race through the mass-rich CSM, possibly creating a luminous type IIn-like object \citep{2020ApJ...892...13S,2020ApJ...891...15G,2022ApJ...933..164G,2024ApJ...962..168S}. 

However, it is also possible that the infall of the low-mass companion could eject Betelgeuse's envelope, leaving behind a close binary of a stripped He core and the companion \citep[e.g.][]{1976IAUS...73...75P,1984Natur.309..598V,1987A&A...183...47D,1999ApJ...521..723K}. Such an object might evolve to something like the observe black-hole low-mass X-ray binaries that contain black holes in tight pairings with K dwarfs \citep{2023arXiv230409368M}.

\section{Discussion}\label{sec:discussion}

\subsection{Betelgeuse's Great Dimming}
Betelgeuse's great dimming of late 2019 appears to have been caused by an anomalous mass ejection \citep{2022ApJ...936...18D}, perhaps launched from the star's turbulent interior \citep{2023ApJ...956...27M}, that left shocks racing through the atmosphere as it expanded \citep{2021A&A...650L..17K,2024A&A...685A.124J}. 

The great dimming occurred within $\sim 10$\% of the nominal phase of optical minimum of the LSP \citep{2022ApJ...936...18D,2023ApJ...956...27M}. And yet, its narrow width of several hundred days precludes it being related purely to the LSP. However, if a burst of mass loss collided with a pre-existing tail or shell trailing behind a companion object, shocking before cooling and forming dust, that could explain the sudden onset of the great dimming. If correct, this might suggest that mass ejection episodes like the one that caused the great dimming are happening continuously in a star like Betelgeuse -- this one just happened to have fortuitous alignment with the companion orbit and us as observers. A strong case for this model in an analogous star has been made by \citet{2021AJ....161...98H}, who trace the kinematics knots in VY CMa's extended atmosphere, and compare to dimming episodes in the star's light curve \citep[also, see]['s conclusions about the connection between the appearance of RW Cephei's disk in interferometric imaging and dimming episodes]{2023AJ....166...78A,2024arXiv240811906A}.

\subsection{A Low-mass, Pre-main Sequence Companion}

A startling feature of Betelgeuse's LSP and implied companion object is its low mass -- and the extreme mass ratio of the system. An immediate consequence of this low mass is that such a star would still be essentially ``forming" even as Betelgeuse itself evolves off the main sequence. Stars in the implied mass range have pre-main sequence lifetimes of $\gtrsim 10^7$~yr, so at the current epoch the companion star is unlikely to have ignited H-burning. Its effective temperature is likely cooler than one would predict at the zero-age main sequence, with our models suggesting $\sim 3800$~K at $\sim 8$~Myr age. 

However, being on the pre-main sequence doesn't necessarily rule such a companion out from being a companion to a massive star. Indeed, its radius is only inflated by a factor of a few relative to its eventual zero-age main sequence size. This means that the companion star might have a radius $\lesssim1.5R_\odot$ instead of $\sim 0.6R_\odot$ -- a difference that is inconsequential relative to the $\sim 9$~au orbital separation. Interestingly, the color similarity of a pre-main sequence companion to the red supergiant primary star is part of what makes low-mass companions especially hard to detect.

\subsection{Betelgeuse's Companion in Context}

This paper has argued that Betelgeuse's LSP can be traced to the presence of a much lower-mass companion star. At face value, this is not particularly surprising -- a large fraction of massive stars are in multiple systems \citep{2023ASPC..534..275O}. Yet the nearly 30\% abundance of LSPs among giant stars in general is surprising \citep{2004ApJ...604..800W}. \citet{2021ApJ...911L..22S} argue based on evidence from infrared secondary transits that binarity can explain most LSPs. Accounting for inclination effects, more or less all giant and supergiant stars would then be expected to host companions, and indeed companions that follow a particular period sequence. \citet{2024arXiv240809089G} have recently pointed out that the phase of optical minimum of the systems varies from tracing the times of conjunction to lying a half-orbit after conjunction -- as is the case with Betelgeuse. Indeed Antares, another of the closest and brightest red supergiants also hosts a LSP with photometric and RV variations \cite{2013AJ....145...38P} that imply a companion with $M_{\rm b} \sin(i) / M_{\rm a} \sim 0.07$. 

Is it possible that so many giants and supergiants have low-mass companions at periods comparable to the LSP sequence? 
Although the during late stages, the stars expand to fill a large fraction of the binary orbit, on the main sequence, these stars would be widely separated. This means that many tighter systems would already have merged \citep[e.g.][]{2020ApJ...900..118N}. With a random inclination distribution, Betelgeuse-like binaries on the main sequence would have a $\sim 0.25$\% chance of eclipsing, with eclipses lasting $\sim 3$~d out of a $\sim 2000$~d orbit. These eclipses might have up to a $(R_{\rm b}/R_{\rm a})^2 \sim 1\%$ photometric amplitude, depending on the observing band. In practice, these sorts of signatures are certainly possible to detect, but are not so obvious that that we can immediately rule out massive main sequence stars having low-mass companions in $\sim$years-period orbits.

 There is larger question of how ubiquitous low mass companions would affect the evolution of Betelgeuse-like massive stars. In such wide orbits, the effects might not be dramatic. Since many massive stars appear to host more-equal, closer companions \citep[e.g.][]{2017PASA...34....1D,2023ASPC..534..275O}, objects like the currently-posited companion to Betelgeuse might be wider, tertiaries orbiting comparatively far from the original, more compact systems. If these tight binaries merged on or near the main sequence, they would continue to evolve until they potentially started to interact with the lower-mass outer companion. Typically, as discussed in Section 4.5, the expectation for these high-mass ratio interactions is a simple merger, perhaps spinning up the envelope of the red supergiant but not leaving otherwise dramatic influence \citep[e.g.][]{2024ApJ...962..168S}. We would not, for example, expect these low mass companions to strip the envelopes of supergiants or to affect the appearance of most supernovae explosions unless their mergers were particularly recent.   Those speculations aside, the key questions of how these systems are assembled, why they have particular orbits, or how they give rise to the precise signatures that we see are all crucial focal points for future study.  We suggest that a detailed understanding of a few objects -- like Betelguese -- may be particularly valuable in  illuminating the features of the broader class of LSPs.

\subsection{Constraints from Future Datasets}

One thing our analysis in this paper makes clear is the immense value of the long baseline of monitoring of Betelgeuse -- in photometry, spectroscopy, and astrometry. More recently, the continuous, high cadence spectroscopy offered by STELLA \citep{2022csss.confE.185G} has been particularly illuminating -- especially in distinguishing stochastic variations from quasi-periodic pulsations and the LSP. 

In addition to continuing these spectroscopic efforts, we suggest that expansions of multi-band photometric monitoring \citep[e.g.][]{2000PASP..112..977R,2022OEJV..233....1O,2022NatAs...6..930T} and new astrometric data will be especially valuable. On the photometric side, the ability to build light curves from UV to Infrared over $\sim$decade time spans would be especially useful in unwrapping possible color shifts with orbital phase due to possible changes in the CSM.

Future, high cadence astrometric monitoring could firm up or rule out our possible detection of an astrometric LSP corresponding to a binary-star model. Higher cadence data would enable more-complex modeling of the various sources of variability -- like we are able to perform in the RV case. These models could, for example, jointly fit an orbital signal with quasi-periodic and stochastic noise due to pulsations and turbulent convection. At present, such a model is far to complex in terms of free parameters for the quantity of data we have. 

With our current astrometric data, as Figure \ref{fig:jitter} shows, a Keplerian model is not preferred in a Bayesian model selection sense over a simple addition of astrometric jitter in quadrature. However, the coincidence of the periodic signals recovered from astrometry alone with the fundamental mode frequency and the RV-derived LSP parameters hint that these signals are present in the current astrometry. We predict that if additional higher-cadence astrometry is collected over the next few LSP cycles, a more complex model consisting of a quasi-periodic component at the fundamental mode plus a Keplerian signal at the LSP will be preferred over the white-noise-only model. We specifically recommend an astrometric monitoring cadence designed to sample both the 416 d fundamental mode, the 365 d parallactic signal, and the 2100 d LSP: $\sim$10-20 pts per year with precision comparable to  the 1995-2005 VLA data. 
Although it has so far not been possible, we also suggest it may be worth investigating whether any constraints can be derived from Gaia despite the fact that Betelgeuse is highly saturated on the spacecraft's detectors. 

With richer datasets in hand, future studies could attempt to model the many sources of Betelgeuse's ongoing variability -- pushing us toward a richer understanding of this enigmatic star. In particular, it may be useful to study the relationship between the astrometry, RVs, and photometry of the primary in order to jointly model and distinguish the signatures of stellar activity and the possible Keplerian motion. For example, by studying phase offsets between RV and photometry across a number of LSP giants and supergiants, \citet{2024arXiv240809089G} have recently taken a step in this direction.

\subsection{Prospects for Direct Detection}
The evidence for a low-mass companion to Betelguese is so far indirect in this paper. We are tracing signatures of a possible companion's effects on Betelgeuse itself. We briefly mention that there may also be cause for optimism about making a direct detection of an orbiting companion. 

At a separation of $\sim 8.5$~au and a distance of $\sim 175$~pc, the orbit would have an angular separation from Betelgeuse of $\sim 48$~mas twice per orbital cycle -- or approximately every three years. With meaningful constraints on a possible inclination and on-sky orientation, a companion location is clearly predicted. However, even with a 48~mas separation, it is important to emphasize that there is an enormous total brightness contrast between Betelgeuse and the companion, and that betelgeuse's radius of 21~mas subtends nearly half of that separation. 

Indeed, using novel speckle-imaging techniques in 1983, \citet{1986ApJ...308..260K} identified bright sources in images that appeared to move between exposures. One of these, at a separation of 60~mas and a position angle of 270-degrees was at approximately the predicted separation, but did not have an orientation compatible with our measure of $i$ and $\Omega$. The speckle was measured to be only 3~mag fainter than Betelgeuse, much brighter than the putative companion.  Unfortunately subsequent observations did not confirm this detection, making it most likely an instrumental effect, and highlighting the challenge of this sort of direct imaging near a bright target.

From space, {\it HST} has been used to directly image Betelgeuse's disk \citep{1996ApJ...463L..29G}. From the ground, adaptive optics have allowed some systems to approach the diffraction limit. The VLT's SPHERE/ZIMPOL has already been applied to successfully resolve and image the disk of Betelgeuse \citep[e.g.][]{2016A&A...585A..28K,2021Natur.594..365M}. At face value the contrast ratio of the companion to primary seems far too large, but with similar effective temperatures, Betelgeuse and its possible companion have similar surface brightness. Unfortunately with a PSF FWHM of $\sim 20$~mas \citep{2016A&A...585A..28K} the companion light is diluted to a surface brightness contrast similar to the original contrast of $\sim 10^{-6}$. The Nancy Grace Roman Space Telescope will have a high contrast coronograph that is nominally focused on slightly larger angular separations $\sim4 \lambda/D \sim 200$~mas, but perhaps there is an observing mode that will be more favorable to an angular separation of $\sim \lambda/D$.   

Interferometric measurements have also been quite productively applied to Betelguese, including the first measurement of a stellar angular diameter \citep{1921ApJ....53..249M}. Interferometry, for example with VLTI, might be useful in this context \citep[e.g. ][]{2016A&A...588A.130M,2023A&A...675A..46C,2024MNRAS.527L..88D}, but it does require some modeling to reconstruct a signal from the visibilities and closure phases.  Betelgeuse's particularly large size on the sky poses a challenge for very-high resolution observatories like CHARA, so it is worth emphasizing that there may be technical challenges associated such a bright, large angular size target.

Spectroscopic diagnostics might also be useful, if the tremendous brightness contrast can be overcome (e.g. Figure \ref{fig:spectra}). In practice, even detecting more massive companions to red supergiants, with larger temperature contrasts  (e.g. O and B stars) either spectroscopically or via photometric indices is challenging \citep[see][for specific examples of selection criteria]{1998ApJS..119...83P,2019ApJ...875..124N, 2021ApJ...908...87N,2022MNRAS.513.5847P}. 
On the other hand, if the companion star is active (as many pre-main sequence stars are) it could have emission lines, perhaps enhanced by its interaction and accretion from Betelgeuse's chromosphere. If they were particularly strong emission lines might be useful diagnostics with distinct kinematic signatures following a companion's orbit. 
Regardless of the technique, observations around the phases of maximum projected on-sky distance -- for example 2024.6 or 2027.5  -- will likely be most useful.

\section{Conclusions}\label{sec:conclusions}

This paper has studied Betelgeuse's LSP, following the hypothesis of \citet{2021ApJ...911L..22S} that red giant and supergiant LSPs might be caused by close companion objects. We build on the long history of spectroscopic, photometric, and astrometric observations of Betelgeuse.  We find that each of these datasets can be explained if Betelgeuse hosts a ``little friend": a much-lower mass companion orbiting at only about 2.3 stellar radii. This proposal is nearly a century old, following up on early investigations into stellar RVs that established Betelgeuse as a possible spectroscopic binary \citep[e.g.][]{1908PASP...20..227P,1910LicOB...6...17C,1911LicOB...6..154C,1916ApJ....44..250L,1928MNRAS..88..660S} with new data and analysis \citep[also see][ for a parallel analysis]{2024arXiv240809089G}. Our key findings are:
\begin{enumerate}
    \item Betelgeuse has a long-documented LSP in RV and photometry (Figure \ref{fig:rvsummary}). We collect historic RVs and characterize this RV signal, as having a tightly-measured period of $P\approx 2109 \pm 9$~d, and an RV semi-amplitude $K\approx 1.5\pm 0.34$~km~s$^{-1}$ (Table \ref{tab:fit_params_rv} and Figure \ref{fig:RVcorner}). We predict the RV and photometric ephemeris in equations \eqref{RVformula} and \eqref{Magformula}. 
    
    \item We search for traces of periodicity in astrometric measurements of Betelgeuse's on-sky position from Hipparcos and radio data. We update the astrometric solution to Betelgeuse on the basis of the latest Hipparcos data release (Table \ref{tab:fit_params} and Figure \ref{fig:plx-comp}). The astrometry favors similar periodicity and time of conjunction as observed in the RV modeling (Figure \ref{fig:window-func}), suggesting that the LSP is detected in Betelgeuse's on-sky motion. However, the astrometric data could also be modeled with a large-amplitude white noise (Figure \ref{fig:jitter}), and more data is needed to confirm our preliminary detection. 
    
    \item Astrometric constraints orient a possible companion edge-on and align the plane of a presumed orbit with the star's spin as reported by \citet{2018A&A...609A..67K} (Figure \ref{fig:angles}). 
    
    \item We suggest a model where a low-mass, $\lesssim 1 M_\odot$, binary companion orbits just outside Betelgeuse's current radius (plausible system parameters in Table \ref{tab:joint}). The companion trails a cool, molecule and dust-rich wake that causes minimum brightness a half-orbit after conjunction \citep[see also][]{2024arXiv240809089G}. Through tidal interaction, the presence of a companion explains Betelgeuse's anomalously rapid spin (Figure \ref{fig:mesa}).
    \item The projected future lifetime of the companion's decaying orbit is short, just $\sim 10^4$~yr, after which it is enveloped by Betelgeuse \citep[a process recently simulated by][]{2024ApJ...962..168S}. 
\end{enumerate}
While it is perhaps surprising that Betelgeuse could have such a close companion, we emphasize that a low mass companion would essentially be hidden in plain sight -- nearly a million times less luminous and of similar color to Betelgeuse itself. Despite this contrast, we offer that the prospects for future constraints and modeling are bright, whether through attempts at direct detection or future RV and astrometric monitoring and modeling that disentangles stellar activity from the orbital signal. 

We close by emphasizing that whether Betelgeuse's LSP ends up being caused by a binary companion, a nonradial pulsation, or some other unknown mechanism remains to be fully resolved, despite the century of observational effort we analyze here. However, the predictions of the binary model are now clear and offer a pathway toward deeper understanding of our nearest red supergiant. 

\vspace{0.5cm}
     We gratefully acknowledge conversations with N. Evans, J. Fuller, J. Goldberg, M. Joyce, P. Kosmin, F. Nail, S. Toonen, G. Torres, and C. Zucker. 
    M.M. is supported by a Clay Postdoctoral Fellowship at the Smithsonian Astrophysical Observatory. 
    J.J.W. and S.B.\ are supported by NASA XRP Grant 80NSSC23K0280. C.D.H. acknowledges financial support from \textit{HST} GO-16655 and GO-16984. 
    E.L.N. is supported by NASA Grants 21-ADAP21-0130 and 80NSSC21K0958.

    Materials to fully reproduce the results of this study are supplied at \url{https://github.com/morganemacleod/BetelgeuseLittleFriend} \citep{macleod_blf} and \url{https://github.com/sblunt/betelgeuse} \citep{blunt_zenodo_bg}, along with other software and resources where noted in the text.  

\software{ {\tt RadVel} \citep{2018PASP..130d4504F} \url{https://github.com/California-Planet-Search/radvel}, {\tt Orbitize!} \citep{2024JOSS....9.6756B,blunt_zenodo_orbitize_bg}, {\tt MESA} \citep{2011ApJS..192....3P,2013ApJS..208....4P,2015ApJS..220...15P,2018ApJS..234...34P,2019ApJS..243...10P,2023ApJS..265...15J} \url{docs.mesastar.org}, {\tt gyre} \citep{2013MNRAS.435.3406T} \url{gyre.readthedocs.io},
{\tt astropy} \citep{astropy:2013, astropy:2018, astropy:2022} \url{astropy.org} }

\clearpage
\appendix
\section{Radial Velocities}

\subsection{Historic RV Sources}\label{sec:RVsources}

\begin{figure}
    \centering
    \includegraphics[width=0.8\linewidth]{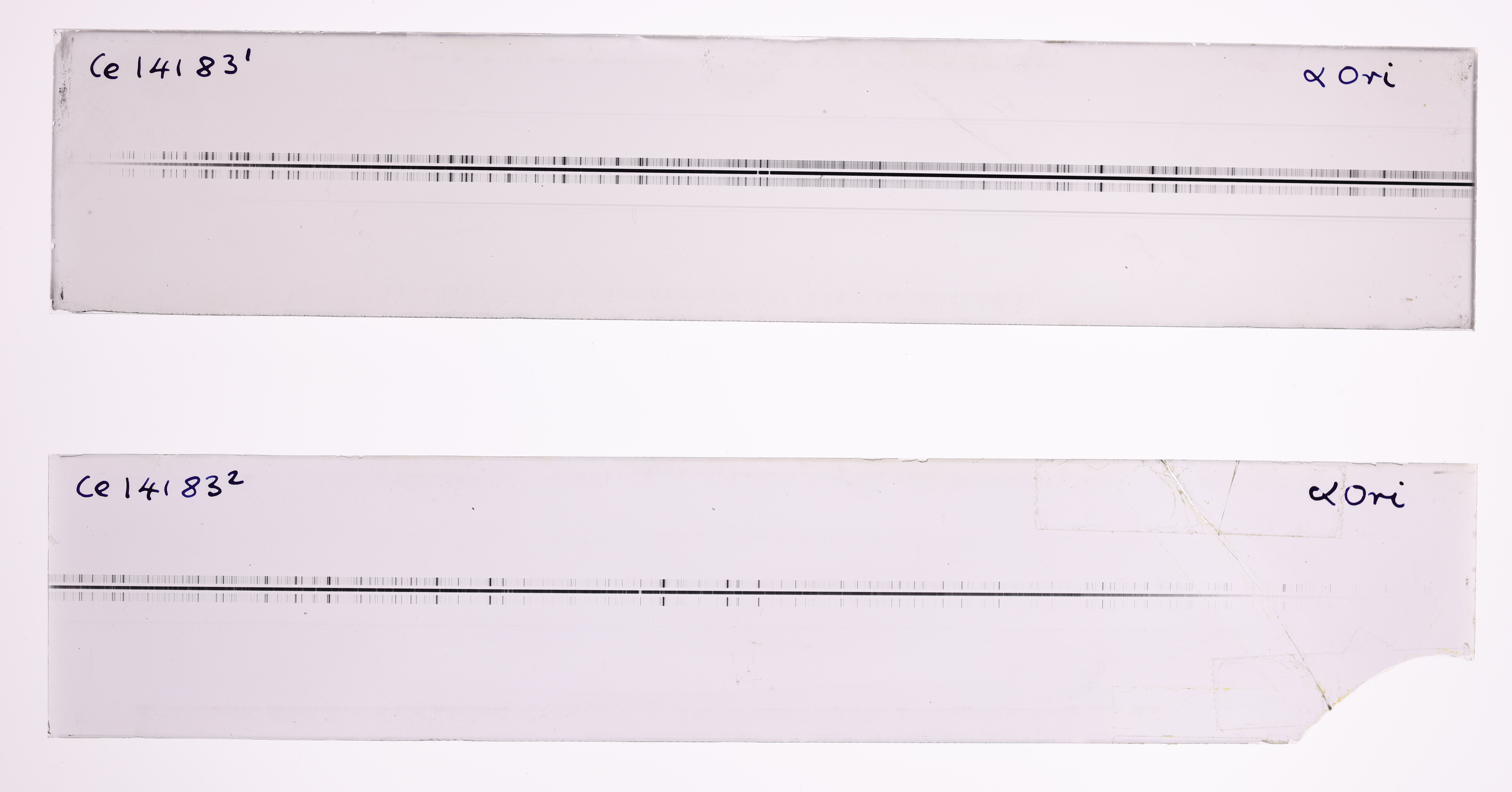}
    \includegraphics[width=0.8\linewidth]{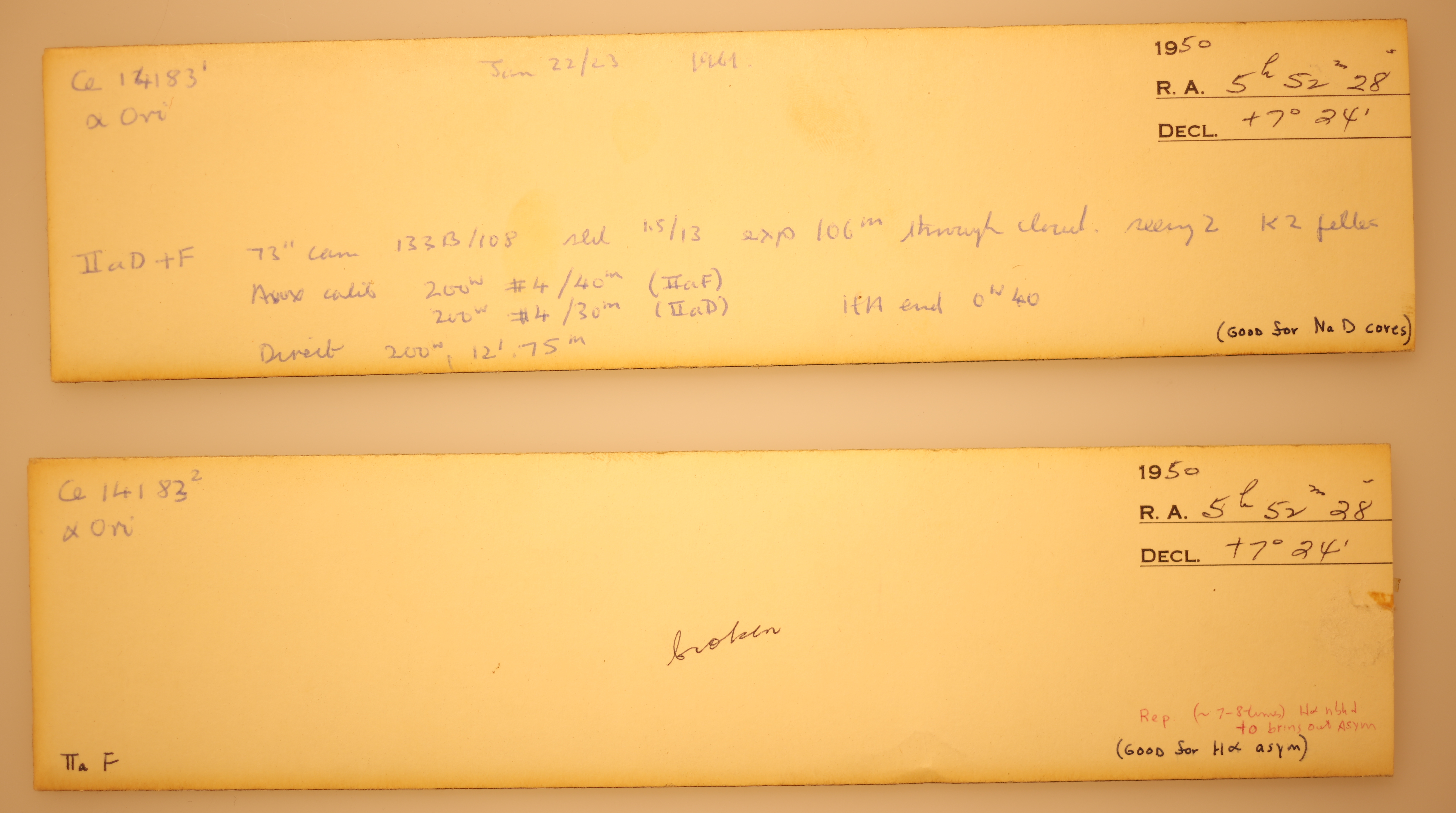}
    \caption{Red and Blue spectral plates number 14183 from the The Carnegie Science Plate Archive, taken with the Coude spectrograph on Mt. Wilson Jan 22, 1961 \citep{1962ApJ...136..844W}. Images are provided courtesy of Carnegie Science.  }
    \label{fig:Carnegie}
\end{figure}

Betelgeuse has been a target of spectroscopic study for well over hundred years. Our analysis relies on this history of observations. In addition, there have been several previous efforts to synthesize available radial velocities of Betelgeuse, and we rely on those in our analysis. We note that there are earlier spectrograms of Betelgeuse \citep[e.g.][]{1868RSPT..158..529H,1892Obs....15..393M}, but velocity discrepancies of up to $\sim 5$~km~s$^{-1}$ between observatories were still typical at this point. Our sources include:
\begin{enumerate}
    \item Lick Observatory Publications, Vol 16, page 81 \citep{1928PLicO..16....1C}. The University of Chicago copy is online at google books at \url{https://www.google.com/books/edition/Publications_of_the_Lick_Observatory_of/f_Mbo2eouOAC?hl=en}.  This publication includes plates from Mt. Hamilton, California, and Santiago, Chile. 
    
    \item Annals of The Cape Observatory, Vol 10, page 54 \citep{1911AnCap..10....1H}. The University of California, Santa Cruz copy is online at google books \url{https://www.google.com/books/edition/Annals_of_the_Cape_Observatory/7ZQ1AQAAIAAJ?hl=en}. This publication reports on measurements from the Cape of Good Hope Observatory, South Africa.

    \item \citet{1908ApJ....27..301K} reports on velocities collected in Bonn in 1905 and 1906. 

    \item \citet{1911AN....187...33B} reports on velocities from Potsdam from 1901 to 1907. 

    \item \citet{1933ApJ....77..110S}  includes an update to the Lick Observatory velocities after 1926 and also reports on Mt Wilson velocities taken 1923-1931. 

    \item \citet{1956ApJ...123..189A} analyze the photospheric and atmospheric lines in spectrograms obtained at Mt. Wilson between 1937 and 1947. 

    \item 
    \citet{1962ApJ...136..844W} reports on plates obtained at Mt Wilson primarily in 1960 and 1961. In addition several plates were analyzed that span between the Adams report and this one. In table 7 of \citet{1962ApJ...136..844W} the plates are listed by their plate number and photospheric RV. We obtained dates through communication with the The Carnegie Science Plate Archive, and an example is shown in Figure \ref{fig:Carnegie}.

    \item \citet{1975ApJ...198..369B} and \citet{1979ApJ...232..485B} report on spectrograms from the Mauna Kea observatory from 1960 into the 1970s. 

    \item \citet{1984PASP...96..366G} includes measurements from the Kitt Peak National Observatory. These were digitized from the figures, and include measurements reported by \citep{1979QJRAS..20..361G}.

    \item 
    \citet{1989AJ.....98.2233S} reports measurements from the National Solar Observatory at Kitt Peak. More measurements were reported in \citet{1998ASPC..154..393U}. We use a table of measurements from 1984 up to 1995 that were communicated  with Andrea Dupree in May, 1995.

    \item Radial velocity observations were made at the Oak Ridge observatory of Betelgeuse from 1996-2005, communicated to Andrea Dupree in April 2005. 

    \item \citet{2008AJ....135.1450G} report velocities averaged over several lines in their Figure 6. These are not calibrated with respect to an absolute offset. We have added a velocity of $+22.5$~km~s$^{-1}$ for consistency with contemporary Oak Ridge measurements. 

    \item \citet{2022csss.confE.185G} Radial velocities from the STELLA robotic observatory in Tenerife, through May 2023 by communication with Andrea Dupree in August 2023.

\end{enumerate}

Text files tabulating these literature results are released along with this publication in order to facilitate reproduction of our calculations. We use rolling 10-day averages of the data, which results in 797 data points. The errorbar allotted to each point is the maximum of the reported measurement error and the standard deviation of data within the 10-day window. For example, in the high-precision and cadence STELLA data, many points go into each 10-day bin and the representation of error comes from their dispersion. In the early glass plates, a typical error is 1-2~km~s$^{-1}$ and there might or might not be multiple measurements within a 10-day period. 

\subsection{Modeling}\label{sec:RVmodelingAppendix}

\begin{figure}[p]
    \centering
    \includegraphics[width=\linewidth]{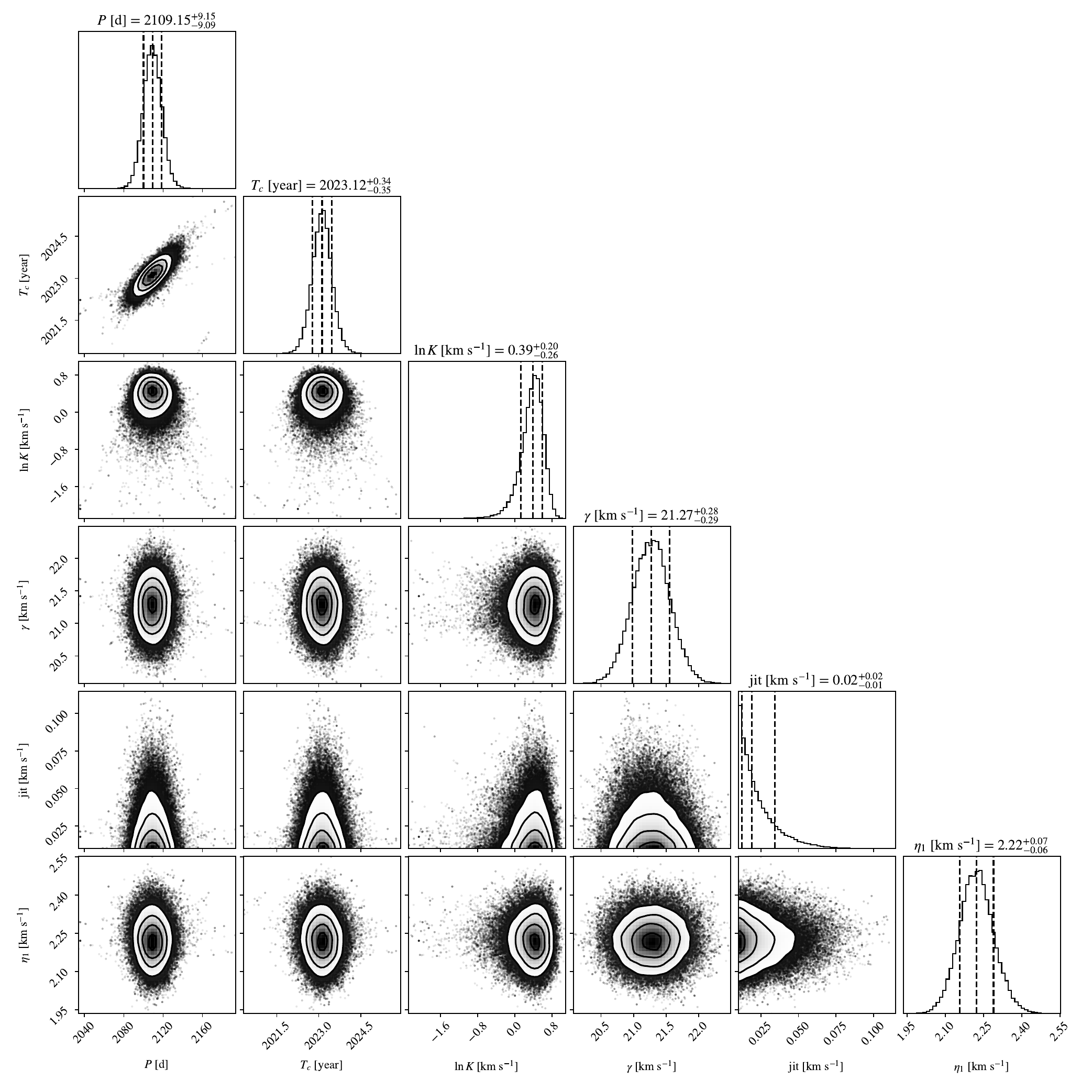}
    \caption{Full corner plot of RV model parameter posterior, the results of which are also tabulated in Table \ref{tab:fit_params_rv}. }
    \label{fig:RVcornerfull}
\end{figure}

In Figure \ref{fig:RVcornerfull} we present the full corner plot of the RV model fit. Of the parameters, the only significant correlations are between $P$ and $T_c$ and between $P$ and $\ln K$. At very low $\ln K$ the amplitude is smaller so a wider range of periods are allowed.

\begin{figure}
    \centering
    \includegraphics[width=0.9\linewidth]{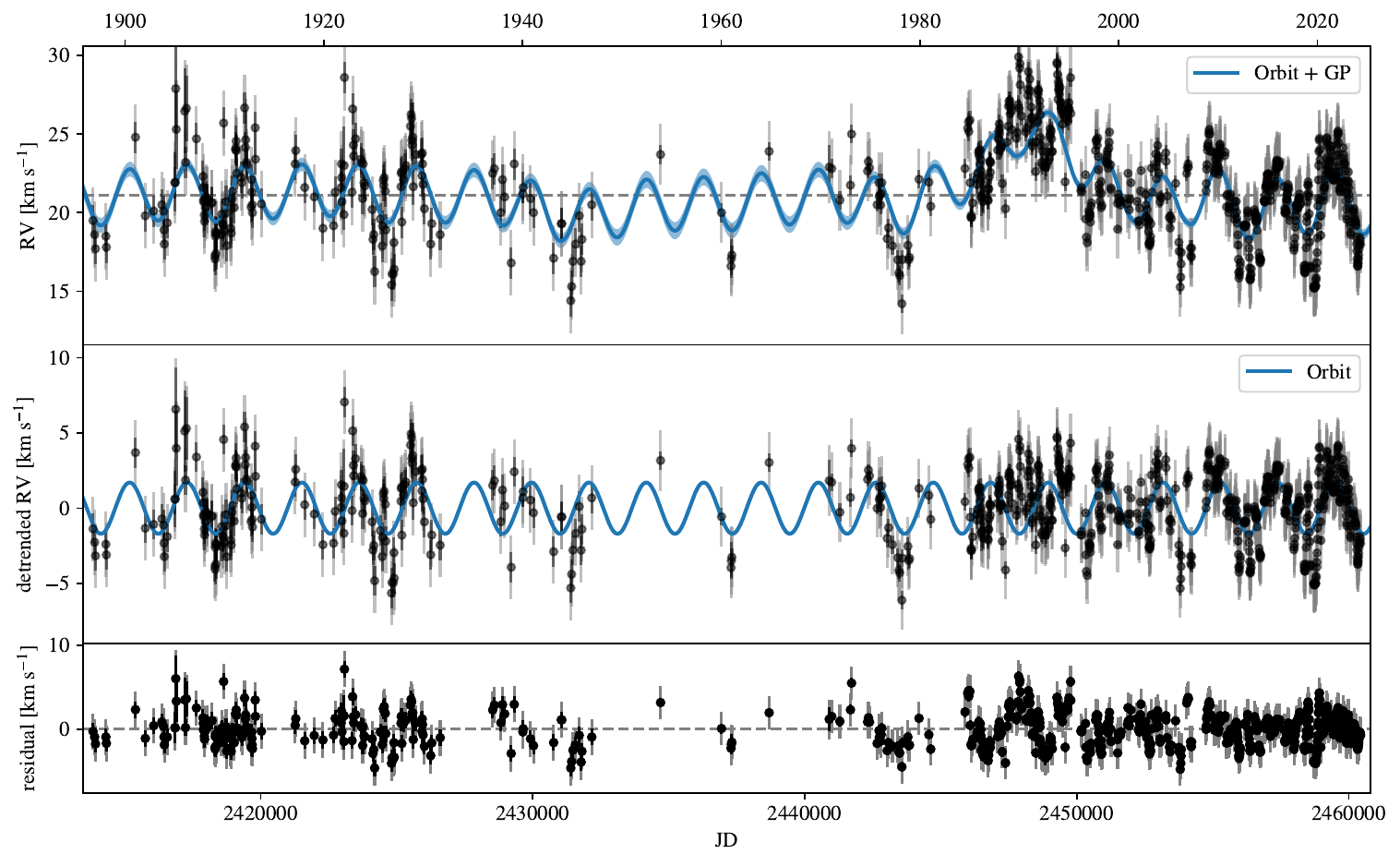}
    \includegraphics[width=0.55\linewidth]{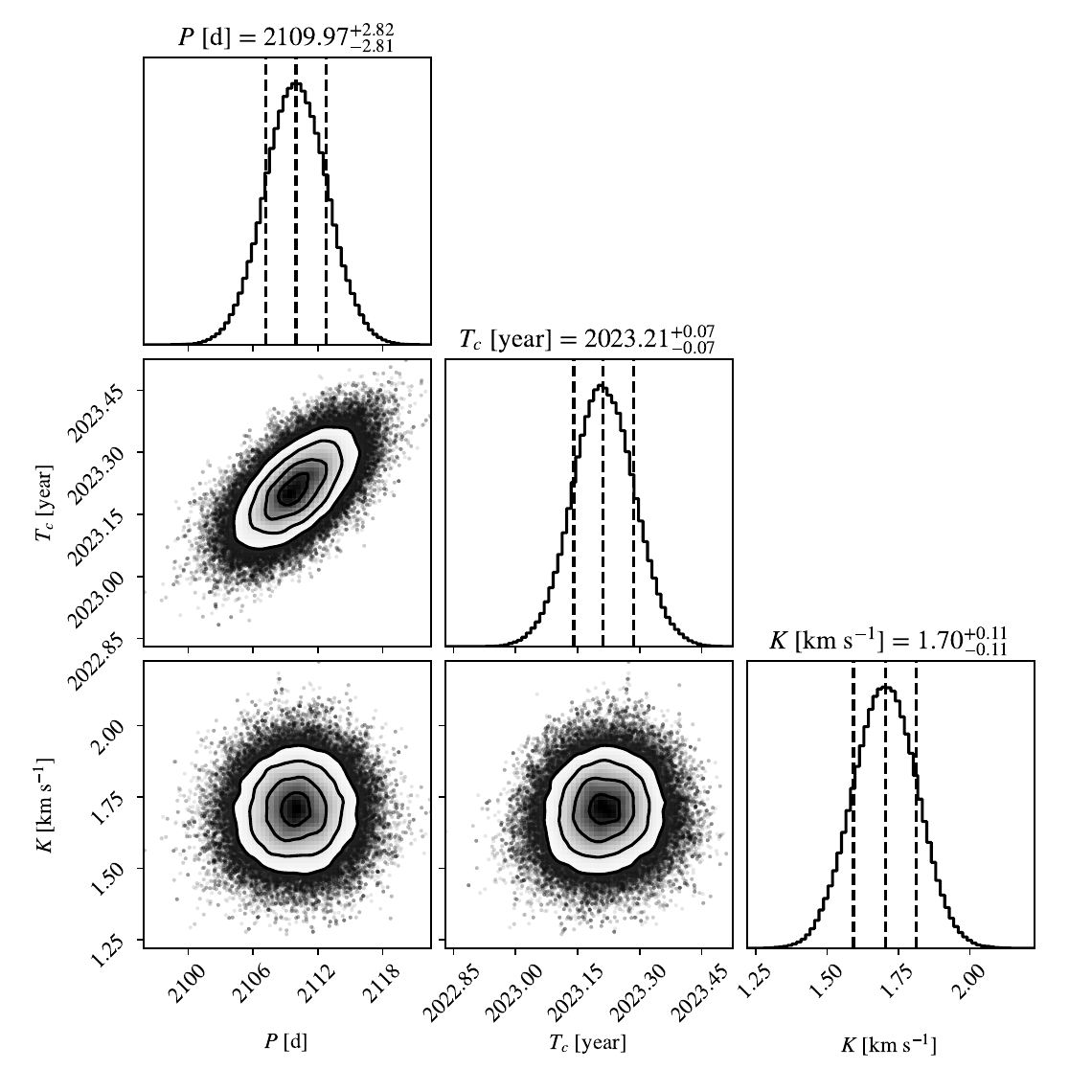}
    \caption{RV model time series and posterior parameters for a fit including a $\eta_2=4000$~d GP fixed to 0.62~km~s$^{-1}$ amplitude (Figure \ref{fig:periodogram}). This model adopts a jitter of $1.87\pm0.06$~km~s$^{-1}$, which accounts for the shorter-term stochastic and pulsational variations. This model fit derives closely overlapping parameters with our default approach of using the GP to model both the short and longer-term variability. However, it is easier, in this realization, to see how the orbital model threads through the detrended RV with only the longest-timescale drift removed. That we can model the noise from convective turbulence and pulsations in two very different manners and derive similar binary posterior results suggests that our results are robust with respect to these modeling choices. }
    \label{fig:RVjit}
\end{figure}

We modified our use of GP models versus white noise ``jitter" in a model variation shown in Figure \ref{fig:RVjit}. In this case, we treat the long-timescale variation with a Exponential GP as before, but with the timescale fixed to $\eta_2 = 4000$~d and the amplitude fixed to $\eta_1 = 0.62$~km~s$^{-1}$, which matches the low-frequency power in the GP to our default fit (see the power spectrum panel of Figure \ref{fig:periodogram}). In this case, the jitter parameter is fitted to $1.87\pm 0.06$~km~s$^{-1}$, and models the pulsations along with any other higher-frequency stellar activity. The GP only models long-term drift in the baseline RV due to wandering convective cells. Our derived model parameters are similar, but with smaller error margins because the jitter model has less flexibility than the shorter-period GP model. This model finds a slightly larger semi-amplitude, $K = 1.70\pm 0.11$~km~s$^{-1}$ than our default model (consistent within approximately $1\sigma$ of the results shown in Table \ref{tab:fit_params_rv}). 

\begin{figure}[p]
    \centering
    \includegraphics[width=\textwidth]{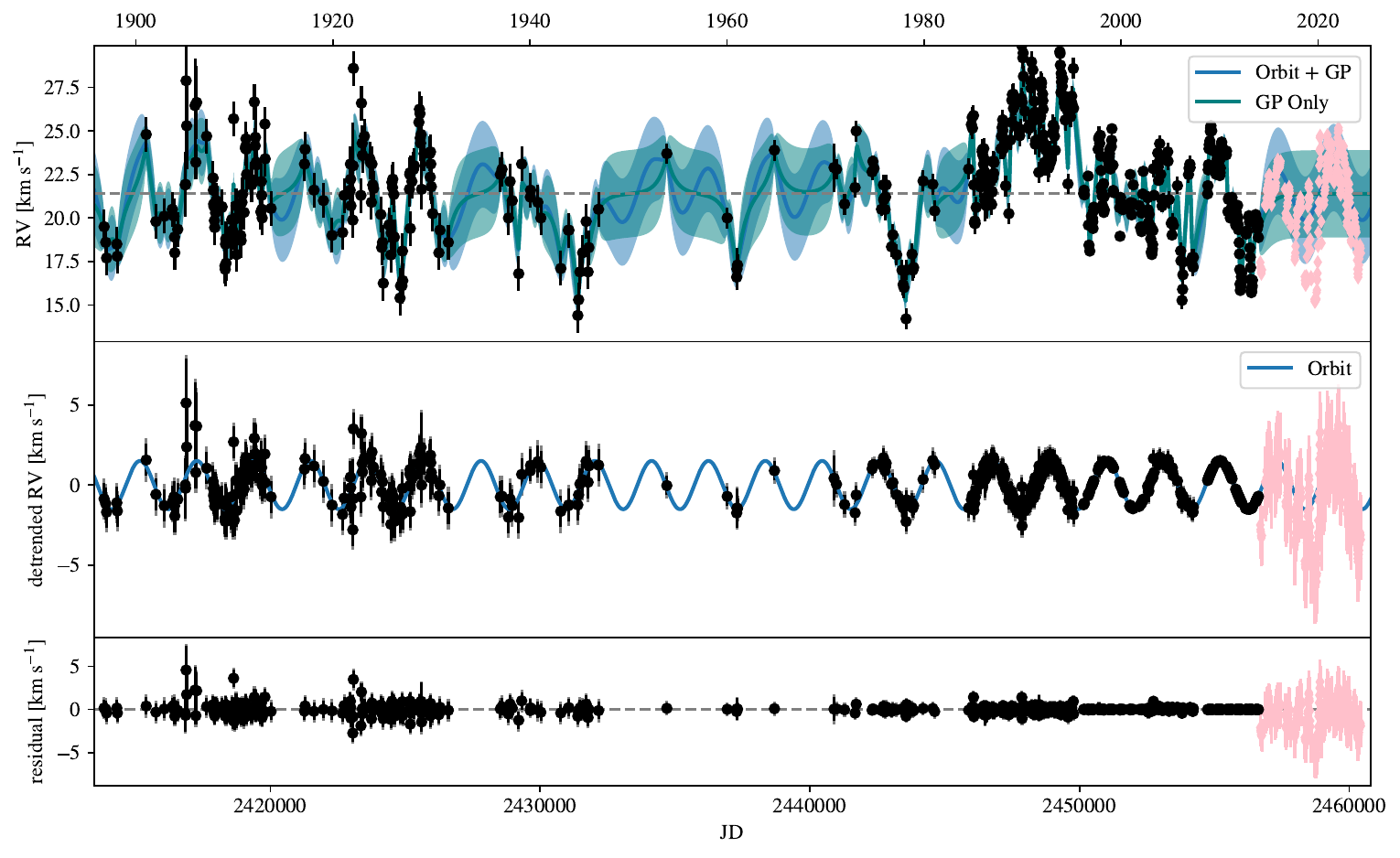}
    \includegraphics[width=0.6\textwidth]{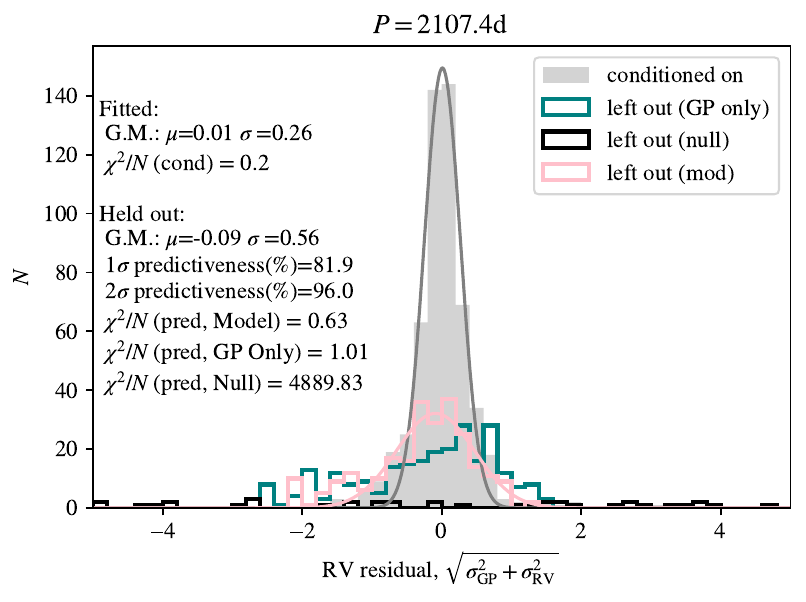}
    \caption{A comparison of fitting in which the final 249 data points (spanning a decade) are held out of the model fit, and the predictiveness of the model over the held-out time span is assessed. We find that a model with an orbit and a GP outperforms a ``GP only" model without an orbit, and this model is predictive extrapolating over decade timescales.  }
    \label{fig:dropsampleGP}
\end{figure}

Figure \ref{fig:dropsampleGP} shows an experiment to test the predictiveness of the RV model. We fit the the dataset up to 2014, dropping the last 249 datapoints that span just over a decade. In Figure \ref{fig:dropsampleGP}, the fitted points are shown in black while the left-out points are shown in pink. We note that we also experimented with random drop-sampling (eg of 20\% of the measurements, fitting on the remaining 80\%) but found that the decade-predictive test was more stringent and more realistic of our modeling goals -- we hope to be able to use a model like this to predict Betelgeuse's future RV behavior. 

The time series shown in Figure \ref{fig:dropsampleGP} is identical to Figure \ref{fig:RV} in format, except that it now includes a model without an orbital component, but with a GP prediction only. We can therefore compare the predictiveness of a model including a GP only to a GP and an orbital component. The histograms of residual in Figure \ref{fig:dropsampleGP} make this model comparison quantitative. As discussed in the main text, a GP model conditioned on the data, tends to overfit because of the sparse sampling relative to the variability time series. We see that $\chi^2/N = 0.2$ for the points that the model is fit to. We also fit a Gaussian to this distribution and find it has a width $\sigma = 0.26$.  Examining the held-out data, we find that the null prediction (constant RV) has $\chi^2 / N \sim 4900$, while a GP only model is much better, with $\chi^2/N \sim 1.0$. The GP + Orbit full model performs best of all, with $\chi^2 / N = 0.63$ over a decade of extrapolation (this can be fit by a gaussian with width 0.56). We find that 82\% of the data points are within $1\sigma$ of the prediction and 96\% are within $2\sigma$. The comparison of these models shows us that though there is an inherent tendency of very-flexible GP models to "thread-through" the available data points when sampling is sparse compared to the intrinsic timescale of the variation, this model still robustly predicts future evolution and represents an improved description of the data as compared to models without an orbital component. 

\begin{figure}[p]
    \centering
    \includegraphics[width=\textwidth]{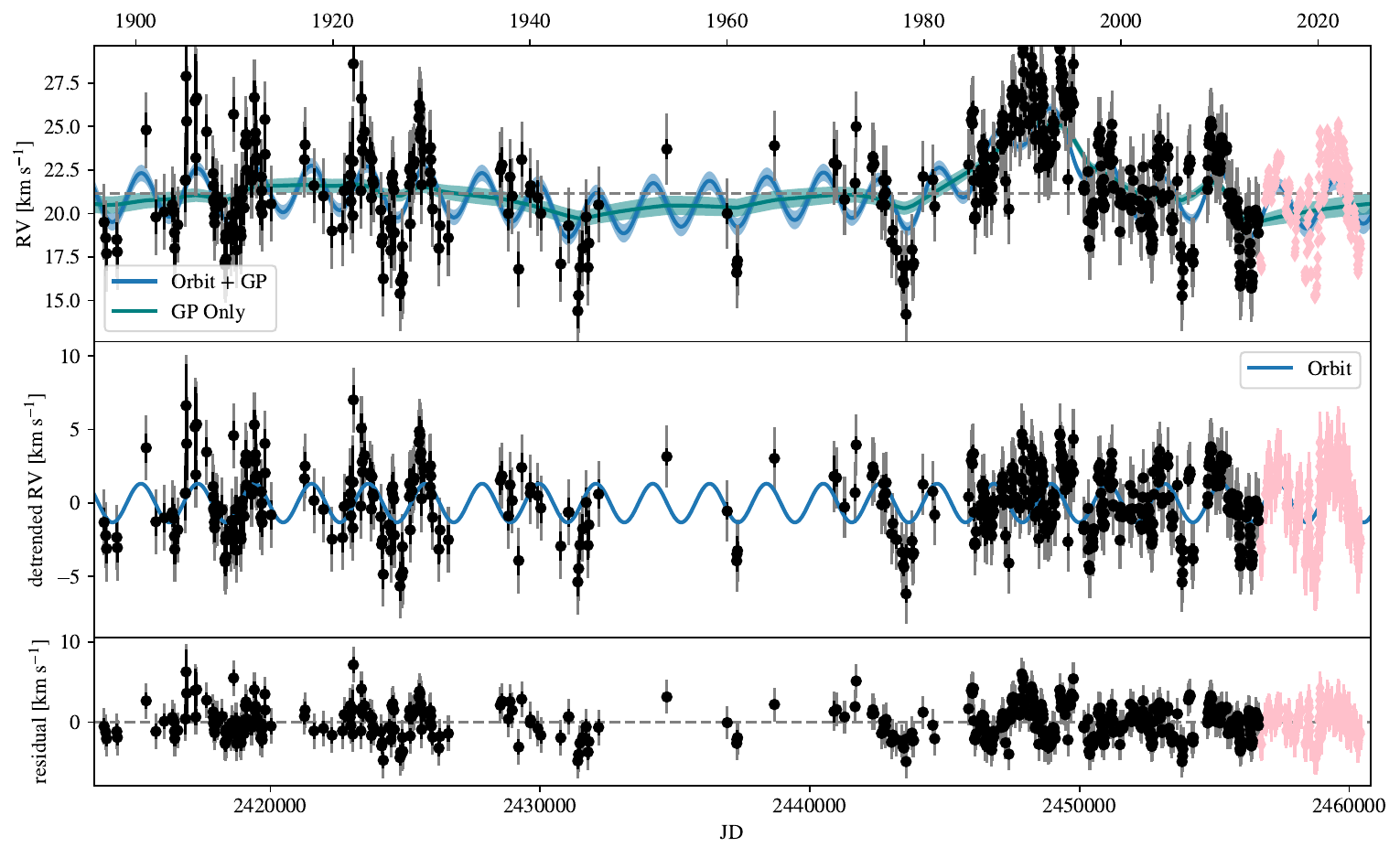}
    \includegraphics[width=0.6\textwidth]{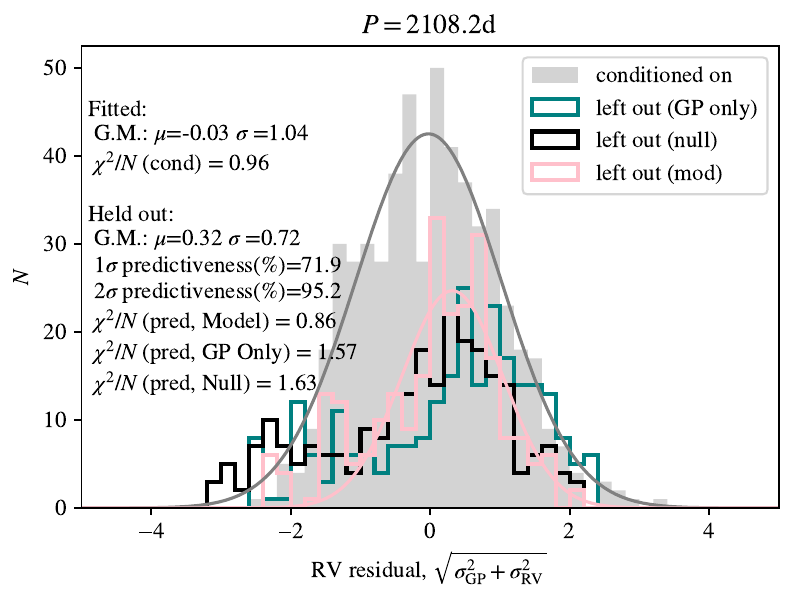}
    \caption{Identical to Figure \ref{fig:dropsampleGP}, except with the model with the longer-timescale $\eta_1 = 4000$~d GP model and white noise.  We again find that a model with an orbit and a GP outperforms a ``GP only" model without an orbit, and this model is predictive extrapolating over decade timescales.  }
    \label{fig:dropsampleJIT}
\end{figure}

Figure \ref{fig:dropsampleJIT} performs the identical drop-sampling model validation applying the long-timescale GP plus white noise model applied in Figure \ref{fig:RVjit}. In this case, the predicted error on each datapoint is inflated by the fitted jitter. Our fit has $\chi^2/N = 0.96$ for the data points it is conditioned on. The full model still performs best on the predicted decade, with $\chi^2/N = 0.86$, while the GP-only prediction has $\chi^2/N = 1.57$, which marginally outperforms the null hypothesis with $\chi^2/N = 1.63$. The reason the null hypothesis performance in this metric is better is because of the larger vaue of $\sigma_{\rm RV}$, which includes both measurement error and jitter. Though the modeling is different, the conclusion here is the same. The full model (including a binary companion) is more predictive over decade timescales than models with pure stellar activity (modeled by a GP kernel) or white noise.

We performed several additional tests. We modeled segments of the data and compared the parameters inferred. We find that the RV period and semi-amplitude inferred are stable across the time series. Secondly, we also modeled the RV time series using rejection-sampling in the {\tt the Joker} \citep{2017ApJ...837...20P}, and infer similar posterior parameter distributions.

These tests of model predictiveness and robustness to modeling choices suggest that:
\begin{enumerate}
    \item the inferences of Table \ref{tab:fit_params_rv} are not strongly sensitive to our modeling approach,
    \item that a model with a binary companion is preferred in describing the data over models without, and 
    \item that these models are predictive in that they can accurately represent  Betelgeuse's future RV time series. 
\end{enumerate} 

\clearpage
\section{Astrometry}

\subsection{Dataset}
We reproduce the dataset of radio observations from \citet{2017AJ....154...11H} in Table \ref{tab:astrometric_data} in the format of right ascension and declination offset relative to the Hipparcos position that is needed to model orbits with {\tt orbitize!}. 

\begin{deluxetable}{ccccc}[p]
    \tablewidth{\textwidth}
    \tablecaption{
    \label{tab:astrometric_data}}
    \tablehead{epoch & $\Delta$ra & $\Delta$dec & $\sigma_{\delta R.A.}$ &  $\sigma_{dec}$ \\ mjd & [mas] & [mas] & [mas] & [mas]}
\startdata
53308.414000 &  343.792994 &  129.992037 &  3.3 &  2.5 \\
53299.630000 &  343.495498 &  130.092037 &  4.8 &  2.0 \\
52862.285000 &  316.720806 &  125.892032 &  2.4 &  4.1 \\
52798.994000 &  305.564684 &  129.492029 & 12.0 &  5.0 \\
52377.200000 &  273.137561 &  106.792024 &  4.9 &  1.8 \\
52323.180000 &  269.121358 &  104.692023 &  3.1 &  1.7 \\
51906.706000 &  238.479212 &   87.792018 &  1.3 &  1.2 \\
51889.138000 &  252.015305 &  100.492020 & 26.0 & 13.0 \\
50901.965003 &  183.144850 &   69.892011 &  5.7 &  3.1 \\
50438.386000 &  148.040252 &   66.792007 &  5.9 &  8.9 \\
47035.075008 &  -79.990888 &  -44.807998 & 15.0 & 15.0 \\
45125.526505 & -192.295861 &  -65.207988 & 30.0 & 30.0 \\
45122.606505 & -201.072016 & -102.207987 & 30.0 & 30.0 \\
56120.979995 &  558.734248 &  194.792098 & 17.5 & 17.5 \\
57087.970002 &  589.971379 &  218.592110 & 22.2 & 22.2 \\
57177.030005 &  594.582562 &  253.792111 & 22.5 & 22.5 \\
57334.709998 &  628.645926 &  233.092125 &  2.0 &  2.0 \\
57616.018007 &  690.822713 &  216.692150 & 14.1 & 14.1 \\
\enddata
\tablecomments{Radio data in \texttt{orbitize!} input format (for convenience), defined as offsets from the published Hipparcos position and epoch (J1991.25). Data are compiled from \cite{Harper:2008a} and \cite{2017AJ....154...11H}.}
\end{deluxetable}

\subsection{Hipparcos Fit Reproduction}

\begin{figure}[p]
    \centering
    \includegraphics[width=0.3\linewidth]{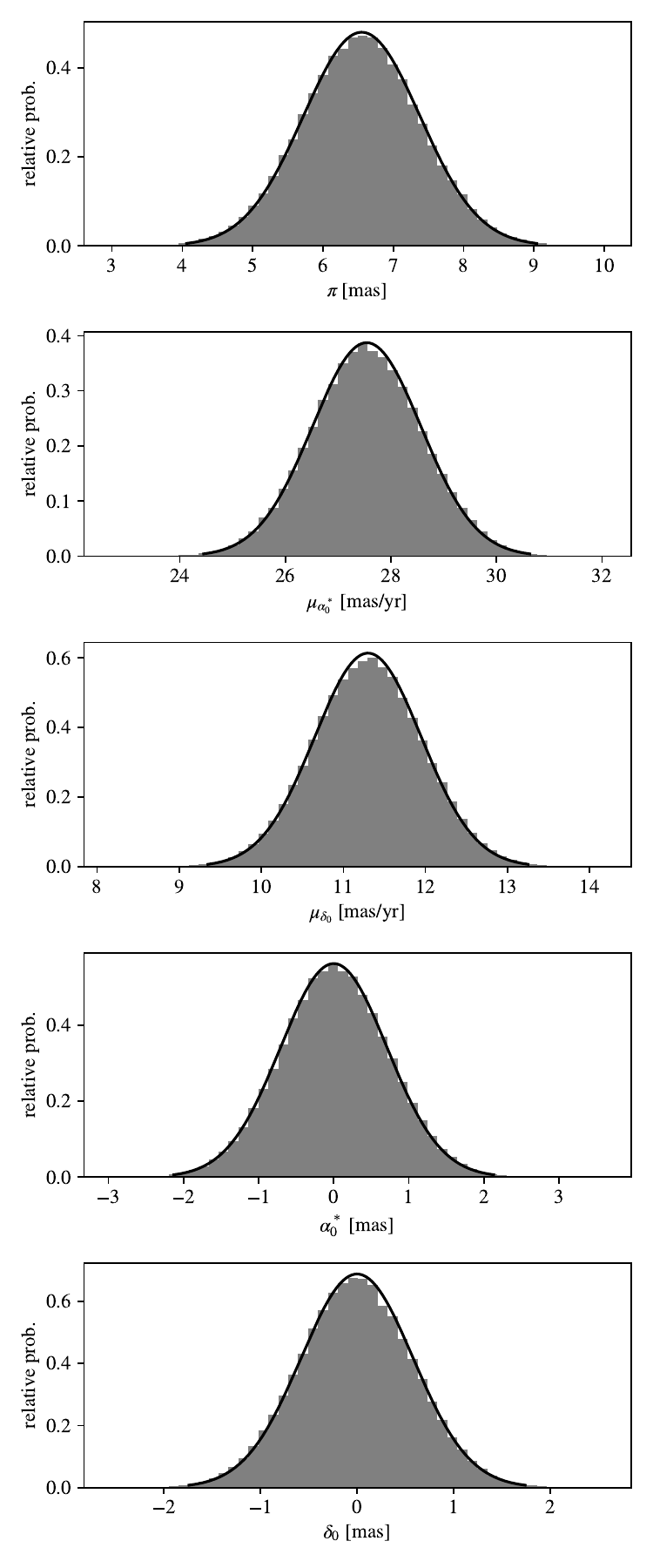}
    \caption{Reproduction of the 5-parameter published Hipparcos fit, following the method of \cite{Nielsen:2020a}. By first reconstructing the raw intermediate astrometric data (IAD) from the available residuals and published astrometric fit parameters, then refitting the 5-parameter solution, we confirm that the reconstructed raw IAD are suitable for orbit fitting. The grey histograms show our 5-parameter fit performed with \texttt{orbitize!}, and the black Gaussians show the expectation from the published 2021 Hipparcos astrometry.}
    \label{fig:hipparcos-reproduction}
\end{figure}

Following \cite{Nielsen:2020a}, we first ensured that we could reproduce the 2021 Hipparcos-2 reduction astrometric fit, both to check that our reconstructed copy of the IAD is suitable for orbit-fitting and to validate our astrometric model-fitting implementation. The results of this exercise are shown in Figure \ref{fig:hipparcos-reproduction}, following section 3.1 of \cite{Nielsen:2020a}.\footnote{We note a typo in Equation 10 of \cite{Nielsen:2020a}, which should read: $f = \left( G \sqrt{\frac{2}{9D}} + 1 - \frac{2}{9D} \right)^{\frac{3}{2}}$.}

The close correspondence of our reproduced 5-parameter fit and the 2021 Hipparcos 5-parameter fit makes us confident in our method.

\subsection{Harper Fit Reproduction}\label{sec:harperreproduction}

\begin{figure}
    \centering
    \includegraphics[width=0.3\linewidth]{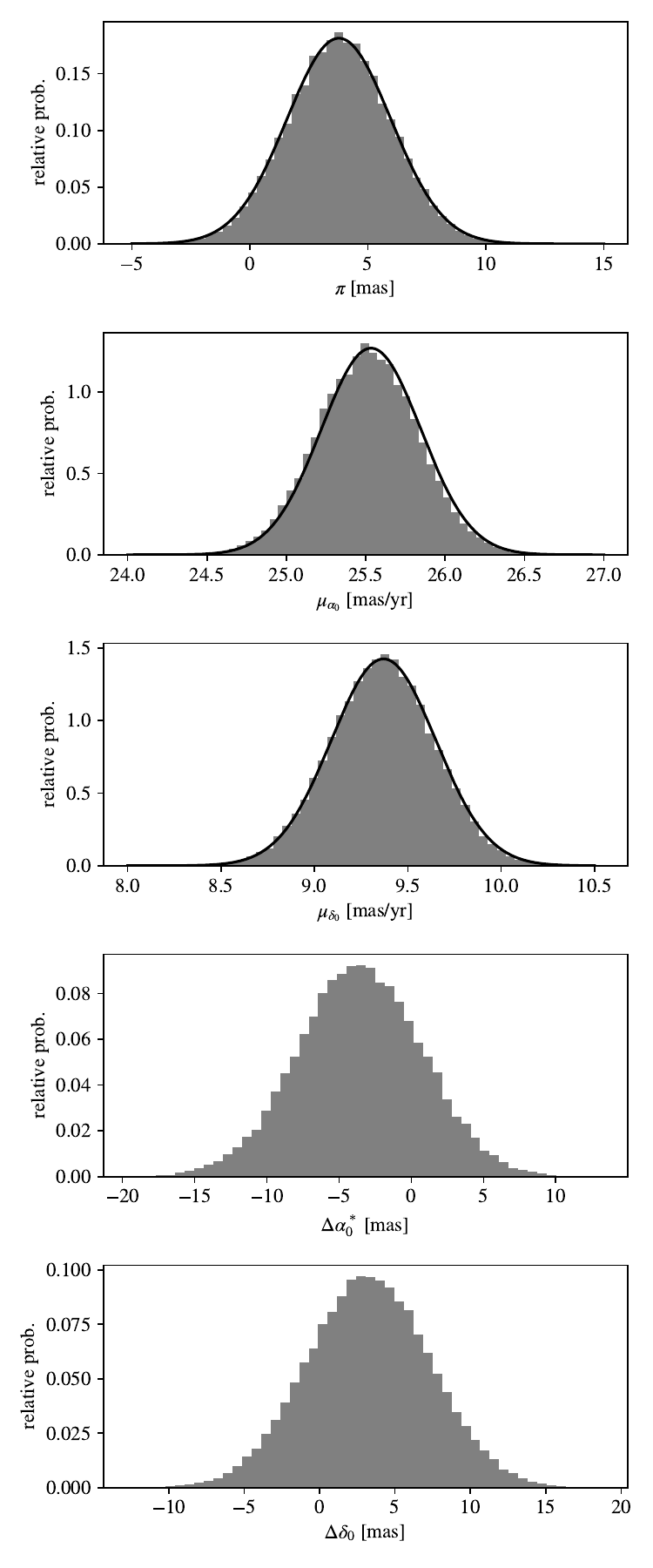}
    \caption{Reproduction of the \cite{2017AJ....154...11H} 5-parameter astrometric fit to only radio astrometric data, with a fixed astrometric jitter value of 2.4 mas added to all observational error bars in quadrature. By reproducing this published fit to only radio data, we validate our implementation of arbitrary absolute astrometry fitting in \texttt{orbitize!}}
    \label{fig:harper-reproduction}
\end{figure}

As an additional check of our implementation, we reproduced a published astrometric fit to only Betelgeuse's radio astrometry. We chose to reproduce the fit displayed in Table 4 of \cite{2017AJ....154...11H}, which added a cosmic jitter term of 2.4 mas in quadrature to all radio epochs, and scaled the resulting astrometric errors by a factor 1.2957671 (G. Harper, private communication) in order to achieve a reduced $\chi^2$ of 1, such that 
\begin{equation}
    \sigma = 1.2957671 \sqrt{\sigma_{\rm obs}^2 + (2.4 \mathrm{mas})^2}
\end{equation}
The results of this test are shown in Figure \ref{fig:harper-reproduction}, which demonstrates that our sampling and implementation of absolute astrometry in {\tt orbitize!} accurately reproduces \citet{2017AJ....154...11H}'s results.

\subsection{Full Standard Fit Corner Plot}

\begin{figure}[p]
    \centering
    \includegraphics[width=\linewidth]{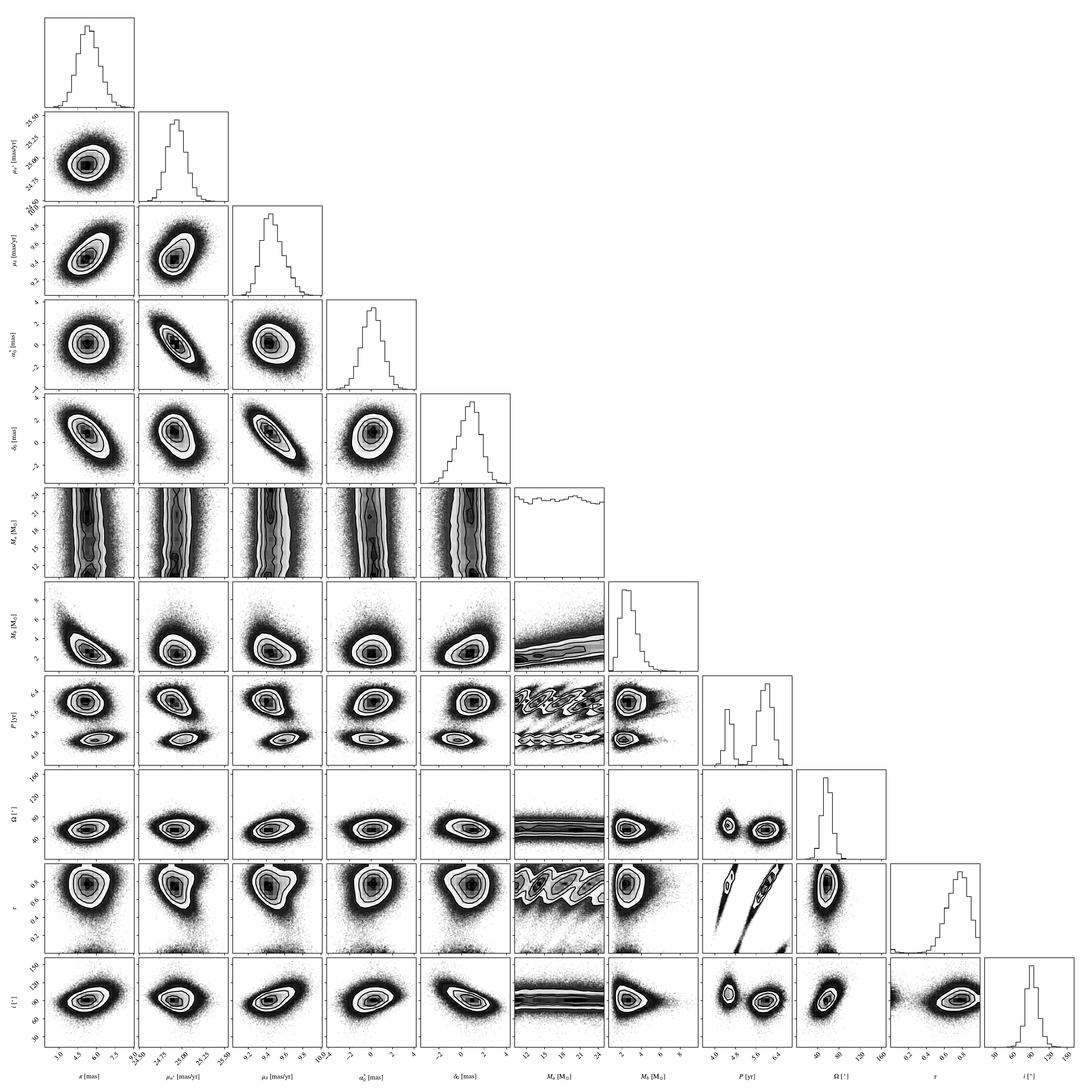}
    \caption{Full corner plot for the astrometric standard fit. The results of this fit are tabulated in Table \ref{tab:fit_params}. }
    \label{fig:corner}
\end{figure}

Figure \ref{fig:corner} shows the full corner plot for the astrometric standard fit, including the reference offsets, parallax, and proper motion in addition to the Keplerian binary model parameters shown in Figure \ref{fig:minimal-corner}. 

\subsection{Method for Astrometric Time Series Plot}
\label{sec:dream-math}

This section describes the construction of the astrometric time series plots (Figure \ref{fig:dreamplot}) in detail. In each plot, the top row shows full astrometric predicitons for 50 random draws from the \texttt{orbitize!} posterior. The middle row shows parallax plus Keplerian predicitons for the same set of 50 posterior samples, and the bottom row shows only Keplerian predictions for the same posterior samples.

\textbf{Overplotting the Hipparcos IAD:} The algorithm for overplotting the Hipparcos IAD is as follows:

\begin{itemize}
    \item Top panel: for a given IAD epoch, compute the tangent point with respect to every model in the posterior. The plotted point is the median of those tangent points, and the thick error bar is the standard deviation of those tangent points. The thin error bar is the projected scan uncertainty: $\epsilon$/$\sin{\phi}$ for R.A.$\cos{\delta_0}$, and $\epsilon$/$\cos{\phi}$ for declination. 
    \item Middle panel: for a given IAD epoch, compute the tangent point with respect to every model in the posterior. Next, compute the proper motion model prediction for each model in the posterior, and subtract them (i.e. for a single posterior orbit, compute the proper motion model prediction, and subtract that from the tangent point for that model. Repeat for every posterior sample). The plotted point is the median of those tangent points, and the thick error bar is their standard deviation. Thin error bars show the same projected observational uncertainties as shown in the top panel.
    \item Bottom panel: for a given IAD epoch, compute the tangent point with respect to every model in the posterior. Next, compute the summed prediction from proper motion and parallax for each model in the posterior, and subtract them. The plotted point is the median of those tangent points, and the thick error bar is their standard deviation. Thin error bars show the same projected observational uncertainties as shown in the top panel.
\end{itemize}

\textbf{Overplotting the Radio Astrometry:} The algorithm for overplotting the radio astrometry is as follows:

\begin{itemize}
    \item Top panel: plot data and error bars as usual.
    \item Middle panel: the plotted points show the median of the data minus the proper motion model prediction. Thick error bars show the standard deviation of this difference. Thin error bars show observational uncertainties. 
    \item Bottom panel: the plotted points show the median of the data minus the model prediction from proper motion and parallax motion.Thick error bars show the std of this difference. Thin error bars show observational uncertainties. 
\end{itemize}


\end{document}